\documentclass[11pt,a4paper]{article}
\usepackage{amsmath}    
\usepackage{amssymb}    
\usepackage{amsfonts}   
\usepackage{mathrsfs}   
\usepackage{dsfont}
\usepackage{euscript}
\usepackage{yfonts}
\usepackage{txfonts}
\usepackage{vmargin}
\usepackage{marvosym}
\usepackage[latin1]{inputenc}
\setmarginsrb{1.8cm}{2cm}{1.8cm}{2cm}{1cm}{1cm}{1cm}{1.6cm}
 \makeatletter
 \@addtoreset{equation}{section}
 \makeatother

\usepackage{tikz}       

\newcommand{\p}{{\parallel}}
\newcommand{\dd}{\mathrm{d}}
\newcommand{\beq}{\begin{equation}}
\newcommand{\enq}{\end{equation}}
\newcommand{\bea}{\begin{eqnarray}}
\newcommand{\eea}{\end{eqnarray}}

\def\br{\begin{remark}\rm\small}
\def\er{\end{remark}}
\def\bt{\begin{theorem}}
\def\et{\end{theorem}}
\def\bcj{\begin{conjecture}}
\def\ecj{\end{conjecture}}
\def\bd{\begin{definition}}
\def\ed{\end{definition}}
\def\bp{\begin{proposition}}
\def\ep{\end{proposition}}
\def\bl{\begin{lemma}}
\def\el{\end{lemma}}
\def\bc{\begin{corollary}}
\def\ec{\end{corollary}}
\def\bh{\begin{hypothesis}}
\def\eh{\end{hypothesis}}

\newtheorem{theorem}{Theorem}[section]
\newtheorem{prop}[theorem]{Proposition}
\newtheorem{cor}[theorem]{Corollary}
\newtheorem{defin}[theorem]{Definition}

\newtheorem{lemma}[theorem]{Lemma}
\newtheorem{hypothesis}{Hypothesis}[section]

\def\ba{\begin{array}}
\def\ea{\end{array}}
\def\det{\operatorname{det}}

\newcommand{\f}[2]{{\ensuremath{%
    \mathchoice%
    {\dfrac{#1}{#2}}
    {\dfrac{#1}{#2}}
    {\frac{#1}{#2}}
    {\frac{#1}{#2}}
}}}
\newcommand{\tf}[2]{\ensuremath{#1/#2}}

\newcommand{\ra}{\rightarrow}

\def\a{\alpha}

\def\be{\beta}

\def\Ga{\Gamma}
\def\de{\delta}

\def\De{\Delta}
\def\eps{\epsilon}

\def\la{\lambda}

\def\sg{\sigma}

\def\Om{\Omega}

\def\vp{\varphi}

\newcommand{\mc}[1]{\ensuremath{\mathcal{#1}}}
\newcommand{\mf}[1]{\ensuremath{\mathfrak{#1}}}
\newcommand{\msc}[1]{\ensuremath{\mathscr{#1}}}

\newcommand{\bs}[1]{\ensuremath{\boldsymbol{#1}}}

\newcommand{\ov}[1]{\ensuremath{\overline{#1}}}
\newcommand{\wt}[1]{\ensuremath{\widetilde{#1}}}

\newcommand{\Int}[2]{\ensuremath{\int\limits_{#1}^{#2}}}
\newcommand{\Oint}[2]{\ensuremath{\oint\limits_{#1}^{#2}}}
\newcommand{\Fint}[2]{\ensuremath{\fint\limits_{#1}^{#2}}}

\newcommand{\sul}[2]{\ensuremath{\sum\limits_{#1}^{#2}}}
\newcommand{\pl}[2]{\ensuremath{\prod\limits_{#1}^{#2}}}

\newcommand{\Cx}{\ensuremath{\mathbb{C}}}

\let\tend=\rightarrow

\newcommand{\Dp}[1]{\ensuremath{\partial_{#1}}}

\newcommand{\intff}[2]{\ensuremath{\left [ \, #1 \,; #2 \, \right ] }}

\newcommand{\intn}[2]{\ensuremath{[\![ \, #1 \,;\, #2 \,]\!]}}

\newcommand{\ex}[1]{\ensuremath{\text{e}^{#1}}}

\newcommand{\norm}[1]{\ensuremath{ || #1 || }}
\newcommand{\Norm}[1]{\ensuremath{ \big|\big| #1 \big|\big| }}

\title{Large-$N$ asymptotic expansion for mean field models with Coulomb gas interaction}
\author{Ga\"etan Borot\footnote{Section de Math\'ematiques, Universit\'e de Gen\`eve, 2-4 rue du Li\`evre, 1211 Gen\`eve 4, Switzerland.}\,\,\footnote{Max Planck Institut f\"ur Mathematik, Vivatsgasse 7, 53111 Bonn, Germany.}\,\,\footnote{Department of Mathematics, MIT, 77 Massachusetts Avenue, Cambridge, MA 02139-4307, USA.} \addtocounter{footnote}{-1}
 \and Alice Guionnet\footnotemark \and Karol K. Kozlowski\footnote{Institut de Math\'ematiques de Bourgogne, CNRS UMR 5584, 9 avenue Alain Savary, BP 47870, 21078 Dijon Cedex, France.}}

\begin{document}

\maketitle

\begin{center}
\textbf{Abstract} \\
\end{center}

We derive the large-$N$, all order asymptotic expansion for a system of $N$ particles with mean-field interactions on top of a Coulomb repulsion at temperature $1/\beta$, under the assumptions that the interactions are analytic, off-critical, and satisfy a local strict convexity assumption.   \\

\vspace{0.5cm}

\section{Introduction}

This article aims at giving a basic framework to study the large-$N$ expansion of the partition function and various observables in the mean-field statistical mechanics of $N$ repulsive particles in $1d$. This is one of the most simple form of interaction between 
particles and constitutes the first case to be fully understood before addressing the problem of more realistic interactions. 

An archetype of such models is provided by random $N \times N$ hermitian matrices, drawn from a measure $\dd M\,e^{-N \mathrm{Tr}\,V(M)}$ \cite{Mehtabook,Loggas}. The corresponding joint distribution of eigenvalues is $$\prod_{i = 1}^N \dd\lambda_i\,e^{-NV(\lambda_i)} \prod_{i < j} |\lambda_i - \lambda_j|^\beta$$ with $\beta = 2$, \textit{ie}. of the form $e^{-E(\lambda)}$ where $E(\lambda)$ includes the energy of a $2d$ Coulomb interaction of eigenvalues, and the effect of an external potential $V$. The large-$N$ behaviour in those models -- and for all values of $\beta > 0$ -- have been intensively studied: they are called \emph{$\beta$-ensembles}. On top of contributions from physics \cite{ACKMe,ACM92,Ake96,E1MM,CE05,CE06,C06}, many rigorous results are available concerning the convergence of the empirical measure when $N$ is large \cite{SaffTotik,Defcours}, large deviation estimates \cite{BookAG}, central limit theorems or their breakdown \cite{Johansson98,Pasturs,Shch2,BG11,BGmulti},
and all-order asymptotic expansion of the partition function and multilinear statistics \cite{APS01,ErcMcL,BG11,Smulticut,BGmulti}. The nature of the expansion depends on the topology of the locus of condensation $S$ of the eigenvalues.  Besides, the asymptotic expansion up to $O(N^{-\infty})$ is fully determined by a universal recursion, called "topological recursion" \cite{E1MM,EORev} (or "topological recursion with nodes" in the multi-cut case \cite{Ecv}), taking as initial data the large-$N$ spectral density and the large-$N$ spectral covariance.

The models we propose to study are generalizations of $\beta$ ensembles, with an arbitrary interaction between eigenvalues (not only pairwise), but assuming pairwise repulsion at short distance approximated by the Coulomb interaction already present in $\beta$ ensembles, see \eqref{law0}. Then, combining tools of complex and functional analysis, we provide techniques showing that the theory for the all-order large-$N$ asymptotic expansion is very similar to the one developed for $\beta$ ensembles. 

In those models, the equilibrium is driven by the balance between Coulomb repulsion and the interactions, whereas the entropy is negligible at leading order. Other types of mean-field models have already been studied in the literature : \cite{Bol86,Bol87} has studied the case where the entropy and smooth interactions balance each other, whereas \cite{Chiyo2000} considered  smooth pairwise interactions only. In both cases, the Coulomb repulsion was absent, and the authors have derived a central limit theorem for fluctuations of linear statistics. The work of \cite{Chiyo2000} represents the infinite temperature case $\beta = 0$ of our model for $r = 2$, and treats only the one-cut regime. For our models, the analysis allows us the derivation of a central limit theorem in the one-cut regime, as well as its analogue -- which includes interference effects -- in the multi-cut regime.

\subsection{The model}
\label{defs1}
\subsubsection{The unconstrained model}

Let $\mathsf{A}\; = \;  \dot \cup_{h=0}^{g} \mathsf{A}_h $ be a closed subset of $\mathbb{R}$ realised as the disjoint union of 
$g$ intervals $\mathsf{A}_h$ -- possibly semi-infinite or infinite. In this paper we focus on the probability measure on 
$\mathsf{A}^N$ defined by:
\beq
\label{law0}
\dd\mu_{\mathsf{A}^N} \; = \;   \frac{1}{Z_{\mathsf{A}^N}}\prod_{i = 1}^N \,\dd\lambda_i
 \prod_{1 \leq i < j \leq N} \big|\lambda_i - \lambda_j\big|^{\beta} \cdot 
 \exp\bigg\{ \frac{N^{2 - r}}{r!} \sum_{1 \leq i_1,\ldots, i_r \leq N}  T(\la_{i_1}, \dots, \la_{i_r} ) \bigg\}\;.
 \enq
We assume $\be>0$ and $Z_{\mathsf{A}^N}$ is the partition function which ensures that 
$\int_{\mathsf{A}^N} \dd\mu_{\mathsf{A}^N} = 1$. The function $T$ represents an $r$-body interaction. 
Without loss of generality, we can assume $T$ to be symmetric in its $r$ variables;
we call it the  \emph{$r$-linear potential}. The scaling in $N$ ensures that it contributes to the same order that the 2-body repulsive Coulomb interaction when $N$ is large. 
A common case is $r = 2$, \textit{ie}. the eigenvalues undergo a pairwise repulsion, which is approximated at short distance by a Coulomb repulsion. The $r$-linear potential can possibly admit a large-$N$ asymptotic expansion of the type
\beq
T(x_1,\dots,x_r) \; = \; \sum_{p \geq 0} N^{-p}\,T^{[p]}(x_1,\dots, x_r) \; ,\nonumber
\enq
where $T^{[p]}$ are symmetric functions on $\mathsf{A}^r$ not depending on $N$. These functions have the same regularity as 
$T$.   

We do stress that the \emph{$r$-linear potential} contains, as a specific example, the case of growing $r$-body interactions, 
namely the substitution
\beq
T(x_1,\dots, x_r)  \; = \; \sul{ J \subseteq \intn{1}{r} }{}   \big( r - |J| \big)! \, T_{|J|} \big( \bs{x}_{J} \big)
\quad \text{with} \quad \left\{  \ba{c}  J=\{j_1,\dots , j_{|J|}\} \\
\bs{x}_{J} = \big(x_{j_1},\dots , x_{j_{|J|}} \big) \ea \right. 
\label{ecriture reduction T a somme pots k corps}
\enq
recast the \emph{$r$-linear} interaction term as 
\beq
 \sul{k=1}{r}\frac{N^{2 - k}}{k!} \sum_{1 \leq i_1,\ldots, i_k \leq N}  T_k(\la_{i_1}, \dots, \la_{i_k} ) \;. \nonumber
\enq
The latter expression has a clear interpretation of a concatenation of $1, 2, \dots, r$ body interactions. In the latter case, it is convenient to include the $2$-body Coulomb repulsion in a total $2$-body interaction:
\beq
T_2^{\mathrm{tot}}(x,y) = \beta\ln|x - y| + T_2(x,y) \; .\nonumber
\enq

For $\beta = 2$, sending $r \rightarrow \infty$ would allow the description a quite general form of a $U(N)$ 
invariant measure on the space of $N \times N$ hermitian matrices. Indeed, for $\beta = 2$, \eqref{law0}  
corresponds to the law of eigenvalues of a $N \times N$ random hermitian matrix $\Lambda$ drawn with (unnormalised) distribution:
\beq
\dd\Lambda\,\exp\bigg\{ \frac{N^{2 - r}}{r!}\,\mathrm{Tr}\,T(\Lambda^{(1)},\ldots,\Lambda^{(r)})\bigg\}\;, \nonumber
\enq
where $\dd\Lambda$ is the Lebesgue measure on the space of Hermitian matrices, and $\Lambda^{(i)}$ is the tensor product of 
$r$ matrices, in which the $i$-th factor is $\Lambda$ and all other factors are identity matrices. In particular, in any model of several random and coupled $N \times N$ hermitian random matrices $\Lambda_1,\ldots,\Lambda_{s}$ which is invariant by simultaneous conjugation of all $\Lambda_i$ by the same unitary matrix, the marginal distribution of $\Lambda_i$ is $U(N)$ invariant, thus of the form \eqref{law0} with, possibly, $r = \infty$ and a different dependence in $N$. Further, for simplicity, we shall restrict ourselves to $r$-linear interactions between eigenvalues with $ r < \infty$. 

\subsubsection{The model with fixed filling fractions}

In the process of studying the unconstrained model in the multi-cut regime, we will need to deal with the so-called
fixed filling fraction model. Let $g \geq 1$, and recall that $\mathsf{A} = \dot{\bigcup}_{h = 0}^{g} \mathsf{A}_{h}$.
Let $N = \sum_{h=0}^{g} N_{h}$ be a partition of $N$ into $g + 1$ integers
and $\mathbf{N} = (N_0,\dots,N_{g})$  a vector built out of the entries of this partition. 
One can associate to such a partition a vector $\bs{\la} \in \mathsf{A}_{\mathbf{N}}\equiv \prod_{h = 0}^{g} \mathsf{A}_{h}^{N_{h}}$
with entries ordered according to the lexicographic order on $\mathbb{N}^2$
\beq
\bs{\la} = \big(  \la_{0,1} , \dots, \la_{0,N_0},\la_{2,1}\dots , \la_{g,1},\dots, \la_{g,N_{g}} \big) \;. \nonumber
\enq
The measure on $\mathsf{A}_{\mathbf{N}}$ associated with this partition reads
\beq
\label{law1}
\dd\mu_{\mathsf{A}_{\mathbf{N}}} \; =  \; \frac{ 1 }{ Z_{\mathsf{A}_{\mathbf{N}}} }
 \prod_{ \bs{i} \in \mc{I}} \Big\{ \mathbf{1}_{\mathsf{A}_{\text{pr}(\bs{i}) }} ( \lambda_{\bs{i}} ) \dd\lambda_{\bs{i}} \Big\} \cdot 
  \prod_{ \bs{i}_1 < \bs{i}_2} \big|\lambda_{\bs{i}_1} \, - \,  \lambda_{\bs{i}_2}\big|^{\beta} 
  \exp\Bigg\{  \frac{N^{2 - r}}{r!} \sum_{\bs{i}_1,\dots, \bs{i_r}\in \mc{I}}
   T\big( \lambda_{ \bs{i}_1 } ,\ldots, \lambda_{ \bs{i}_r } \big) \Bigg\}\;.
\enq
Above, $\bs{i}_k$ are elements of 
\beq
\mc{I} \; = \; \Big\{ (a,N_a) \; : \; a\in \intn{0}{g} \Big\}  \;, \nonumber
\enq
$<$ is the lexicographic order on $\mc{I}$ and $\text{pr}$ is the projection on the first coordinate. Note that the relation between the partition function of the unconstrained model and the fixed filling fraction model is:
\beq
Z_{\mathsf{A}^N} = \sum_{N_0 + \cdots + N_{g}  = N} \frac{N!}{\prod_{h = 0}^{g} N_h!}\,Z_{\mathsf{A}_\mathbf{N}} \; . \nonumber
\enq

\subsubsection{Observables}

In this section, $\mu_{\mc{S}}$ denotes the measure and $Z_{\mc{S} }$ the partition function in any of the two models, \textit{viz}.
$\mc{S}= \mathsf{A}^{N}$ or $\mathsf{A}_{\mathbf{N}}$. The empirical measure is the random probability measure:
\beq
L_N = \frac{1}{N}\sum_{ \bs{i} \in \mc{I}_{\mathcal{S}} }  \delta_{ \lambda_{\bs{i}} }  \qquad 
\text{with} \quad \mc{I}_{ \mathsf{A}^{N} } \;  =  \; \intn{1}{N} \quad \text{and} \quad 
\mc{I}_{\mathsf{A}_{\mathbf{N}} } \; =  \; \mc{I} \; .\nonumber
\enq
We introduce the Stieltjes transform of the $n$-th order moments of the unnormalised empirical measure, called \emph{disconnected correlators}:
\beq
\label{discon}\widetilde{W}_n(x_1,\ldots, x_n) \; = \; N^n\,
\mu_{\mathcal{S}}\Big[\prod_{i = 1}^{n}\Int{}{} \f {\dd L_N(s_i)}{x_i - s_i}\Big]\;. 
\enq
They are holomorphic functions of $x_i \in \mathbb{C}\setminus \mathsf{A}$. When $N$ is large, for reasons related to concentration of measures, it is more convenient to consider the Stieltjes transform of the $n$-th order cumulants of the unnormalised empirical measure, called \emph{correlators}:
\beq
\label{defcow} W_n(x_1,\ldots, x_n) \; = \; 
 \partial_{t_1}\cdots\partial_{t_n} \; \ln Z_{\mc{S}}\big[T \rightarrow \wt{T}_{t_1,\dots, t_n} \big]_{ |_{t_i = 0}}   
\enq
with
\beq
 \wt{T}_{t_1,\dots, t_n}(\xi_1,\dots,\xi_r) \; =\; 
 T(\xi_1,\dots, \xi_r) \; +\;  \f{(r - 1)!}{ N } \,\sum_{i = 1}^n \sul{a=1}{r} \frac{t_i}{x_i - \xi_a} \;,
\label{definition tilde T}
\enq
and we have explicitly insisted on the functional dependence of $Z_{\mc{S}}$ on the $r$-linear potential. 

If $J$ is a set, exactly as in \eqref{ecriture reduction T a somme pots k corps} we denote by $\bs{x}_J$
the $|J|$-dimensional vector whose components are labelled by the elements of $J$. 
The above two types of correlators are related by:
\beq
\widetilde{W}_n(x_1,\ldots,x_n) \; = \;  
\sum_{s = 1}^n \sum_{\substack{ \intn{1}{n}=\\ J_1 \dot{\cup}\cdots\dot{\cup}J_s }} 
\prod_{i = 1}^s W_{|J_i|}( \bs{x}_{J_i} )\;. \nonumber
\enq
Above, the sum runs through all partitions of the set $\intn{1}{n}$ into $s$ non-empty, disjoint sets $J_{\ell}$.

We do stress that the knowledge of the correlators for a smooth family of potentials $\{ T_{t} \}$ indexed by some continuous
variable $t$ determines the partition function up to an integration constant. Indeed, let $\mu_{\mc{S}}^{T_t}$
denote the probability measure in any of the two models and in the presence of the $r$-linear interaction $T_t$. 
Then, one has
\beq
\partial_t \ln Z_{\mathcal{S}} \big[ T \tend T_{t} \big] \; = \; \f{ N^{2} }{ r! }\,
\mu^{T_t}_{\mathcal{S}}\bigg[ \Int{}{} \partial_t T_t(s_1,\ldots,s_r) \prod_{i = 1}^r \dd L_N(s_i) \bigg]\;. \nonumber
\enq
If $\partial_{t}T_{t}$ is analytic in a neighbourhood of $\mathsf{A}^r$, we can rewrite:
\beq
\partial_t \ln Z_{\mathcal{S}}\big[ T \tend T_{t} \big] \; = \; \f{ N^{ 2 -r } }{ r! }\,\oint_{\mathsf{A}^r}
 \partial_t T_{t}(\xi_1,\ldots,\xi_r)
 \,\widetilde{W}_r\big[ T \tend T_{t} \big](\xi_1,\ldots,\xi_r) \prod_{i = 1}^r \frac{\dd\xi_i}{2{\rm i}\pi}\;. \nonumber
\enq
In both cases, the superscript $T_{t}$ denotes the replacement of the $r$-linear potential by the $t$-dependent one $T_{t}$. 

\subsection{Motivation}

In the context of formal integrals, which is accurate for combinatorics, \eqref{law0} describes the generating series of discrete surfaces obtained by gluing along edges discrete surfaces of any topology and with up to $r$ polygonal boundaries. The enumeration of maps carrying any of the classical statistical physics models: self-avoiding loop configurations \cite{K89}, spanning forests \cite{Sporti}, the Potts model \cite{KazPotts,BE99,ZinnPotts}, the Ising model \cite{KazakovO1}, the $6$-vertex model \cite{Kos6}, \ldots{} arise as special cases. There exists by construction a $1/N$ expansion, and it was shown in \cite{BEO} for $r = 2$, and \cite{Bstuff} for arbitrary $r$ that the large-$N$ expansion of the correlators are given by the topological recursion "with initial conditions".

In the context of convergent integrals, the present article gives conditions under which the existence of the large-$N$ expansion can be established. For $\beta = 2$, combined with the results of \cite{BEO} for $r = 2$ and \cite{Bstuff} for arbitrary $r$, it shows that the large-$N$ expansion is governed in the one-cut regime or in the multi-cut regime with fixed filling fractions by the topological recursion. When $\beta \neq 2$, the answer should also be given by the $\beta$-topological recursion of \cite{CE06}, although this has only been demonstrated in the case $r = 1$. The universality of this recursion goes therefore far beyond the case $r = 1$ where it was discovered \cite{E1MM}. Given this result, the heuristics of \cite{Ecv} in the multi-cut case are obviously valid for arbitrary $r$, and we justify them rigorously here, following the strategy of \cite{BGmulti} for $\beta$ ensembles ($r = 1$).  Let us also mention that a certain class of $r = 2$ models have been studied from the point of view of local universality in \cite{Venker1}.

The models we are studying are encountered for instance in $3$-dimensional topology: the computation of torus knot invariants and of the Chern-Simons partition function in certain Seifert manifolds is of the form \eqref{law0} for $r = 2$ \cite{MarinoCSM},  and it was claimed in \cite{GarMar} that the $\mathrm{SU}(N)$ Chern-Simons partition function of $3$-manifolds obtained by filling of a knot in $\mathbb{S}_3$ should be described by \eqref{law0} for $r = \infty$. For related reasons, \eqref{law0} with $r = 2$ is also relevant in topological strings and supersymmetric gauge theories, see \textit{e.g.} \cite{Sulko}, although our results would have to be generalised to complex-valued $T$ in order to be applied to such problems. Here are a few $r = 2$ examples to which our methods apply:
\begin{itemize}
\item[$\bullet$] With some natural assumptions on the function $f$:
$$ T_2^{\mathrm{tot}}(x,y) = \beta\Big\{\ln |x - y| + \ln |f(x) - f(y)|\Big\} \;.$$
For $\beta = 2$, the corresponding random matrix model is determinantal. For $f(x) = x^{\theta}$, these are the biorthogonal ensemble studied in \cite{Muttalib,Borodinbi}. 
\item[$\bullet$] The sinh interactions: for $\psi(x) = \mathrm{sinh}(x/2)$, 
$$T_2^{\mathrm{tot}}(x,y) = \beta\,\ln\big|\psi(x - y)\big|\;.$$
\item[$\bullet$] The $U(N)$ Chern-Simons partition function spherical and euclidean Seifert spaces: for a $d$-uple of positive integers $(p_1,\ldots,p_d)$ so that $2 - d + \sum_{i = 1}^d 1/p_i \geq 0$,
$$
T_2^{\mathrm{tot}}(x,y) = (2 - d)\ln\big|\psi(x - y)\big| + \sum_{i = 1}^d \ln\big|\psi[(x - y)/p_i]\big|\;.$$
\item[$\bullet$] The $q$-deformed interactions, non-compact case: for $|q| < 1$,
$$ T_2^{\mathrm{tot}}(x,y) = (e^{x - y},q)_{\infty}(e^{y - x},q)_{\infty},\qquad (z,q)_{\infty} = \prod_{k = 0}^{\infty}(1 - zq^{k})$$
\item[$\bullet$] The $(q,t)$-deformed interactions, compact case: for $|q| < 1$, $t \in ]-1,1[$ and $\mathsf{A}\subseteq [0,1]$,
$$ T_2^{\mathrm{tot}}(x,y) = \frac{(e^{2{\rm i}\pi(x - y)},q)_{\infty}\,(e^{2{\rm i}\pi(y - x)},q)_{\infty}}{(t\,e^{2{\rm i}\pi (x - y)},q)_{\infty}\,(t\,e^{2{\rm i}\pi (y - x)},q)_{\infty}} $$
\item[$\bullet$] The $O(n)$ model: for $|n| \leq 2$ and $\mathsf{A} \subseteq \mathbb{R}_+$,
$$T_2^{\mathrm{tot}}(x,y) = \beta\Big\{2\ln|x - y| - n\ln(x + y)\Big\}\;.$$
The previous examples were related to $A_{N - 1}$ root systems. The analogue of the simplest Coulomb interaction 
for $BC_N$ root systems is the $O(-2)$ model, and the sinh, $q$ and $(q,t)$ cases also have a natural $O(n)$ deformation. They are relevant in $\mathrm{SO}$ or $\mathrm{Sp}(N)$ Chern-Simons theory, but also in condensed matter. For instance, the two-body interaction:
$$
T_2^{\mathrm{tot}}(x,y) = (\beta/2)\Big\{\ln|x - y| + \ln\big|\psi(x - y)\big| + \ln|x + y| + \ln\big|\psi(x + y)\big|\Big\}
$$
has been shown to occur between transmission eigenvalues in metallic wires with disorder \cite{Beena}.
\end{itemize}
All of the above examples do satisfy the strict convexity property (\textit{cf}. Hypothesis~\ref{h2} for its details) and thus fall into the scope of our method. As a matter of fact, one can define $(q,t)$-deformed interactions associated to any pair of root systems, and they intervene in the orthogonality measures of Macdonald polynomials \cite{MacDoqt}.

\subsection{Main result}
\label{gammaapp}

Our main result is Theorem~\ref{T1} is an all order expansion for the partition function of our model  that we emphasise below. This in particular  allows the study of fluctuations of linear statistics in \S~\ref{fluctulin}.

\begin{theorem}
\label{T1bis}Assume Hypothesis~\ref{h1}, $T$ holomorphic in a neighbourhood of $\mathsf{A}^r$, and $\mu_{{\rm eq}}$ in the $(g + 1)$-cut regime and off-critical. Then the partition function function in the $\mathsf{A}^{N}$
model admits the asymptotic expansion:
\bea
Z_{\mathsf{A}^{N}} & = & N^{(\beta/2)N + \gamma}\,
\exp\Bigg(\sum_{k \geq  -2}^{} N^{-k}\,F^{[k]}_{\bs{\epsilon}^{\star}} \Bigg) \nonumber \\
\label{theofr}& & \times \Bigg\{\sum_{m \geq 0} \sum_{\substack{\ell_1,\ldots,\ell_m \geq 1 \\ k_1,\ldots,k_m \geq -2 \\ \sum_{i = 1}^m \ell_i + k_i  > 0}} \frac{N^{-\sum_{i = 1}^{m} (\ell_i + k_i)}}{m!}\Bigg(\bigotimes_{i = 1}^m \frac{F_{\bs{\epsilon}^{\star}}^{[k_i],(\ell_i)}}{\ell_i!}\Bigg)\cdot \nabla_{\bs{v}}^{\otimes (\sum_{i = 1}^m \ell_i)}\Bigg\}\Theta_{-N\bs{\epsilon}^{\star}}\Big( F_{\bs{\epsilon}^{\star}}^{[-1],(1)}\,\Big|\,F_{\bs{\epsilon}^{\star}}^{[-2],(2)}\Big)\;.
\eea
\end{theorem}
The various terms appearing in Theorem~\ref{T1bis} will be defined in Sections~\ref{S7}-\ref{S8}.
Here, we briefly comment on the structure of the asymptotic expansion. $F^{[k],(\ell)} \in (\mathbb{R}^{g})^{\otimes \ell}$ are tensors independent of $N$. $\Theta_{\bs{\nu}}(\bs{w}|\bs{T})$ is the Siegel theta function depending on a $g$-dimensional vector $\bs{w}$ and $\bs{T}$ is a definite positive quadratic form in $\mathbb{R}^g$. $\nabla$ is the gradient operator acting on the variable $\bs{w}$. This Theta function is $\mathbb{Z}^{g}$-periodic function of the vector $\bs{\nu}$. Since it is evaluated to $\bs{\nu} = -N\bs{\epsilon}^{\star}$ in \eqref{theofr}, the partition function enjoys a pseudo-periodic behaviour in $N$ at each order in $1/N$. We mention that the definite positive quadratic form in the Theta function is evaluated at
\beq
\bs{T} = F_{\bs{\epsilon}^{\star}}^{[-2],(2)} = -\mathrm{Hessian}_{\bs{\epsilon} = \bs{\epsilon}^{\star}}\,\mathcal{E}[\mu_{{\rm eq}}^{\bs{\epsilon}}]\;. \nonumber
\enq
The exponent $\gamma = \sum_{h = 0}^{g}\gamma_h$ only depends on $\beta$ and the nature of the edges, it was already determined in \cite{BGmulti}:
\begin{itemize}
\item[$\bullet$] $\gamma_h = \frac{3 + \beta/2 + 2/\beta}{12}$ if the component $\mathsf{S}_h$ of the support has two soft edges ;
\item[$\bullet$] $\gamma_h = \frac{\beta/2 + 2/\beta}{6}$ if it has one soft edge and one hard edge ;
\item[$\bullet$] $\gamma_h = \frac{-1 + \beta/2 + 2/\beta}{4}$ if it has two hard edges.
\end{itemize}
Note that, in the $1$-cut regime ($g = 0$), the Theta function is absent and we retrieve a $1/N$ expansion (established in Corollary~\ref{cc1}).

\subsection{Method and outline}

We stress that in general \eqref{law0} is not an exactly solvable model even for $\beta = 2$ -- with the exceptional of the aforementioned biorthogonal ensembles -- so the powerful techniques of orthogonal polynomials and integrable systems cannot be used. In principle, at $\be=2$ one could analyse the integral within the 
method developed in \cite{Koz}. For such a purpose, one should first carry out the Riemann--Hilbert analysis of a general multiple integral with $T = T_1$ (\textit{cf}. \eqref{ecriture reduction T a somme pots k corps}) and then implement the multideformation procedure developed in \cite{Koz}. Here, we rather rely on a priori concentration of measures properties, and the analysis of the Schwinger-Dyson equations of the model
what allows us, in particular, to treat uniformly the case of general $\be$.

In Section~\ref{S2}, we establish the convergence of the empirical measure 
$L_N = \frac{ 1 }{N}\sum_{ \bs{i} \in \mc{I}_{\mc{S}} } \delta_{\lambda_{\bs{i}}} $ in the unconstrained model ($\mc{S}= \mathsf{A}^N$), and in the model with fixed filling fractions ($\mc{S}= \mathsf{A}_{\mathbf{N}}$), to an equilibrium measure $\mu_{\mathrm{eq}}$. We study the properties of $\mu_{\mathrm{eq}}$ in \S~\ref{SDeasy}, showing that the results of regularity of $\mu_{\mathrm{eq}}$ compared to the Lebesgue measure, and squareroot behaviors at the edges -- which are well-known for pure Coulomb two-body repulsion -- continue to hold in the general setting. We give a large deviation principle in \S~\ref{Lares} allowing, as a particular case, a restriction to $\mathsf{A}$ compact.

We prove in Section~\ref{S3} the concentration of the empirical measure around $\mu_{\mathrm{eq}}$, modulo the existence of an adapted functional space $\mathcal{H}$. Such adapted spaces are constructed in Section~\ref{S4}, thanks to the existence of the inverse of a linear operator $\mathcal{T}$. This inverse is explicitly constructed in Appendix~\ref{App2} by invoking 
functional analysis arguments. All these handlings lead to rough a priori bounds on the correlators. In Section~\ref{S5}, we improve those bounds by a bootstrap method using the Schwinger-Dyson equations in the fixed filling fraction model, and obtain the asymptotic expansion of the correlators in this model. The bootstrap method is based on the existence of a continuous inverse for a linear operator $\mathcal{K}$, which relies on basic results of Fredholm theory reminded in Appendix~\ref{App1}. In Section~\ref{S6}, we deduce the asymptotic expansion of the partition function in the fixed filling fraction model by performing an interpolation to a model with $r = 1$, for which we can use the result of \cite{BGmulti} relating the partition function to asymptotics of Selberg $\beta$ integrals, again by interpolation. In the one-cut regime, this concludes the proof. In the multi-cut regime, we prove that the coefficients of expansion depend smoothly on the filling fractions. This allows in Section~\ref{S7} to establish the asymptotic expansion of the partition function for the unconstrained model in the multi-cut regime, and to study the convergence in law of fluctuations of linear statistics in Section~\ref{fluctulin}.

\subsection{Notations and basic facts}
\label{not}

\subsubsection*{Functional analysis}

\begin{itemize}

\item[$\bullet$] $L^p(\mathsf{X})$ is the space of real-valued mesurable functions $\varphi$ on $\mathsf{X}$ such that $|\varphi|^p$ is integrable. Unless specified otherwise, the space $\mathsf{X}$ will be endowed with its canonical measure (Lebesgue measure for a subset of $\mathbb{R}$, curvilinear measure for a Jordan curve, etc.).
\item[$\bullet$] $\mathcal{F}$ denotes the Fourier transform which, defined on $L^1(\mathsf{A})$, reads 
$\mc{F}[\vp](\bs{k}) \; = \; \int_{\mathsf{A}} \ex{{\rm i} \bs{k}\cdot \bs{x}} \vp(\bs{x})\,\dd^{r}\bs{x}.$
%
%
%
%
%
%
\item[$\bullet$] $H^s(\mathbb{R}^r)$ is the Sobolev space of functions $\varphi \in L^2(\mathbb{R}^r)$ such that:
$$
\p \varphi \p_{H^{s}}\,\,=\,\,\Int{\mathbb{R}^r}{} \big|\mathcal{F}[\varphi](\mathbf{k})\big|^2 
 \Big( 1  + \sum_{i = 1}^r k_i^2 \Big)^s\dd^{r}\mathbf{k}  \; < \; +\infty \; .
$$
\item[$\bullet$] More generally, $W^{1;p}(\mathsf{A})$ denotes the space of measurable functions $\vp$ on $\mathsf{A}$ such that 
$$
\norm{ \vp }_p \; = \; \norm{\vp}_{ L^{p}(\mathsf{A}) } \; + \; \norm{ \vp^{\prime} }_{ L^{p}(\mathsf{A}) } \; < \; + \infty   \;. 
$$
\item[$\bullet$] If $b > 0$ and $f$ is a real-valued function defined on a subset $\mathsf{X}$ of a normed vector space, we agree upon:
$$
\kappa_{b}[\varphi] = \sup_{x,y \in \mathsf{X}} \frac{|\varphi(x) - \varphi(y)|}{|x - y|^{b}} \in [0,+\infty]\;.
$$
In the case when $\mathsf{X} \subset \Cx^p$, $|\cdot |$ will stand for the sup-norm. The space of $b$-H\"older functions corresponds to:
$$\mathrm{Ho}_{b}(\mathsf{X}) = \big\{\varphi \in \mathcal{C}^0(\mathsf{X}),\quad \kappa_{b}[\varphi] < +\infty\big\}\;.$$
and the space of Lipschitz functions to $\mathrm{Ho}_{1}(\mathsf{X})$.
\end{itemize}

\subsubsection*{Complex analysis}

\begin{itemize}
\item[$\bullet$] If $\mathsf{A}$ is a compact of $\mathbb{R}$ and $m \geq 1$, $\mathscr{H}^{m}(\mathsf{A})$ denotes the space of holomorphic functions $f$ in $\mathbb{C}\setminus\mathsf{A}$, so that $f(x) \in O(1/x^{m})$ when $x \rightarrow \infty$. If $f$ is a function in $\mathbb{C}\setminus\mathsf{A}$, we denote $f\cdot \mathscr{H}^m(\mathsf{A}) = \{f\cdot \varphi,\quad \varphi \in \mathscr{H}^m(\mathsf{A})\}$.
\item[$\bullet$] We can define similarly a space $\mathscr{H}^m(\mathsf{A},r)$ for functions of $r$ variables. 
In that case, the asymptotics in each variables take the form $f(x_1,\dots,x_r) \in O(1/x_p^{m})$, with a $O$
that is uniform with respect to the other variables satisfying $d(x_k, A) > \eta$, for some $\eta>0$. 
\item[$\bullet$] If $\Gamma$ is a Jordan curve (hereafter called contour) surrounding $\mathsf{A}$ in $\mathbb{C}\setminus\mathsf{A}$, we denote $\mathrm{Ext}(\Gamma) \subseteq 
\mathbb{C}\setminus\mathsf{A}$ the unbounded connected component of $\mathbb{C}\setminus\Gamma$, and $\mathrm{Int}(\Gamma)$ the other connected 
component. If $\Gamma$ and $\Gamma'$ are two contours, we say that $\Gamma'$ is exterior to $\Gamma$ if $\Gamma' \subseteq \mathrm{Ext}(\Gamma)$ and we denote $ \Ga \subset \Ga^{\prime}$. We denote $\Gamma[1]$ an arbitrary contour in $\mathbb{C}\setminus\mathsf{A}$ exterior to $\Gamma$, and more generally $(\Gamma[i])_{i \geq 0}$ with $\Gamma[0] = \Gamma$ an arbitrary sequence of contours in $\mathbb{C}\setminus\mathsf{A}$ so that $\Gamma[i + 1]$ is exterior to $\Gamma[i]$. $\Gamma[-1]$ denotes a contour interior to $\Gamma$, etc.

\item[$\bullet$] We can equip $\mathscr{H}^m(\mathsf{A})$ (resp. $\mathscr{H}^m(\mathsf{A},r)$) with the norm:
$$
\p \varphi \p_{\Gamma} = \sup_{x \in \Gamma} |\varphi(x)| = \sup_{x \in \mathrm{Ext}(\Gamma)} |\varphi(x)|
\qquad \big( \text{resp.}  \; \;  
\p \varphi \p_{\Gamma^r} = \sup_{x \in \Gamma^r} |\varphi(x)|  \big)\;.
$$
%
%
%
%
%
%
%
%
%
%
%
\item[$\bullet$] Given a contour $\Ga$ in $\Cx$ and a holomorphic function $f$ on $\Cx\setminus\Ga$, we denote
by $f_{\pm}$ its boundary values (if they exist) when a point $z \in \Cx\setminus\Ga$  approaches a point $x \in \Ga$
from the $+$ (\textit{ie}. left) side  or $-$ (\textit{ie}. right) side of $\Ga$ and non-tangentially to $\Ga$. The convergences of 
$f(z)$ to $f_{\pm}(x)$ will be given in terms of a norm ($L^p, \mc{C}^{0},\ldots$) appropriate to the nature of $f_{\pm}$. 

\subsubsection*{Probability}

\begin{itemize}
\item[$\bullet$] $\mathbf{1}_{\mathsf{X}}$ denotes the indicator function of a set $\mathsf{X}$.
\item[$\bullet$] $\mathcal{M}^1(\mathsf{A})$ denotes the space of probability measures on $\mathsf{A}$. 
$\mathcal{M}^0(\mathsf{A})$ denotes the set of differences of finite positive measure with same mass.
\item[$\bullet$] $\mathcal{C}^0_b(\mathsf{A})$ denotes the space of bounded continuous functions on $\mathsf{A}$. $\mathcal{M}^{1}(\mathsf{A})$ and $\mathcal{M}^0(\mathsf{A})$ are endowed with the weak-* topology, which means that:
$$
\lim \mu_{n} = \mu_{\infty}\qquad \Longleftrightarrow \qquad \forall f \in \mathcal{C}_{b}^0(\mathsf{A}),\quad\lim_{n \rightarrow \infty} \Bigg(\Int{\mathsf{A}}{} f(x)\,\dd\mu_{n}(x)\Bigg) = \Int{\mathsf{A}}{} f(x)\,\dd\mu_{\infty}(x)\;.
$$
If $\mathsf{A}$ is compact, Prokhorov theorem ensures that $\mathcal{M}^1(\mathsf{A})$ is compact for this topology.
\item[$\bullet$] If $\nu \in \mathcal{M}^0(\mathsf{A})$, the Vasershtein norm is defined as:
$$
\p \nu \p = \sup_{\substack{\varphi \in \mathrm{Ho}_{1}(\mathsf{A}) \\ \kappa_1[\varphi] \leq 1}} \Big|\Int{\mathsf{A}}{} \varphi(x)\dd\nu(x) \Big|
$$
\item[$\bullet$] Given the representation as a disjoint union 
$\mathsf{A} =  \dot \cup_{h=0}^{g} \mathsf{A}_h $
and $\bs{\epsilon} = \big(\eps_0,\dots, \eps_{g} \big)$ a $g+1$ dimensional vector with entries consisting of non-negative real numbers summing up to $1$, we denote $\mathcal{M}^{ \bs{\epsilon} }(\mathbf{A})$ the set of probability measures $\mu$ on $\mathsf{A}$ such that $\mu[\mathsf{A}_h] = \epsilon_h$, $h=0,\dots, g$. 
We recall that $\mathcal{M}^{ \bs{\epsilon} }(\mathbf{A})$ is a closed, convex subset of $\mathcal{M}^1(\mathsf{A})$.
\item[$\bullet$] If $\mathsf{X}$ is a union of segments or a Jordan curve, $\ell(\mathsf{X})$ denotes its length.

\item[$\bullet$] The notation $O(N^{-\infty})$ stands for $O(N^{-k})$ for any $k \geq 0$.

\item[$\bullet$] $c,C$ denote constants whose values may change from line to line.

\end{itemize}

\end{itemize}

\section{The equilibrium measure}
\label{S2}

In this section, we assume:
\begin{hypothesis}
\label{h1}
\begin{itemize}
\item[$\phantom{\bullet}$] 
\item[$\bullet$] (Regularity) $T \in \mathcal{C}^0(\mathsf{A}^r)$.
\item[$\bullet$] (Confinement) If $\pm \infty \in \mathsf{A}$, we assume the existence of a function $f$ so that $T(x_1,\ldots,x_r) \leq - (r - 1)! \sum_{i = 1}^r f(x_i)$, when $|\bs{x}|$ is large enough, and:
$$
\liminf_{x \rightarrow \pm \infty} \frac{f(x)}{\beta\ln |x|} > 1\;.
$$
\item[$\bullet$] In the fixed filling fraction model, let $\bs{\eps}=(\eps_0,\dots, \epsilon_g) \in [0,1]^{g+1}$ be such that 
$\sum_{h = 0}^g \epsilon_h = 1$, and $\bs{N}=(N_0,\dots,N_g)$ be a vector of integers whose components depend on $N$  and satisfy the constraint $\sum_{h = 0}^g N_h = N$ and $N_h/N  \rightarrow \epsilon_h$.
\item[$\bullet$] (Uniqueness of the minimum) The energy functional $\mathcal{E}$ has a unique global minimum on $\mathcal{M}^1(\mathsf{A})$ (in the unconstrained model), or on $\mathcal{M}^{\bs{\epsilon}}(\mathbf{A})$ (in the fixed filling fraction model).
\end{itemize}
\end{hypothesis}

\subsection{Energy functional}
\label{energf}

We would like to consider the energy functional:
\beq
\label{defE}\mathcal{E}[\mu] = - \Int{\mathsf{A}^r}{} \Big(\frac{T(x_1,\ldots,x_r)}{r!} + \frac{\beta}{r(r - 1)}\sum_{1 \leq i \neq j \leq r} \ln|x_i - x_j|  \Big)\prod_{i = 1}^r \dd\mu(x_i)\;.
\enq
Because of the singularity of the logarithm, $\mathcal{E}$ assumes value in $\mathbb{R}\cup\{+\infty\}$, and it is well-known that $\mathcal{E}$ is lower semi-continuous. Let us introduce the level sets:
\beq
E_{M} = \Big\{\mu \in \mathcal{M}^1(\mathsf{A}),\quad \mathcal{E}[\mu] \leq M\Big\}\;,\qquad 
E_{<\infty} = \Big\{\mu \in \mathcal{M}^1(\mathsf{A}),\quad \mathcal{E}[\mu] < \infty\Big\}\;. \nonumber
\enq
We know that $E_{<\infty}$ is not empty. Let $\mathcal{M}'$ be a closed subset of $\mathcal{M}^1(\mathsf{A})$ which intersect $E_{<\infty}$. By standard arguments \cite{SaffTotik,Defcours,BAG}, $\mathcal{E}$ has compact level sets $E_M$ in $\mathcal{M}'$, has a minimising measure $\mu^*$ on $\mathcal{M}'$, and $\mathcal{E}[\mu^*]$ is finite. $\mathcal{M}'$ can be either $\mathcal{M}^1(\mathsf{A})$ or $\mathcal{M}^{\bs{\epsilon}}(\mathbf{A})$, and Hypothesis~\ref{h1} guarantees in either case that $\mu^*$ is unique. Exactly as in \cite{BAG} (see also \cite{BookAG}), we can prove the following large deviation principle:
\begin{theorem} \label{ldl}Assume Hypothesis~\ref{h1}. Then the law of 
 $L_N = \frac{1}{N}\sum_{i \in \mc{I}_{\mathcal{S} }} \delta_{\lambda_i}$ under the probability measure  \eqref{law0} (resp. 
 \eqref{law1})
 satisfies a large deviation principle  on $\mathcal M^1(\mathsf{A})$ (resp. $\mathcal{M}^{\bs{\epsilon}}(\mathbf{A})$) with speed $N^2$ and 
 good rate function $\mathcal E-\inf_{\mathcal M^1(\mathsf{A})}\mathcal E$ (resp. $\mathcal E-\inf_{\mathcal M^{\epsilon}(\mathbf{A})}\mathcal E$) .
\end{theorem}

\subsection{Convergence of the empirical measure}
As a consequence of the previous large deviation principle, we can state the following convergence;
\begin{theorem}
\label{tc0}Assume Hypothesis~\ref{h1} in the unconstrained model, \textit{ie}. $\mathcal{S} = \mathsf{A}^N$. When $N \rightarrow \infty$, $L_N = \frac{1}{N}\sum_{i = 1}^N \delta_{\lambda_i}$ under the law $\mu_{\mathsf{A}^N}$ converges almost surely and in expectation to the unique minimiser of $\mu_{\mathrm{\rm eq }}$ of $\mathcal{E}$ on $\mathcal{M}^1(\mathsf{A})$. $\mu_{\mathrm{\rm eq }}$ has a compact support, denoted $\mathsf{S}$. It is characterised by the existence of a constant $C$ such that:
\beq
\label{juje}\forall x \in \mathsf{A},\qquad \beta \Int{\mathsf{A}}{} \ln|x - \xi|\,\dd\mu_{\mathrm{\rm eq }}(\xi) \; + \;  
\Int{\mathsf{A}^{r - 1}}{} \frac{T(x,\xi_2,\ldots,\xi_r)}{(r - 1)!}\,\prod_{i = 2}^r \dd\mu_{\mathrm{\rm eq }}(\xi_i) \leq C\;,
\enq
with equality $\mu_{\mathrm{\rm eq }}$-almost surely.
\end{theorem}

\begin{theorem}
\label{tc1} Assume Hypothesis~\ref{h1} in the model with fixed filling fractions, \textit{ie}. $\mathcal{S} = \mathsf{A}_{\mathbf{N}}$. Then, $L_N = \frac{1}{N}\sum_{\bs{i} \in \mathcal{I}}^N \delta_{\lambda_{\bs{i}}}$ under the law $\mathcal{\mu}_{A_N}$ converges almost surely and in expectation to the unique minimiser $\mu_{\mathrm{eq}}$ of $\mathcal{E}$ on $\mathcal{M}^{\bs{\epsilon}}(\mathbf{A})$. $\mu_{\mathrm{\rm eq }}$ has a compact support, denoted by $\mathsf{S}$. It is characterised by the existence of constants $C_h^{\bs{\epsilon}}$ such that:
\beq\label{juje2}
\forall h \in \intn{0}{g},\qquad \forall x \in \mathsf{A}_h,\qquad \beta \Int{\mathsf{A}}{} \ln|x - \xi|\,\dd\mu_{\mathrm{\rm eq }}(\xi) + \Int{\mathsf{A}^{r - 1}}{} \frac{T(x,\xi_2,\ldots,\xi_r)}{(r - 1)!}\,\prod_{i = 2}^r \dd\mu_{\mathrm{\rm eq }}(\xi_i) \leq C_{h}^{\bs{\epsilon}}\;,
\enq
with equality $\mu_{\mathrm{\rm eq }}$-almost surely.
\end{theorem} 

In either of the two models, we define the \emph{effective potential} as:
\beq
\label{Teff} T_{\mathrm{eff}}(x) = \beta \Int{\mathsf{A}}{} \ln|x - \xi|\,\dd\mu_{\mathrm{\rm eq }}(\xi) 
\; +  \; \Int{\mathsf{A}^{r - 1}}{} \frac{T(x,\xi_2,\ldots,\xi_r)}{(r - 1)!}\,\prod_{i = 2}^r \dd\mu_{\mathrm{\rm eq }}(\xi_i) - \left\{\begin{array}{l} C \\ \mathbf{1}_{\mathsf{A}_h}(x)C_h^{\bs{\eps}} \end{array}\right. .
\enq
if $x \in \mathsf{A}$, and $T_{\mathrm{eff}}(x) = -\infty$ otherwise. It is thus non-positive and vanishes $\mu_{\mathrm{\rm eq }}$-almost surely.

\vspace{2mm}

We will wait until Section~\ref{S711} and Proposition~\ref{smoothen} to establish that if $\mathcal{E}$ has a unique global minimum on $\mathcal{M}^{1}(\mathsf{A})$, and if we denote $\epsilon_h^{\star} = \mu_{{\rm eq}}[\mathsf{A}_h]$, then for $\bs{\epsilon}$ close enough to $\bs{\epsilon}^{\star}$, $\mathcal{E}$ has a unique minimiser over $\mathcal{M}^{\bs{\epsilon}}(\mathbf{A})$. In other words, Hypothesis~\ref{h1} for the unconstrained model will imply Hypothesis~\ref{h1} for the model with fixed filling fractions close to $\bs{\epsilon}^{\star}$. Although the full Proposition~\ref{smoothen} is stated for $T$ holomorphic, the aforementioned statement is valid under weaker regularity, e.g. $T \in \mathcal{C}^m(\mathsf{A}^r)$ with $m > \min(3,2r)$.

\subsection{Regularity of the equilibrium measure}
\label{SDeasy}

In this section, we shall be more precise about the regularity of equilibrium measures, using the first Schwinger-Dyson equation.

\begin{lemma}
\label{thregup} Assume Hypothesis~\ref{h1} and $T \in \mathcal{C}^m(\mathsf{A})$ with $m \geq 2$. Then, $\mu_{\mathrm{eq}}$ has a $\mathcal{C}^{m - 2}$ density on $\mathring{\mathsf{S}}$. Let $\alpha \in \partial\mathsf{S}$.
\begin{itemize}
\item[$(i)$] If $\alpha \in \partial\mathsf{A}$ (hard edge), then 
$ \f{ \dd \mu_{\mathrm{\rm eq }} }{  \dd x }(x) \in O\big(|x - \alpha|^{-1/2}\big)$ when $x \rightarrow \alpha$.
\item[$(ii)$] If $\alpha \notin \partial\mathsf{A}$ (soft edge) and $T \in \mathcal{C}^3(\mathsf{A}^r)$, then 
$\f{ \dd \mu_{\mathrm{\rm eq }} }{  \dd x }(x) \in O\big(|x - \alpha|^{1/2}\big)$ when $x \rightarrow \alpha$.
\end{itemize}
\end{lemma}

\begin{lemma}
\label{thregana}Assume Hypothesis~\ref{h1} and $T$ holomorphic in a neighbourhood of $\mathsf{A}$ in $\mathbb{C}$. Then, $\mathsf{S}$ is a finite union of segments which are not reduced to a point, and the equilibrium measure takes the form:
\beq
 \dd \mu_{\mathrm{\rm eq }} (x) \; =  \;
\label{slus}\frac{\mathbf{1}_{\mathsf{S}}(x)\,\dd x}{2\pi}\,M(x)\,\sigma_0(x)\,\prod_{\alpha \in \partial\mathsf{S}\setminus\partial\mathsf{A}} |x - \alpha |^{1/2} 
\prod_{\alpha \in \partial\mathsf{S}\cap\partial\mathsf{A}} |x - \alpha |^{-1/2}\;, 
\enq
where $M$ is holomorphic and positive (a fortiori nowhere vanishing) on $\mathsf{A}$, and $\sigma_0(x)$ is a polynomial assuming non-negative values on $\mathsf{S}$.
\end{lemma}

\noindent \textbf{Proof of Lemma~\ref{thregup}.} As soon as $T \in \mathcal{C}^1(\mathsf{A}^r)$, we can derive a Schwinger-Dyson for the model $\mu_{\mathcal{S}}$. It is an exact equation, which can be proved by integration by parts, or by expressing the invariance of the integral $Z_{\mathcal{S}}$ by change of variables preserving $\mathsf{A}$. It can be written: for any $x \in \mathbb{C}\setminus\mathsf{A}$,
\bea
\label{Sdeaq}
\mu_{\mathcal{S}}\Bigg[N\Int{\mathsf{A}}{} \partial_{\xi}\Big(\frac{(1 - \beta/2)\sigma_{\mathsf{A}}(\xi)}{x - \xi}\Big)\,\dd L_N(\xi) 
\; + \; 
 N^2\Int{\mathsf{A}^2}{} \frac{\beta}{2(\xi_1 - \xi_2)}\Big(\frac{\sigma_{\mathsf{A}}(\xi_1)}{x - \xi_1} - \frac{\sigma_{\mathsf{A}}(\xi_2)}{x - \xi_2}\Big)\dd L_N(\xi_1)\dd L_N(\xi_2) & & \nonumber \\
 + N^2 
\Int{\mathsf{A}^r}{} \frac{\partial_{\xi_1} T(\xi_1,\xi_2,\ldots,\xi_r)}{(r - 1)!} \,\frac{\sigma_{\mathsf{A}}(\xi_1)}{x - \xi_1} \prod_{i = 1}^r \dd L_N(\xi_i)\Bigg] & = & 0\;.
\eea
Here, we have defined
$$\sigma_{\mathsf{A}}(x) = \prod_{a \in \partial\mathsf{A}} (x - a)\;.$$
We insist that it takes the same form for the unconstrained model $\mathcal{S} = \mathsf{A}^N$ and in the model with fixed filling fractions $\mathcal{S} = \mathsf{A}_{\mathbf{N}}$, see \S~\ref{defs1} for the definitions. We do not attempt to recast this Schwinger-Dyson equation in the most elegant form ; this is the matter of Section~\ref{S5}.

For any fixed $x \in \mathbb{C}\setminus \mathsf{A}$, the functions against which the empirical measure are integrated are continuous. Therefore, since $L_N$ converges to $\mathrm{\mu}_{\mathrm{\rm eq }}$ (Theorem~\ref{tc0} or \ref{tc1}), the first term is negligible in the large $N$ limit, and we obtain:
\beq
\label{joee}
\Int{\mathsf{S}^2}{} \frac{\beta}{2(\xi_1 - \xi_2)}\Big(\frac{\sigma_{\mathsf{A}}(\xi_1)}{x - \xi_1} - \frac{\sigma_{\mathsf{A}}(\xi_2)}{x - \xi_2}\Big)\dd \mu_{\mathrm{\rm eq }}(\xi_1)\dd\mu_{\mathrm{\rm eq }}(\xi_2) \; + \;
 \Int{\mathsf{S}^r}{} \frac{\partial_{\xi_1} T(\xi_1,\ldots,\xi_r)}{(r - 1)!} \,\frac{\sigma_{\mathsf{A}}(\xi_1)}{x - \xi_1} \prod_{i = 1}^r \dd \mu_{\mathrm{\rm eq }}(\xi_i) = 0\;.
\enq
This equality only involves analytic functions of $x \in \mathbb{C}\setminus\mathsf{S}$ and is established for
 $x \in \mathbb{C}\setminus\mathsf{A}$. Thus is also valid for $x \in \mathbb{C}\setminus\mathsf{S}$. The first term can be rewritten partly in terms of the Stieltjes transform of the equilibrium measure:
\beq
\label{minsuq}
\frac{\beta}{2}\,\sigma_{\mathsf{A}}(x)\,W_{\mathrm{\rm eq }}^2(x) + U(x) + P(x) = 0\;,
\enq
with:
\bea
W_{\mathrm{\rm eq }}(x) & = & \Int{\mathsf{S}}{} \frac{\dd\mu_{\mathrm{\rm eq }}(\xi)}{x - \xi}\;, \nonumber \\
U(x) & = & \Int{\mathsf{S}^{r}}{} \sigma_{\mathsf{A}}(\xi_1)\,
\frac{\partial_{\xi_1} T(\xi_1,\ldots,\xi_r)}{(r - 1)!\,(x - \xi_1)}\,\prod_{i = 1}^{r} \dd\mu_{\mathrm{\rm eq }}(\xi_i)\;, \nonumber \\
P(x) & = & \Int{\mathsf{S}^2}{} \frac{\beta}{2(\xi_1 - \xi_2)}
\Big(\frac{\sigma_{\mathsf{A}}(\xi_1) - \sigma_{\mathsf{A}}(x)}{x - \xi_1} - 
\frac{\sigma_{\mathsf{A}}(\xi_2) - \sigma_{\mathsf{A}}(x)}{x - \xi_2}\Big)\dd\mu_{\mathrm{\rm eq }}(\xi_1)\dd\mu_{\mathrm{\rm eq }}(\xi_2)\;.\nonumber
\eea
Since $\sigma_{\mathsf{A}}(x)$ is a polynomial (of degree $g + 1$), $P(x)$ is also a polynomial. Since $T \in \mathcal{C}^2(\mathsf{A}^r)$, $U(x)$ admits continuous 
$\pm$ boundary values when $x \in \mathring{\mathsf{S}}$. Therefore, $\sigma_{\mathsf{A}}(x)\,W_{\mathrm{eq}}^2(x)$ -- and a fortiori $W_{\mathrm{eq}}(x)$ -- also admits continuous $\pm$ boundary values when 
$x \in \mathring{\mathsf{S}}$. Then, \eqref{minsuq} at $x \in \mathring{\mathsf{S}}$ leads to:
\beq
\label{qudara}\sigma_{\mathsf{A}}(x)\Big(W_{\mathrm{eq};\pm}^2(x) - V'(x)W_{\mathrm{eq};\pm}(x) + \frac{\widetilde{P}(x)}{\sigma_{\mathsf{A}}(x)}\Big) = 0
\enq
with:
\bea
\label{vdef} V(x) & = & -\frac{2}{\beta}\,\Int{\mathsf{S}^{r - 1}}{} \frac{T(x,\xi_2,\ldots,\xi_r)}{(r - 1)!} \prod_{i = 2}^{r} \dd\mu_{\mathrm{\rm eq }}(\xi_i)\;, \\
\label{pdef} \widetilde{P}(x) & = & \frac{2}{\beta} P(x) \; - \; 
 \Int{\mathsf{S}}{} \frac{\sigma_{\mathsf{A}}(x)\,V^{\prime}(x) - \sigma_{\mathsf{A}}(\xi)\,V^{\prime}(\xi)}{(x - \xi)}
 					\,\dd\mu_{\mathrm{\rm eq }}(\xi) \; .
\eea
Since we assume $T \in \mathcal{C}^2(\mathsf{A}^r)$, we also find $V \in \mathcal{C}^2(\mathsf{S})$, hence $\widetilde{P} \in \mathcal{C}^0(\mathsf{S})$. We also remind that the equilibrium measure is given in terms of its Stieltjes transform by:
\beq
\label{froma}2{\rm i} \pi \f{ \dd\mu_{\mathrm{\rm eq }} }{ \dd x } (x)  \; = \; 
 W_{\mathrm{\rm eq };-}(x ) \, - \,  W_{\mathrm{\rm eq };+}(x ) \; .
\enq
Therefore, solving the quadratic equations \eqref{qudara} for $W_{\mathrm{\rm eq };\pm}(x)$, we find:
\beq
\label{vrar}\f{ \dd\mu_{\mathrm{\rm eq }} }{ \dd x } (x) \;  = \;  \frac{\bs{1}_{\mathsf{S}}(x)}{2\pi}\,
			\sqrt{\frac{ 4\widetilde{P}(x) \, - \, \sigma_{\mathsf{A}}(x)\,\big(V'(x)\big)^2  }{\sigma_{\mathsf{A}}(x)}} 
			 \; .
\enq
From \eqref{vrar}, we see that the only possible divergence of $\tf{ \dd\mu_{\mathrm{\rm eq }} }{ \dd x }$ is at 
$\alpha \in \mathsf{S}\cap\partial\mathsf{A}$, and the divergence is at most a $O((x - \alpha)^{-1/2})$, hence $(i)$. If $\alpha \in \partial\mathsf{S}\setminus\partial\mathsf{A}$, we have $\sigma_{\mathsf{A}}(\alpha) \neq 0$ but the density of the equilibrium measure must vanish at $\alpha$. If $T \in \mathcal{C}^3(\mathsf{A}^r)$, we find that $\widetilde{P} \in \mathcal{C}^1(\mathsf{A})$ and thus the quantity inside the squareroot is $\mathcal{C}^1$. So, it must vanish at least linearly in $\alpha$, which entails $(ii)$. \hfill $\Box$

\vspace{0.2cm}

\noindent \textbf{Proof of Lemma~\ref{thregana}.} Let $\Omega$ be an open neighbourhood of $\mathsf{A}$ such that $T$ is holomorphic in $\Omega^r$. Then, $V(x)$ and $\widetilde{P}(x)$ defined in \eqref{vdef}-\eqref{pdef} are well-defined, holomorphic functions of $x \in \Omega$. So, the limiting Schwinger-Dyson equation \eqref{joee} can be directly recast for any $x \in \Omega \setminus \mathsf{A}$:
\beq
W_{\mathrm{\rm eq }}^2(x) - V'(x)\,W_{\mathrm{\rm eq }}(x) + \frac{\widetilde{P}(x)}{\sigma_{\mathsf{A}}(x)} = 0\;. \nonumber
\enq
Its solution is:
\beq
\label{516}
W_{\mathrm{\rm eq }}(x) = \frac{V'(x)}{2} \pm \f{1}{2}
\sqrt{\frac{\sigma_{\mathsf{A}}(x)\,\big(V'(x)\big)^2 - 4\widetilde{P}(x)}{\sigma_{\mathsf{A}}(x)}}\;.
\enq
By continuity of $W_{\mathrm{\rm eq }}(x)$, the sign is uniformly $+$ or uniformly $-$ in each connected component of $\Omega$. From \eqref{froma}, the equilibrium measure reads:
\beq
\label{517}
\f{ \dd\mu_{\mathrm{\rm eq }} }{ \dd x }(x) = 
\pm \frac{\mathbf{1}_{\mathsf{S}}(x)\,\dd x}{2\pi}\sqrt{R(x)}\;,\qquad 
R(x) = \frac{\sigma_{\mathsf{A}}(x)\,\big(V'(x)\big)^2 - 4\widetilde{P}(x)}{-\sigma_{\mathsf{A}}(x)}\;.
\enq
The support $\mathsf{S}$ is the closure of the set of $x \in \mathsf{A}$ for which the right-hand side is positive. The 
function $R$ is meromorphic in $\Omega\cup\mathsf{A}$ and real-valued on $\mathsf{A}$ ; further, 
its only poles are simple and all located in $\Dp{}A$. Hence, given a compact $\Omega' \subseteq \Omega$ neighbourhood of $\mathsf{A}$, $R$ can be recast as $R=R_0\cdot M^2$. 
In such a factorisation, $R_0(x)$ is a rational function having the same poles and zeroes as $R(x)$
on $\Om^{\prime}$ while $M^2$ is a holomorphic function on $\Om$ that is nowhere vanishing on $\Om^{\prime}$
and that keeps a constant sign on $\mathsf{A}$. We shall denote its square root by $M$. 
According to the formula \eqref{517}, $R_0(x)$ can only have simple poles that occur at the edges of $\mathsf{A}$. 
Thence, the edges of $\mathsf{S}$ must be either its poles or its zeroes. 
Therefore, we may factorise further the zeroes of even order and write:
\beq
\f{ \dd\mu_{\mathrm{\rm eq }} }{ \dd x }(x) = \frac{\mathbf{1}_{\mathsf{S}}(x)\,M(x)}{2\pi}\,\sigma_0(x)\,
\prod_{\alpha \in \partial\mathsf{S}\setminus\partial\mathsf{A}} |x - \alpha|^{1/2} \prod_{\alpha \in \partial\mathsf{S}\cap\partial\mathsf{A}} |x - \alpha|^{-1/2}\;. \nonumber
\enq
for some polynomial $\sigma_0(x)$. Since $\tf{ \dd\mu_{\mathrm{\rm eq }} }{ \dd x }$ is a density, $\sigma_0(x)$ has constant sign on $\mathsf{S}$. If we require that it is non-negative and has dominant coefficient $\pm 1$, $\sigma_0(x)$ is uniquely determined. \hfill $\Box$

\vspace{0.2cm}

\begin{defin}
\label{offcsq} We speak of a $(g_0 + 1)$-cut regime when $\mathsf{S}$ is the disjoint union of $g_0 + 1$ segments, and we write:
\beq
\mathsf{S} = \dot{\cup}_{h = 0}^{g_0} \mathsf{S}_h\;,\qquad \mathsf{S}_h = [\alpha_h^-,\alpha_h^+]\;. \nonumber
\enq
We speak of an off-critical regime when $\sigma_0(x) = 1$.
\end{defin}

\section{Concentration around equilibrium measures}
\label{S3}

\subsection{Large deviation for the support of the spectrum}
\label{Lares}
Exactly as in Borot-Guionnet \cite{BG11,BGmulti}, we can prove:
\begin{lemma}
\label{uuu3} Assume Hypothesis~\ref{h1}. We have large deviation estimates: for any $\mathsf{F} \subseteq \mathsf{A}$ closed and $\Omega \subseteq \mathsf{A}$ open,
\bea
\limsup_{N\ra\infty}\frac{1}{N}\ln \mu_{\mathcal{S}}\big[\exists i\quad\lambda_i \in \mathsf{F}\big] & \le & 
\sup_{x \in \mathsf{F}} T_{\mathrm{eff}}(x)\;, \nonumber \\
\liminf_{N\ra\infty}\frac{1}{N}\ln \mu_{\mathcal{S}}\big[\exists i\quad\lambda_i \in \Omega\big] & \ge & 
\sup_{x \in \Omega} T_{\mathrm{eff}}(x)\;. \nonumber
\eea
$-T_{\mathrm{eff}}(x)$ defined in \eqref{Teff} is thus the rate function. \hfill $\Box$
\end{lemma}


\vspace{0.2cm}

\noindent It is natural to supplement the conclusion of this lemma with an extra assumption:
\begin{hypothesis}
\label{hcc}(Control of large deviations) $T_{\mathrm{eff}}(x) < 0$ outside $\mathsf{S} = \mathrm{supp}\,\mu_{\mathrm{\rm eq }}$.
\end{hypothesis}

Lemma \ref{uuu3} along with Hypothesis \ref{hcc} allows one the simplification of the form of $\mathsf{A}$. 
First of all, we can always assume the domain of integration $\mathsf{A}$ to be 
compact. Indeed, a non-compact domain $\mathsf{A}$ would only alter the answer obtained for the correlators or the partition function in the case of the compact domain  $\mathsf{A}_{[M]} = \mathsf{A} \cap \intff{-M}{M}$ 
with $M$ sufficiently large, by exponentially small in $N$ terms. Second, when the 
control of large deviations holds, for the price of the same type of exponentially small in N corrections
(see \cite[Proposition 2.2 and 2.3]{BG11} for more precise statements),
we may restrict further the domain of integration to any fixed $\mathsf{A}^{\prime} \subseteq \mathsf{A}$ such that 
$\mathsf{A}^{\prime} \setminus \mathsf{S}$ is as small as desired. 
For instance, in the $(g_0 + 1)$ cut regime, one can always restrict $\mathsf{A}$ to be a disjoint union of $(g + 1) = (g_0 + 1)$ closed compact intervals $\mathsf{A}_h^{\prime}\cap \mathsf{A}
$, such that $\mathsf{A}_h^{\prime}$ contains  an open  neighbourhood of $\mathsf{S}_h$ in $\mathsf{A}_h$ for any $h \in \intn{0}{g_0}$. 

Therefore, from now on, we shall always assume $\mathsf{A}$ to be a disjoint union of $(g + 1)$ closed compact intervals $\mathsf{A}_h$, such that $\mathsf{S}_h \subseteq \mathsf{A}_h$ for any $h \in \intn{0}{g}$ as above.
In particular, we will not continue distinguishing $g$ from $g_0$.

\subsection{Pseudo-distance and adapted spaces}

\label{s1u}
In view of showing concentration of the empirical measure around the equilibrium measure $\mu_{{\rm eq}}$ in either of the two models, we add two assumptions:
\begin{hypothesis}
\label{h2}
(Local strict convexity) For any $\nu \in \mathcal{M}^0(\mathsf{A})$,
\beq
\mathcal{Q}[\nu] = - \beta  \,\Int{\mathsf{A}^2}{} \ln|x - y|\dd\nu(x)\dd\nu(y) \; - \;  \Int{\mathsf{A}^r}{} \frac{T(x_1,\ldots,x_r)}{(r - 2)!}\,\dd\nu(x_1)\dd\nu(x_2)\prod_{i = 3}^r \dd\mu_{\mathrm{\rm eq }}(x_i)
\label{definition fonctionnelle Q}
\enq
is non-negative, and vanishes iff $\nu = 0$.
\end{hypothesis}
We observe that for any measure with zero mass:
\beq
\label{Qcdef}\mathcal{Q}[\nu] = \beta\,\mathcal{Q}_{C}[\nu] + \mathcal{Q}_{T}[\nu]\;,\qquad \mathcal{Q}_{C}[\nu] = -\Int{\mathsf{A}^2}{} \ln|x_1 - x_2|\dd\nu(x_1)\dd\nu(x_2) = \Int{0}{\infty} \frac{
\big|\mathcal{F}[\nu](k)\big|^2}{k}\,\dd k \in [0,+\infty]\;, 
\enq
whereas the other part $\mathcal{Q}_{T}$ is always finite since $T \in \mathcal{C}^0(\mathsf{A}^r)$ and $\mathsf{A}$ is compact. Therefore, $\mathcal{Q}[\nu]$ is well-defined and takes its values in $\mathbb{R}\cup\{+\infty\}$. Hypothesis~\ref{h2} requires it to assume values in $[0,+\infty]$.

\begin{defin}
\label{h3}
A vector subspace $\mathcal{H} \subseteq \mc{C}^{0}\big(\mathbb{R})$ is then called an adapted space if 
there exists a norm $\p \cdot \p_{\mathcal{H}}$ on $\mc{H}$ and a continuous function $\chi_{\mathsf{A}}$ which assumes values $1$ on $\mathsf{A}$ and $0$ outside of a compact, and such that:
\begin{itemize}
\item[$\bullet$] there exists $c_0 > 0$ such that:
$$
\forall \nu \in \mathcal{M}^0(\mathsf{A}),\qquad \forall \varphi \in \mathcal{H},\qquad \Big|\int_{\mathsf{A}} \varphi(x)\,\dd\nu(x)\Big| \leq c_0\,\mathcal{Q}^{1/2}[\nu]\,\p \varphi \p_{\mathcal{H}}\;.
$$
\item[$\bullet$] there exists $c_1 > 0$ and an integer $m \geq 0$ called the growth index such that, for any $k \in \mathbb{R}$, 
the function ${\rm e}_{k}(x) = \chi_{\mathsf{A}}(x)\,e^{{\rm i}kx}$ belongs to $\mathcal{H}$ and one has 
$\p {\rm e}_{k} \p_{\mathcal{H}} \leq c_1\,\big( |k|^{m} + 1\big)$
\end{itemize}
\end{defin}

We will show in Section~\ref{S4} how to construct an adapted space $\mathcal{H}$ provided Hypothesis~\ref{h2} holds. 
We will often encounter multilinear statistics, and we will use both Vaserstein norm and $\mathcal{Q}$ in their estimation. The following technical lemma will appear useful in the following.

\begin{lemma}
\label{l322}Let $l,l' \geq 0$ be integers, $l'' \leq l'$ be another integer, and $m' > (m - 1)l'' + 2l' + l$. Then, 
given an adapted space $\mc{H}$, for any $\varphi \in \mathcal{C}^{m'}(\mathsf{A}^{l + l'})$, any $\nu_1,\ldots,\nu_k \in \mathcal{M}^{0}(\mathsf{A})$ and $\mu_1,\ldots,\mu_l \in \mathcal{M}^1(\mathsf{A})$, one has the bounds:
\beq
\label{boun}\Bigg|\Int{\mathsf{A}^{l + l'}}{} \varphi(x_1,\ldots,x_{l + l'})
 \prod_{i = 1}^{l'} \dd\nu_i(x_i) \prod_{j = l' + 1}^{l + l'} \dd\mu_j(x_j)\Bigg| \leq c_0\,C_{l,l',l''}[\varphi]\,\prod_{i = 1}^{l''} \mathcal{Q}^{1/2}[\nu_i]\,\prod_{i = l'' + 1}^{l} \p \nu_i \p
\enq
for some finite non-negative constant $C[\varphi]$.
\end{lemma}
\noindent \textbf{Proof.} We may extend $\varphi \in \mathcal{C}^{m'}(\mathsf{A}^{l' + l})$ to a function $\widetilde{\varphi} \in \mathcal{C}^{m'}(\mathbb{R}^{l' + l})$. Then, we may write:
\bea
X[\varphi] & = & \Int{\mathsf{A}^{l' + l}}{} \varphi(x_1,\ldots,x_{l' + l})\,\prod_{i = 1}^{l'} \dd\nu_i(x_i) \prod_{j = l' + 1}^{l' + l} \dd\mu_j(x_j) \nonumber \\
& = & \Int{\mathbb{R}^{l' + l}}{} \widetilde{\varphi}(x_1,\ldots,x_{l' + l})\,\prod_{i = 1}^{l'} \dd\nu_i(x_i) \prod_{j = l' + 1}^{l'+l} \dd\mu_j(x_j) \nonumber \\
& = & \Int{\mathbb{R}^{l' + l}}{} \frac{\dd^{l' + l}\bs{k}}{(2\pi)^{l' + l}}\,\mathcal{F}[\widetilde{\varphi}](\bs{k})\,\prod_{i = 1}^{l''} \Bigg(\Int{\mathbb{R}}{} {\rm e}_{k_i}(x_i)\,\dd\nu_i(x_i)\Bigg)\,\prod_{i = l'' + 1}^{l'} \Bigg(\Int{\mathbb{R}}{} e^{{\rm i}k_jx_j}\dd\nu(x_j)\Bigg)\prod_{j = l' + 1}^{l' + l} \Bigg(\Int{\mathbb{R}}{} e^{{\rm i}k_jx_j}\,\dd\mu_j(x_j)\Bigg)\;, \nonumber
\eea
where we could introduce the function ${\rm e}_k(x) = \chi_{\mathsf{A}}(x)\,e^{{\rm i}kx}$ since $\nu_i$ are supported on $\mathsf{A}$. We will bound the first group of integrals for $i \in \intn{1}{l''}$ by 
$c_0\,\mathcal{Q}^{\tf{1}{2}}[\nu_i] \cdot \p {\rm e}_{k_i} \p_{\mathcal{H}}$, the second group for $i \in \intn{l'' + 1}{l'}$
with the Vaserstein norm by $|k_i|\cdot\p \nu_i \p$, and the remaining by the obvious bound $1$ since $\mu_j$ are probability measures.
\beq
|X[\varphi]| \leq c_0^{l''}\, \Bigg(\prod_{i = 1}^{l''} \mathcal{Q}^{1/2}[\nu_i] \prod_{i = l'' + 1}^{l'} \p \nu_i \p \Bigg) \int_{\mathbb{R}^{l' + l}} \frac{\dd^{l + l'}\bs{k}}{(2\pi)^{l + l'}}\,\big|\mathcal{F}[\widetilde{\varphi}](\bs{k})\big|\,\prod_{i = 1}^{l''} \p {\rm e}_{k_i} \p_{\mathcal{H}} \prod_{i = l'' + 1}^{l'} |k_i| \;. \nonumber 
\enq
Since $\widetilde{\varphi}$ is $\mathcal{C}^{m'}$, the integral in the right-hand side will converge at least for $m' > l + 2l' + (m - 1)l''$, and gives the constant $C_{l,l',l''}[\varphi]$ in \eqref{boun}. \hfill $\Box$

\subsection{Concentration results} 
The next paragraphs are devoted to the proof of:
\begin{theorem}
\label{t1}
Assume Hypothesis \ref{h1}-\ref{hcc}-\ref{h2}, an adapted space with growth index $m$, and $T \in \mathcal{C}^{m'}(\mathsf{A}^{r})$ with $m' > 2m + r + 1$. Denote $\mu_{\rm{eq}}$ the equilibrium measure in one of the two models \eqref{law0} or \eqref{law1} and $\tilde L_N^u$ the regularization of $L_N$ defined in Section~\ref{subsecreg}.
 There exists constants $c>0$ and $C, C^{\prime}$, such that:
\beq
\mu_{\mc{S}}\big[\mathcal{Q}[\widetilde{L}_{N}^{u} - \mu_{\mathrm{\rm eq }}]^{\frac{1}{2}} \geq t \big] \leq \exp\Big\{CN\ln N - \frac{N^2 t^2}{4} \Big\}  +C'\exp\{ -c N^2\}\,. \nonumber
\enq
\end{theorem}
As in \cite{BGmulti} we  easily derive
\begin{cor}
\label{c1} Under the same assumptions, let $b > 0$. There exists finite constants  $C ,C'$  and $c>0$ such that, for $N$ large enough, for any $\varphi \in \mathcal{H}\cap\mathrm{Ho}_{b}(\mathsf{A})$, we have:
\beq
\mu_{\mc{S}}\left[\Big|\int \varphi(\xi)\dd(L_N - \mu_{\mathrm{\rm eq }})(\xi)\Big| \geq 
\frac{c\kappa_{b}[\varphi]}{N^{2b}} + c_0 t \p\mathcal{\varphi}\p_{\mathcal{H}}\right] 
\leq \exp\Big\{CN\ln N - \frac{1}{4}N^2t^2\Big\} +C'\exp\{ -c' N^2\}\,. \nonumber
\enq
\end{cor}

As a special case, we can obtain a rough a priori control on the correlators:
\begin{cor}
\label{c2} Let $w_N = \sqrt{N\ln N}$ and $\psi_{x}(\xi) = \mathbf{1}_{\mathsf{A}}(\xi)/(x - \xi)$. For $N$ large enough, and there exists $c,c_1 > 0$ such that:
\beq
\big|W_1(x) - NW_{\mathrm{eq}}(x)\big| \leq \frac{c}{Nd(x,\mathsf{A})^2} + c_1 \p \psi_{x} \p_{\mathcal{H}} w_N  \nonumber
\enq
Similarly, for any $n \geq 2$ and $N$ large enough, there exists $c_n > 0$ such that:
\beq\label{apriori}
\big|W_{n}(x_1,\ldots,x_n)\big| \leq c_n \prod_{i = 1}^n\Big[\frac{c}{Nd(x,\mathsf{A})^2} + c_1 \p \psi_{x} \p_{\mathcal{H}} w_N\Big]
\enq
\end{cor}
We recall that we are in a $(g + 1)$-cut regime with $g \geq 1$ and that $\mathsf{A}_h$ is a partition of $\mathsf{A}$ in $(g + 1)$ segments so that $\mathsf{A}_h$ is a neighbourhood of $\mathsf{S}_h$ in $\mathsf{A}$. For any configuration 
$(\lambda_1,\ldots,\lambda_N) \in \mathsf{A}^N$, we denote $\wt{N}_h$ the number of $\lambda_i$'s in $\mathsf{A}_h$, and 
$\wt{\mathbf{N}} = (\wt{N}_0,\dots, \wt{N}_g)$. Let 
\beq
N \bs{\epsilon}^{\star} \; = \; \big( N \eps_{0}^{\star},\dots, N \eps_{g}^{\star} \big) \qquad \text{with} \qquad 
 \eps_{\mathrm{eq}}^{\star} \; = \; \Int{ \mathsf{S}_h }{} \dd \mu_{\text{\rm eq }}(\xi)  \;.  \nonumber
\enq
We can derive an estimate for large deviations of $\mathbf{N}'$ away from $N\bs{\epsilon}^{\star}$:
\begin{cor}
\label{c3} Assume a $(g + 1)$-cut regime with $g \geq 1$ and let $\wt{\mathbf{N}}$ be as above. Then, there exists a positive constant $C$ such that, for $N$ large enough and uniformly in $t$:
\begin{equation}
\mu_{\mathsf{A}^N}\Big[ \big| \wt{\mathbf{N}} - N\bs{\epsilon}^{\star}\big| > t\,\sqrt{N\ln N}\Big] \leq \exp\{N\ln N(C - t^2)\}. \nonumber
\end{equation}
\end{cor}

\subsection{Regularization of $L_N$}\label{subsecreg}

We cannot compare directly $L_N$ to $\mu_{\mathrm{\rm eq }}$ with $\mathcal{Q}$, because of the logarithmic singularity in $\mathcal{Q}_{C}$ and the atoms in $L_N$. Following an idea of Ma\"ida and Maurel-Segala \cite{MMEMS}, we associate to any configurations of points $\lambda_1 < \cdots < \lambda_N$ in $\mathsf{A}$, another configuration $\widetilde{\lambda}_1 < \cdots < \widetilde{\lambda}_{N}$ by the formula:
\beq
\widetilde{\lambda}_1 = \lambda_1,\qquad \widetilde{\lambda}_{i + 1} = \widetilde{\lambda}_{i} + \max(\lambda_{i + 1} - \lambda_{i},N^{-3}). \nonumber
\enq
It has the properties:
\beq
\forall i \neq j,\qquad |\widetilde{\lambda}_i - \widetilde{\lambda}_j| \geq N^{-3},\qquad |\lambda_i - \lambda_j| \leq |\widetilde{\lambda}_i - \widetilde{\lambda}_j|,\qquad |\widetilde{\lambda}_i - \lambda_i|  \leq (i - 1)N^{-3} \nonumber
\enq
Let us denote $\widetilde{L}_N = \frac{1}{N}\sum_{i = 1}^N \delta_{\widetilde{\lambda}_i}$ the new counting measure, and $\widetilde{L}_N^u$ its convolution with the uniform measure on $[0,N^{-7/2}]$. Let us define a regularised version of the energy functional $\mathcal{E}^{\Delta} = (\beta/2)\,\mathcal{E}_{C}^{\Delta} + \mathcal{E}_{T}$ with:
\bea
\mathcal{E}_{C}^{\Delta}[\mu] & = & -\int_{x_1 \neq x_2} \ln|x_1 - x_2|\dd\mu(x_1)\dd\mu(x_2)\;,\nonumber \\
\mathcal{E}_{T}[\mu] & = & - \int_{\mathsf{A}^{r}} \frac{T_r(x_1,\ldots,x_r)}{r!}\,\prod_{i = 1}^r \dd\mu(x_i)\;.
\nonumber
\eea
As in  \cite{MMEMS} (see also \cite{BGmulti}), we have:
\begin{lemma}
\beq
\mathcal{E}_{C}^{\Delta}[L_N] - \mathcal{E}_{C}^{\Delta}[\widetilde{L}_{N}^u] \geq  \frac{1}{3N^4} - \frac{7\ln N}{2N}. \nonumber
\enq
\end{lemma}
It is then straightforward to deduce that
\begin{cor}
\label{cTlip} There exists constants $c,c' > 0$ such that:
\beq
\mathcal{E}^{\Delta}[L_N] - \mathcal{E}[\widetilde{L}_N^u] \geq c\frac{\ln N}{N} + c'\,\frac{\kappa_{1}[T]}{N^{3}}\;. \nonumber
\enq
\end{cor}
We also have:
\begin{lemma}
\label{lemu} There exists $c > 0$ such that, for any $f \in \mathrm{Ho}_{b}(\mathsf{A})$, we have:
\beq
\left|\int_{\mathsf{A}} f(\xi)\dd(L_N - \widetilde{L}_N^{u})(\xi)\right| \leq \frac{c\,\kappa_{b}[f]}{N^{2b}} \nonumber
\enq
\end{lemma}
%
%
%
%
%
%
%
%
%
%

\subsection{Concentration of $\widetilde{L}_N^u$ (Proof of Theorem~\ref{t1}).}\label{S3.5}

We would like to estimate the probability of large deviations of $\widetilde{L}_N^u$ from the equilibrium measure $\mu_{\mathrm{\rm eq }}$. We first need a lower bound on $Z_{\mathcal{S}}$ similar to that of \cite{BAG}, obtained by localizing the ordered eigenvalues at distance $N^{-3}$ of the quantiles $\lambda_{i}^{\mathrm{\rm cl }}$ of the equilibrium measure $\mu_{\mathrm{\rm eq }}$, which are defined by:
\beq
\lambda_{i}^{\mathrm{\rm cl }} = \inf \bigg\{x \in \mathsf{A},\qquad \int_{-\infty}^{x} \dd\mu_{\mathrm{\rm eq }}(x) \geq i/N \bigg\}. \nonumber
\enq
\begin{lemma}
\label{lowL} Assume Hypothesis~\ref{h2} with $T \in \mathcal{C}^3(\mathsf{A})$. Then, there exists a finite constant $c$ so that:
\beq
Z_{\mathcal{S}} \geq \exp\Big\{-cN\ln N - N^2\mathcal{E}[\mu_{\mathrm{\rm eq }}]\Big\}.\nonumber
\enq 
\end{lemma}
\textbf{Proof.} According to Lemma~\ref{thregup}, $T \in \mathcal{C}^3(\mathsf{A})$ implies that $\mu_{\mathrm{eq}}$ has a $\mathcal{C}^1$ density in the interior of $\mathsf{S}$, and behaves at most like the inverse of a squareroot at $\partial\mathsf{S}$. This ensures the existence of $c_0 > 0$ such that
\beq
\label{lwb}|\lambda_{i + 1}^{\mathrm{\rm cl }} - \lambda_{i}^{\mathrm{\rm cl }}| \geq c_0N^{-2}
\enq
for any $i \in \intn{0}{N}$, where by convention $\lambda_{0}^{\text{cl}} = \mathrm{min}\,\{x \; : \; x \in \mathsf{S} \}$ 
and $\lambda_{N + 1}^{\text{cl}} = \max \,\{x \; : \; x \in \mathsf{A} \}$. The proof of the lower bound for $Z_{\mathcal{S}}$ is similar 
to \cite{BGmulti}, we redo it here for sake of being self-contained. It can be obtained by restricting the integration over the 
configurations $\big\{ \bs{\lambda} \in \mathcal{S},\quad |\lambda_i - \lambda_i^{\mathrm{cl}}| \leq N^{-3} \big\}$, where 
$\mathcal{S} =  \mathsf{A}^N$ or $=\mathsf{A}_{\mathbf{N}}$ depending on the model. 
For any such $\bs{\la}$, one has:
\beq
|\lambda_i-\lambda_j|\ge |\lambda_i^{cl}-\lambda_j^{cl}| \Big( 1 - \f{2}{c_0 N} \Big) \qquad \big|T(\lambda_{i_1},\ldots,\lambda_{i_r})-T(\lambda_{i_1}^{\rm cl},\ldots,\lambda_{i_r}^{\rm cl})\big|\le  \frac{\kappa_1[T]}{N^3}\nonumber
\enq
and this implies that:
\beq
\label{ei1}Z_N\ge (1 - N^{-1})^{N(N - 1)\beta/2}\exp\Big(\frac{\kappa_1[T]}{r!\,N}\Big)\,N^{-3N}\,\prod_{1 \leq i < j \leq N} |\lambda_{i}^{\rm cl }-\lambda_{j}^{\rm cl }|^\beta \exp\Big(\frac{N^{2-r}}{r!} \sum_{1 \leq i_1,\ldots,i_r \leq N} T(\lambda^{\rm cl }_{i_1},\ldots,\lambda_{i_r}^{\rm cl })\Big)
\enq
Then, for any $i,j$ such that $j + 1 \leq i - 1$, we have, by monotonicity of the logarithm,:
\beq
\ln |\lambda_{i}^{\rm cl }-\lambda_{j}^{\rm cl }|  \geq N^2 \int_{\lambda_{i - 1}^{\mathrm{cl}}}^{\lambda_{i}^{\mathrm{cl}}}\int_{\lambda_{j}^{{\rm cl}}}^{\lambda_{j + 1}^{{\rm cl}}} \ln|\xi_1 - \xi_2|\,\dd\mu_{\mathrm{eq}}(\xi_1)\,\dd\mu_{\mathrm{eq}}(\xi_2)\nonumber
\enq
For the remaining pairs $\{i,j\}$, we rather use the lower bound \eqref{lwb}, and we find after summing over pairs:
\beq
\label{ei2} \beta \sum_{i < j} \ln |\lambda_{i}^{\rm cl }-\lambda_{j}^{\rm cl }| \geq -\beta N^2\mathcal{E}_{C}[\mu_{\mathrm{eq}}] + c_1\,N\ln N 
\enq
If $\varphi\,:\,\mathsf{A} \rightarrow \mathbb{R}$ is a function with finite total variation 
$\mathrm{TV}[\varphi]$ we can always decompose it as the difference of two increasing functions, the total variation of each of them being $\mathrm{TV}[\varphi]$. And, if $\varphi_{>}$ is an increasing function:
\beq
\frac{1}{N}\sum_{i = 0}^{N-1} \varphi_{>}(\lambda_i^{{\rm cl}})   \, \leq  \, 
 \Int{\mathsf{A}}{} \varphi_{>}(\xi)\,\dd\mu_{\mathrm{eq}}(\xi) \,  \leq \, 
  \frac{1}{N}\sum_{i = 1}^{N} \varphi_{>}(\lambda_i^{{\rm cl}}) 
\enq

Therefore:
\beq
\Bigg|\frac{1}{N}\sum_{i = 1}^N \varphi(\lambda_i^{{\rm cl}}) - \Int{\mathsf{A}}{} \varphi(\xi)\,\dd\mu_{\mathrm{eq}}(\xi)\Bigg| \leq 
\frac{2 \mathrm{TV}[f]}{N} \nonumber
\enq
This can be generalised for functions defined in $\mathsf{A}^r$ by recursion, and we apply the result to $T$, which is $\mathcal{C}^1$, hence is of bounded total variation with $\mathrm{TV}[T] \leq \ell(\mathsf{A})\,\kappa_1[T]$:
\beq
\label{ei3}\Bigg|\frac{1}{N^{r}} \sum_{1 \leq i_1,\ldots,i_r \leq N} T(\lambda_{i_1}^{{\rm cl}},\ldots,\lambda_{i_r}^{{\rm cl}}) - \Int{\mathsf{A^r}}{} T(\xi_1,\ldots,\xi_r)\,\prod_{i = 1}^r \dd\mu_{\mathrm{eq}}(\xi_i)\Bigg| \leq \frac{c_2\,\kappa_1[T]}{N}
\enq
Combining \eqref{ei2}-\eqref{ei3} with \eqref{ei1}, we find the desired result. \hfill $\Box$

\vspace{0.2cm}

\noindent Now, the density of probability measure in either of the models \eqref{law0} or \eqref{law1} can be written:
\beq
\dd\mu_{\mathcal{S}} = \Big[\prod_{ \bs{i} \in \mc{I}_{\mc{S}} } \dd\lambda_i\Big]\,\exp\Big\{-N^2\mathcal{E}^{\Delta}[L_N]\Big\}. \nonumber
\enq
With the comparison of Corollary~\ref{cTlip}, we find that, for $N$ large enough:
\beq
\dd\mu_{\mathcal{S}} \leq \Big[\prod_{ \bs{i} \in \mc{I}_{\mc{S}} }  \dd\lambda_i \Big]\,\exp\big\{c N\ln N - N^2\mathcal{E}[\widetilde{L}_N^u]\big\}. \nonumber
\enq
We can then compare the value of the energy functional at $\widetilde{L}_N^u$ and $\mu_{\mathrm{\rm eq }}$ by a Taylor-Lagrange formula to order $3$. The existence of the order $3$ Fr\'echet derivative of $\mathcal{E}$ is here guaranteed since $\mathcal{E}$ is polynomial. Setting $\nu_N = \widetilde{L}_N^u - \mu_{\mathrm{\rm eq }}$, we find:
\beq
\label{TaylL}\mathcal{E}[\widetilde{L}_N^u] = \mathcal{E}[\mu_{\mathrm{\rm eq }}] - \Int{\mathsf{A}}{} T_{\mathrm{eff}}(\xi)\dd\nu_N(\xi) + \frac{1}{2}\mathcal{Q}[\nu_N] + \mathcal{R}_{3}[\nu_N]
\enq
and we compute from the definition of $\mathcal{E}$:
\bea
\mathcal{R}_{3}[\nu_N] & = & \Int{0}{1} \frac{\dd t\,(1 - t)^2}{2}\,\mathcal{E}^{(3)}[(1 - t)\mu_{\mathrm{\rm eq }} + t\widetilde{L}_N^u]\cdot(\nu_N,\nu_N,\nu_N)\;, \nonumber \\
\label{Fre3}\mathcal{E}^{(3)}[\mu]\cdot(\nu^1,\nu^2,\nu^3) & = & - \Int{\mathsf{A}^{r - 3}}{} \frac{T(\xi_1,\ldots,\xi_r)}{(r - 3)!}\,\prod_{i = 1}^3 \dd\nu^{i}(\xi_i) \prod_{j = 4}^{r} \dd\mu(\xi_j)\;.
\eea
Since the $\nu^i$ have zero masses, $\mathcal{E}^{(3)}[\mu]\cdot(\nu^1,\nu^2,\nu^3)$ vanishes if there are only $1$ or $2$
body interactions. In other words, the remainder $\mathcal{R}_{3}[\nu]$ is only present in the case where there are at least 
$r \geq 3$ body interactions. Since $T_{\mathrm{eff}}(x)$ is non-positive and vanishes $\mu_{\mathrm{\rm eq }}$-almost surely, we have for the linear term:
\beq
-\Int{\mathsf{A}}{} T_{\mathrm{eff}}(\xi)\dd\nu_N(\xi) = -\Int{\mathsf{A}}{} T_{\mathrm{eff}}(\xi)\,\dd \widetilde{L}_N^u(\xi) \geq 0.\nonumber
\enq
Therefore, combining with the lower bound of Lemma~\ref{lowL}, we find:
\beq
\label{intsa}\frac{\dd\mu_{\mathcal{S}}}{Z_{\mathcal{S}}} \leq \Big[\prod_{i = 1}^N \dd\lambda_i\Big]\exp\Bigg\{cN\ln N  -\frac{N^2}{2}\big(\mathcal{Q}[\nu_N] + 2 \mathcal{R}_{3}[\nu_N]\big)\Bigg\} \;. 
\enq
By using Lemma~\ref{l322} with $(l,l',l'') = (r - 3,3,2)$ and the fact that $T$ is $\mathcal{C}^{m'}$ for
 $m' > 2m + r + 1$, we get, for some $T$-dependent constant $C$:
\beq
\big|\mathcal{R}_{3}[\nu_N]\big| \leq  C[T]\,\mathcal{Q}[\nu_N]\,\p \nu_N \p\;.\nonumber
\enq
Note that, in the above bound, we have used the existence of an adapted space, as will be inferred in Section \ref{S4}.
So, if we restrict to configurations realizing the event $\{ \p\nu_N\p \le \varepsilon\} $ for some fixed but small enough $\varepsilon > 0$, we will have $\big|\mathcal{R}_3[\nu_N]\big| \leq \mathcal{Q}[\nu_N]/4$. Integrating \eqref{intsa} on this event, we find:
\beq
\mu_{\mc{S}}\Big[ \big\{\mathcal{Q}^{1/2}[\nu_N] \ge t \big\} \cap\{ \p\nu_N\p \le \varepsilon\} \big] 
\leq \exp\Big\{CN\ln N - \frac{N^2 t^2}{4}\Big\}\,. \nonumber
\enq
On the other end, since $\big\{ \nu \in \mathcal{M}^0(\mathsf{A}),\quad \p\nu-\mu_{\rm eq}\p \ge \varepsilon\big\}$ is a closed set which does not contain $\mu_{\rm eq}$, and since $L_N-\widetilde L_N^u$ converges
to zero uniformly for the weak-* topology as $N$ goes to infinity, uniformly on configurations of $\la_i$'s according to Lemma~\ref{lemu}, we find by the large deviation principle of Theorem~\ref{ldl} that
there exists a positive constant $c_\varepsilon$ such that
 $$\mu_{\mc{S}}\Big[\big\{ \p\nu_N\p \ge \varepsilon\big\} \Big] \leq e^{-c_\varepsilon N^2}\,.$$ 
This concludes the proof of Theorem~\ref{t1}. \hfill $\Box$

\subsection{Proof of Corollaries~\ref{c1},~\ref{c2}-\ref{c3}}

\subsubsection*{Proof of Corollary~\ref{c1}.}

We decompose:
\beq
\Int{\mathsf{A}}{} \varphi(\xi)\dd(L_N - \mu_{\mathrm{eq}})(\xi) = \Int{\mathsf{A}}{} \varphi(\xi)\dd\nu_N(\xi) + \Int{\mathsf{A}}{} \varphi(\xi)\dd(L_N - \widetilde{L}_N^{u})(\xi) \nonumber
\enq
As will be shown in Section \ref{S4}, Hypothesis~\ref{h2} ensures the existence of an adapted space, \textit{viz}. 
\beq
\Bigg|\Int{\mathsf{A}}{} \varphi(\xi)\dd\nu_N(\xi)\Bigg| \leq c_0\,\mathcal{Q}^{1/2}[\nu_N]\,\p \varphi \p_{\mathcal{H}} 
\qquad \text{with}  \quad \nu_N \, = \, \wt{L}_N^u \, - \, \mu_{\text{eq}}   \nonumber
\label{definition nuN}
\enq
and the second term is bounded by Lemma~\ref{lemu}. Therefore:
\bea
\mu_{\mathcal{S}}\Bigg[\Big|\Int{\mathsf{A}}{} \varphi(\xi)\dd(L_N - \mu_{\mathrm{eq}})(\xi)\Big| \geq 
\frac{c\kappa_b[\varphi]}{N^{2b}} + c_0 t\,\p \varphi \p_{\mathcal{H}}\Bigg] & \leq & \mu_{\mathcal{S}}\big[\mathcal{Q}[\nu_N]^{1/2} \geq t\big] \nonumber \\
\eea
so that the conclusion follows from Theorem \ref{t1}.
\hfill $\Box$

\subsubsection*{Proof of Corollary~\ref{c2}.}

We have set $\psi_{x}(\xi) = \mathbf{1}_{\mathsf{A}}(\xi)/(x - \xi)$, and we have:
\bea
N^{-1}\,W_1(x)- W_{\mathrm{eq}}(x) & = & \mu_{\mathcal{S}}\Big[\int \psi_x(\xi) \dd(L_N-\dd\mu_{\mathrm{eq}})(\xi)\Big] \nonumber \\
& = & 
\mu_{\mathcal{S}}\Big[ \int \psi_{x}(\xi)\dd\nu_N(\xi) + \int \psi_{x}(\xi)\dd(L_N - \widetilde{L}_N^{u})(\xi) \Big] \;,
\label{bornage du correlateur a un point}
\eea
where $\nu_N$ is as defined in \eqref{definition nuN}. The function $\psi_x$ is Lipschitz with constant 
$\kappa_1[\psi_x] = d^{-2}(x,\mathsf{A})$. In virtue of Lemma~\ref{lemu}, it follows that:
\beq
\label{isq0}\Bigg|\Int{\mathsf{A}}{} \psi_{x}(\xi)\dd(L_N - \widetilde{L}_N^u)(\xi)\Bigg| \leq \frac{c}{N^2d^2(x,\mathsf{A})}  \;. 
\enq
In what concerns the first term in \eqref{bornage du correlateur a un point}, we set $t_N = 4\sqrt{|C|\ln N/N}$ in Theorem~\ref{t1} to find, for $N$ large enough:
\bea
\label{isq}
\mu_{\mathcal{S}}\bigg[ \left| \int \psi_{x}(\xi)\dd(\widetilde{L}_N^u - \mu_{\mathrm{eq}})(\xi)\right| \bigg] 
& \leq & c_0 t_N \p \psi_{x} \p_{\mathcal{H}} + \f{2 \mu_{\mathcal{S}}\big[\mathcal{Q}^{1/2}[\nu_N]\,\geq t_N \big] }{ d(x,\mathsf{A}) } \\
& \leq &  t_N \p \psi_{x} \p_{\mathcal{H}} + \frac{c''\,e^{-3|C|\,N\ln N}}{d(x,\mathsf{A})} \nonumber
\eea
And, for $x$ bounded independently of $N$, and $N$ large enough, the last term in \eqref{isq} is $o(N^{-2}d^{-2}(x,\mathsf{A}))$. So, combining \eqref{isq0} and \eqref{isq}, we obtain the existence of constants $c,c_1 > 0$ such that:
\beq
\big|N^{-1}W_1(x) - W_{\mathrm{eq}}(x)\big| \leq \frac{c}{N^2d^2(x,\mathsf{A})} + c_1\,\p \psi_{x} \p_{\mathcal{H}} \sqrt{\frac{\ln N}{N}} \nonumber
\enq
Multiplying by $N$ entails the result.

For $n \geq 2$, $N^{-n}W_n(x_1,\ldots,x_n)$ is the $\mu_{\mathcal{S}}$ expectation value of a homogeneous polynomial of degree $n$ 
having a partial degree at most 1 in the quantities $\int_{\mathsf{A}} \psi_{x_i}(\xi_i)\dd(L_N - \mu_{\mathrm{eq}})(\xi_i)$ and $\mu_{\mathcal{S}}\big[\int_{\mathsf{A}} \psi_{x_i}(\xi_i)\dd(L_N - \mu_{\mathrm{eq}})(\xi_i)\big]$.
The coefficients of this polynomial are independent of $N$. A similar reasoning shows that, for any subset $I$ of $\intn{1}{n}$, and if $x_i$ is bounded independently of $N$ and $N$ large enough:
\beq
\mu_{\mathcal{S}}\Bigg[\prod_{i \in I} \Int{\mathsf{A}}{} \psi_{x_i}(\xi_i)\dd(L_N - \mu_{\mathrm{eq}})(\xi_i)\Bigg] \leq \prod_{i \in I} \Big[\frac{c}{N^2d^2(x_i,\mathsf{A})} + c_1\,\p \psi_{x} \p_{\mathcal{H}} \sqrt{\frac{\ln N}{N}}\Big] \nonumber
\enq
Multiplying back by $N^{ |I| }$ gives the desired result. \hfill $\Box$

\subsubsection*{Proof of Corollary~\ref{c3}.}

We have $\wt{N}_h - N\epsilon_{h}^{\star} = N\int_{\mathsf{A}} \mathbf{1}_{\mathsf{A}_h}(\xi)\dd(L_N - \mu_{\mathrm{eq}})(\xi)$.
Let us choose $(\mathsf{A}'_h)_{0 \leq h \leq g}$ to be a collection of pairwise disjoint segments, such that $\mathsf{A}_h'$ is a neighbourhood of $\mathsf{S}_h$ in $\mathsf{A}_h$, and denote $\mathsf{A}' = \bigcup_{h = 0}^{g} A_h'$. We would like to consider the model $\mu_{\mathcal{S}}$ or $\mu_{\mathcal{S}'}$ where eigenvalues are integrated over $\mathsf{A}$, or over $\mathsf{A}'$. More precisely, 
\begin{itemize}
\item in the unconstrained model, $\mathcal{S} = \mathsf{A}^N$ and $\mathcal{S}' = \big( \mathsf{A}'\big)^N$. 
\item  in model with fixed filling fractions subordinate to the vector $\mathbf{N} = (N_0,\ldots,N_{g})$, $\mathsf{A}$ 
is already partitioned as $\bigcup_{h = 0}^{g} \mathsf{A}_{h}$, and this induces a partition 
$\mathbf{A}' = (\mathsf{A}_{0}\cap\mathsf{A}', \dots,\mathsf{A}_{g}\cap\mathsf{A}' )$, and we define 
$\mathcal{S} = \mathsf{A}_{\mathbf{N}}$ and $\mathcal{S}' = \mathsf{A}'_{\mathbf{N}}$.
\end{itemize}
In either case, we stress that $\wt{N}_h = \wt{N}^{\mathcal{S}}_h$ is computed for the model $\mathcal{S}$, and we have:
\beq
\label{tisane}
\mu_{\mc{S}} \Big[ \big|\wt{N}_h - N\epsilon_{h}^{\star}\big|  \Big] \leq N \mu_{\mathcal{S}}\big[\exists\,\bs{i},\quad \lambda_{\bs{i}} \in \mathsf{B} = \mathsf{A}\setminus \overset{\circ}{\mathsf{A}} \, \!'\big]  \; + \; 
 \Big| \Oint{\Gamma'_h}{} \frac{\dd\xi}{2{\rm i}\pi}\,\big[W_1^{\mathcal{S}'}(\xi) - NW_{\mathrm{eq}}(\xi)\big] \Big|\;,
\enq
where $\Gamma'_h$ is a contour surrounding $\mathsf{A}'_h$ in $\mathbb{C}\setminus\mathsf{A}'$. The large deviations of Lemma~\ref{uuu3} give:
\beq
\mu_{\mathcal{S}}\big[\exists\,\bs{i},\quad \lambda_{\bs{i}} \in \mathsf{B}\, \!'\big] \leq 
\exp\big\{N\mathop{\mathrm{sup}}_{x \in \mathsf{B}\, \!' }\,T_{\mathrm{eff}}(x)\big\}\;, \nonumber
\enq
and by construction of $\mathsf{A}'_h$ and Hypothesis~\ref{hcc}, the sup is negative. We also observe that $W_{\mathrm{eq}}$ is the same in the models $\mu_{\mathcal{S}}$ and $\mu_{\mathcal{S}'}$. Therefore, we can use Corollary~\ref{c2} for the model $\mu_{\mathcal{S}'}$ to evaluate the large deviations of the second term, and obtain the existence of $C' > 0$ so that, for $N$ large enough:
\beq
\mu_{\mathcal{S}}\Big[| \wt{N}_h - N\epsilon_{h}^{\star}| \geq t\,\sqrt{N\ln N}\Big] \leq e^{N\ln N(C' - t^2)}\;.
\nonumber
\enq
\hfill $\Box$

\section{Construction of adapted spaces}
\label{S4}
\subsection{Example: translation invariant two-body interaction}

The construction of adapted spaces as described in Definition \ref{h3} can be easily addressed in the case of two-body interactions ($r = 2$) depending only of the distance. We shall consider in this paragraph:
\beq
T(x,y) = u(x - y) + \frac{v(x) + v(y)}{2}\;. \nonumber
\enq
The functional $\mathcal{Q}$ introduced in \S~\ref{s1u} takes the form:
\beq
\mathcal{Q}[\nu] = \int q(x - y)\,\dd\nu(x)\dd\nu(y) = \Int{\mathbb{R}}{} \mathcal{F}[q](k)\,\big|\mathcal{F}[\nu](k)\big|^2\,\dd k\;,\qquad \text{with} \quad  q(x) = -\beta\,\ln |x| - u(x)\;. \nonumber
\enq

\begin{lemma}
\label{ksiq} Assume $u \in \mathcal{C}^1(\mathbb{R})$ is such that $k \mapsto \mathcal{F}[q](k)$, with $\mathcal{F}[q]$ understood in the sense of distributions is continuous on $\mathbb{R}^{*}$ and positive everywhere, and $|k|^b\,\mathcal{F}[q](k) \geq c$ for some $c > 0$ and $b \geq 1$ when $|k|$ is small enough. Then, $\mathcal{H} = \iota_{\mathsf{A}}(H^{b/2}(\mathbb{R}))$ equipped with its Sobolev norm, and growth index $m = 0$ is an adapted space. Here, $\iota_{\mathsf{A}}$ is the operation of restriction of the domain of definition to $\mathsf{A}$.
\end{lemma}
\textbf{Proof.} Let $\nu \in \mathcal{M}^0(\mathsf{A})$ and $\varphi \in H^{b/2}(\mathbb{R})$. Then
\bea
\Bigg|\Int{\mathsf{A}}{} \varphi(x)\,\dd\nu(x)\Bigg| & = & \Bigg| \Int{\mathbb{R}}{} \mathcal{F}[\varphi](-k)\cdot \mathcal{F}[\nu](k)\,\dd k\Bigg| \nonumber \\
& \leq & \Bigg(\Int{\mathbb{R}}{} \mathcal{F}[q](k)\,\big|\mathcal{F}[\nu](k)\big|^2\,\dd k\Bigg)^{1/2}\,\Bigg(\Int{\mathbb{R}}{} \frac{\big|\mathcal{F}[\varphi](k)\big|^2}{\mathcal{F}[q](k)}\,\dd k\Bigg)^{1/2} \nonumber \\
& \leq & \mathcal{Q}^{1/2}[\nu]\,\Bigg(\Int{\mathbb{R}}{} |k|^b\,\big|\mathcal{F}[\varphi](k)\big|^2\,\frac{\dd k}{|k|^b\,\mathcal{F}[q](k)}\Bigg)^{1/2}\;. \nonumber
\eea
We observe that $|k|^b\,\mathcal{F}[q](k) = \be |k|^{b - 1} -  |k|^b\,\mathcal{F}[u](k)$. Since $u$ is $\mathcal{C}^1$ and $b \geq 1$, we have $|k|^{b}\,\mathcal{F}[q](k) \geq 1$ when $|k| \rightarrow \infty$. And, by assumption, we have $|k|^b\,\mathcal{F}[q](k) > c$ when $|k|$ is small enough. Since, furthermore, $\mathcal{F}[q](k)>0$, there exists $c' > 0$ so that $|k|^b\,\mathcal{F}[q](k) > c'$ for any $k$, and:
\beq
\Bigg|\Int{\mathsf{A}}{} \varphi(x)\,\dd\nu(x)\Bigg| \leq 
\frac{1}{ \sqrt{c'}}\,\mathcal{Q}^{1/2}[\nu]\,\p \varphi \p_{H^{b/2}}\;. \nonumber
\enq
Eventually, for some $\varepsilon > 0$, we can always find a function $\chi_{\mathsf{A}} \in \mathcal{C}^{(b + 1 + \varepsilon)/2}(\mathbb{R})$ with compact support and assuming values $1$ on $\mathsf{A}$. Then:
\beq
\mathcal{F}[{\rm e}_{k_0}] = \mathcal{F}[\chi_{\mathsf{A}}](k + k_0)\;,\qquad \lim_{|k| \rightarrow \infty} |k|^{(b + 1 + \varepsilon)/2}\,\mathcal{F}[\chi_{\mathsf{A}}](k) = 0\;. \nonumber
\enq
Therefore, $\p {\rm e}_{k} \p_{H^{b/2}}$ remains bounded when $k \in \mathbb{R}$. \hfill $\Box$

In many applications, $\mathcal{F}[q](k)$ can be explicitly computed. In such cases it is relatively easy to check its positivity 
 and extract its $k \rightarrow 0$ behaviour, which is related to the growth of $q(x)$ when $|x| \rightarrow \infty$.

\subsection{Space adapted to $\mathcal{Q}$}
\label{heh}

We now present a general construction of adapted spaces thanks to functional analysis arguments. It applies in particular to $2$-body interactions which are not translation invariant (i.e. do not have a simple representation in Fourier space) or to more general $r$-body interactions.

\begin{theorem}
Let $q > 2$ and assume that $\mc{Q}$ defined as in \eqref{definition fonctionnelle Q} satisfies to Hypothesis~\ref{h2}. Then, the space 
$ W^{1;q}(\mathsf{A})$ equipped with its Sobolev norm $\p \cdot \p_q$ is an adapted space. It has a
growth index $m=1$. 
\end{theorem}
\noindent \textbf{Proof.} Let $2> p \geq 1$ be conjugate to $q$ (\textit{viz}. $1/p + 1/q =1$) 
and define the integral operator $\underline{\mc{T}}: L^{p}_0(\mathsf{A}) \mapsto L^p_0(\mathsf{A})$ with 
\beq
L_0^p(\mathsf{A}) \; = \; \bigg\{ f \in L^{p}(\mathsf{A}) \; : \; \Int{\mathsf{A}}{}f(x) \, \dd x \, = \, 0 \bigg\} \nonumber
\enq
by
\beq
\underline{\mc{T}}[f](x) \; = \; - \be \Int{ \mathsf{A} }{} \ln|x-y|  f(y) \, \dd y  \; - \;   \Int{ \mathsf{A} }{} 
\tau(x,y) \, f(y) \, \dd y \; + \; 
\Int{ \mathsf{A}^2 }{} \Big( \be  \ln|x-y| \; + \;   \tau(x,y) \Big) f(y) \, \dd y  \dd x  \;, \nonumber
\enq
where 
\beq
\tau(x,y) \; = \; \Int{ \mathsf{A}^{r-2} }{} \frac{T\big( x , y, x_1, \dots, x_{r-2}\big)}{(r - 2)!} \, \pl{a=1}{r-2} \dd \mu_{\text{eq}}(x_a)  \;. \nonumber
\enq
The functional $\mc{Q}$ is strictly positive definite by Hypothesis \ref{h2}. Given $\vp \in L^{r}_0(\mathsf{A})$, 
$\vp \not= 0$ and $r>1$, it is clear
from the fact that $\mathsf{A}$ is compact that 
$\underline{\mc{T}}[\vp] \in  L^{r}_0(\mathsf{A})$ as well. Hence, 
\beq
\Int{A}{} \vp(x) \underline{\mc{T}}[\vp](x)\,\dd x \; = \; \mc{Q}\big[ \nu_{\vp} \big]  \; > \; 0 \qquad \text{with} \quad 
\dd \nu_{\vp} = \vp(x)\,\dd x \in \mc{M}^{0}(\mathsf{A}) \;. \nonumber
\enq
In particular,  $\underline{\mc{T}}$ defines a continuous positive definite self-adjoint operator on $L^{2}_0(\mathsf{A})$. 
By functional calculus on its spectrum, one can define any power of the operator $\underline{\mc{T}}$
as an operator on $L^{2}_0(\mathsf{A})$. Further, observe that given $\varphi \in L^{2}_0(\mathsf{A})$ and 
any $1 \leq  p < 2$
\beq
\label{jih}\Int{ \mathsf{A} }{} \underline{\mc{T}}^{\f{1}{2}}[\varphi] (x)\,\underline{\mc{T}}^{\f{1}{2}}[\varphi] (x)\,\dd x \; = \;
\Int{ \mathsf{A} }{} \varphi(x)\,\underline{\mc{T}}[\varphi] (x)\,\dd x  \; \leq \; \norm{ \varphi }_{L^p(\mathsf{A})}\,
\norm{ \underline{\mc{T}}[\varphi] }_{L^q(\mathsf{A})}  \; \leq  \; C  \norm{ \varphi }_{L^p(\mathsf{A})}^2 
\enq
where $q$ is conjugated to $p$. To get the last inequality, we have used that $\p \mathcal{T}[\varphi] \p_{L^q(\mathsf{A})} \leq C\,\p \mathcal{T}[\varphi]\|_{L^{\infty}(\mathsf{A})}$ since $\mathsf{A}$ is compact, and since the kernel 
 $ \mathscr{T}$ of $\mathcal{T}$ is in $L^{q}$ uniformly -- provided $q < \infty$ -- so if 
$1/p + 1/q = 1$ we can bound:
\beq
\p \,\underline{\mathcal{T}}[\varphi] \,\p_{L^{\infty}(\mathsf{A})} \leq C\,\mathrm{sup}_{x \in \mathsf{A}} \p \mathscr{T}(x,\bullet) \p_{L^{q}(\mathsf{A})}  \,\p \varphi \p_{L^p(\mathsf{A})}\,.\nonumber
\enq
From \eqref{jih}, we deduce that $\underline{\mc{T}}^{\f{1}{2}}$ extends into a continuous operator 
$\underline{\mc{T}}^{\f{1}{2}} \; = \; L^{p}_0(\mathsf{A}) \mapsto L^{2}_0(\mathsf{A})$ for any $1 \leq p < 2$. 
Further, it is established in the appendix, see Proposition \ref{theop}, that 
$\underline{\mc{T}}^{-1} \; : \; W^{1;q}(\mathsf{A}) \mapsto L^{p}_0(\mathsf{A})$ is continuous. Thus, 
given $\varphi_1 \in L^{2}_0(\mathsf{A})$ and $ \varphi_2 \in W^{1;q}(\mathsf{A})$, one has that 
\bea
\Big|  \Int{ \mathsf{A} }{} \varphi_1(x)\cdot\varphi_2(x)\,\dd x \Big| &  = & 
\Big| \Int{ \mathsf{A} }{} \varphi_1(x)\cdot\underline{\mc{T}}^{\f{1}{2}} \circ \underline{\mc{T}}^{\f{1}{2}}\Big[ \underline{\mc{T}}^{-1}[\varphi_2] \Big](x)\,\dd x   \Big|
=
\Big| \Int{ \mathsf{A} }{} \underline{\mc{T}}^{\f{1}{2}}[\varphi_1](x)\cdot\underline{\mc{T}}^{\f{1}{2}}\Big[ \underline{\mc{T}}^{-1}[\varphi_2] \Big](x)\,\dd x   \Big|  \nonumber \\
& \leq & \Big( \mc{Q}[\nu_{\varphi_1}] \Big)^{\f{1}{2} }\,\bigg( \Int{ \mathsf{A} }{}   \underline{\mc{T}}^{\f{1}{2}}\Big[ \underline{\mc{T}}^{-1}[\varphi_2] \Big](x)\cdot  \underline{\mc{T}}^{\f{1}{2}}\Big[ \underline{\mc{T}}^{-1}[\varphi_2] \Big](x)\,\dd x  \bigg)^{\f{1}{2}} \nonumber
\eea
In the first line, we have used that $ \underline{\mc{T}}^{-1} \circ \underline{\mc{T}}^{-1} = \text{id}_{ W^{1;q}(\mathsf{A}) }$. 
In order to obtain the second equality, we have used that 
$\underline{\mc{T}}^{\f{1}{2}}\Big[ \underline{\mc{T}}^{-1}[\varphi_2] \Big] \in L^{2}(\mathsf{A})$, $\underline{\mc{T}}^{\f{1}{2}}$ 
is a continuous self-adjoint operator on $L^{2}(\mathsf{A})$ and that $\vp_{1} \in L^{2}(\mathsf{A})$. 

For $x \not \in \mathsf{A}$, let $\sg_{\mathsf{A}}(z) \; = \; \big(\prod_{a \in \Dp{} \mathsf{A} } (x-a)\big)^{1/2}$, where the square root is taken so that $\sg_{\mathsf{A}}(x) \sim x^{g+1}$ when $x \tend \infty$. 
It is readily checked on the basis of the explicit expression given in \eqref{ecriture inverse de operateur mathcal T}, that 
 for any $\varepsilon>0$, the function 
\beq
\Phi_{\varepsilon}( x ) \; = \; | \sg^{\tf{1}{2}}_{\mathsf{A},+}|^{\varepsilon}(x) \,\underline{\mc{T}}^{-1}[\varphi_2](x) \; - \; 
 \Int{  \mathsf{A} }{} | \sigma_{\mathsf{A},+}^{\tf{1}{2}}|^{\varepsilon}(\xi)\,\underline{\mc{T}}^{-1}[\varphi_2](\xi)\,\dd \xi \nonumber
\enq
 belongs to $L^{2}_0(\mathsf{A})$ and converges in $L^{p}_0(\mathsf{A})$, $1 \leq p <2$, to $\underline{\mc{T}}^{-1}[\phi]$. Thus
\begin{multline*}
\Int{ \mathsf{A} }{}   \underline{\mc{T}}^{\f{1}{2}}\Big[ \underline{\mc{T}}^{-1}[\varphi_2] \Big](x)\cdot\underline{\mc{T}}^{\f{1}{2}}\Big[ \underline{\mc{T}}^{-1}[\varphi_2] \Big](x)\,\dd x
\; = \; \Int{ \mathsf{A} }{}  \lim_{\varepsilon \tend 0^+}\Big\{
 \underline{\mc{T}}^{\f{1}{2}}\big[\Phi_{\varepsilon} \big](x)\cdot\underline{\mc{T}}^{\f{1}{2}}\big[ \Phi_{\varepsilon} \big](x)  \Big\}\,\dd x  \\
 \; = \; \lim_{\varepsilon \tend 0^+}\bigg\{ 
 \Int{ \mathsf{A} }{} \underline{\mc{T}}^{\f{1}{2}}\big[\Phi_{\varepsilon} \big](x)\cdot\underline{\mc{T}}^{\f{1}{2}}\big[ \Phi_{\varepsilon} \big](x)\,\dd x \bigg\}
\; = \; \lim_{\varepsilon \tend 0^+}\bigg\{ 
 \Int{ \mathsf{A} }{} \Phi_{\varepsilon}(x)\cdot\underline{\mc{T}}\big[ \Phi_{\varepsilon} \big](x)\,\dd x \bigg\} 
\; = \; \Int{ \mathsf{A} }{} \Phi_{0}(x)\cdot\underline{\mc{T}}\big[ \Phi_{0} \big](x)\,\dd x \; . 
\end{multline*}
Above we have used the continuity of $\underline{\mc{T}}^{\f{1}{2}}$ on $L^{p}_0(\mathsf{A})$,
the dominated convergence, the self-adjointness of $\underline{\mc{T}}^{\f{1}{2}}: L^{2}_0(\mathsf{A}) \tend L^{2}_0(\mathsf{A})$
and, finally, dominated convergence and the fact that $ \underline{\mc{T}}\big[ \Phi_{\varepsilon} \big](x) \in L^{\infty}_0(\mathsf{A})$
uniformly in $\eps$. 
 
Thus, all in all, we have shown that
\beq
\Big|  \Int{ \mathsf{A} }{} \varphi_1(x)\cdot\varphi_2(s)\,\dd x \Big| \; \leq \; \Big( \mc{Q}[\nu_{\vp_1}] \Big)^{\f{1}{2} }\,
 \bigg( \Int{ \mathsf{A} }{}  \varphi_2(x)\cdot\underline{\mc{T}}^{-1}[\varphi_2](x)\,\dd x  \bigg)^{\f{1}{2}} \nonumber
\enq
For any $\varphi_2 = \phi \in  W^{1;q}(\mathsf{A})$ and $\nu \in \mc{M}^0(\mathsf{A})$ such that $\mc{Q}^{1/2}[\nu]$, we apply this bound to $\varphi_{1;m} = (\nu * G_m)/\dd x \in L^2_0(\mathsf{A})$ for $G_m$ a centered Gaussian distribution with variance $1/m$. Then, noticing that
$$
\lim_{m \rightarrow \infty} \mathcal{Q}^{1/2}[\varphi_{1;m}] = \mathcal{Q}^{1/2}[\nu],
$$
by an argument similar to \cite[Lemma 2.2]{ZBenarous}, we obtain:
\beq
\Big|  \Int{ \mathsf{A} }{} \phi(x)\,\dd \nu(x) \Big| \;  \leq \; 
\mc{Q}^{1/2}[\nu]\,\bigg( \Int{ \mathsf{A} }{}  \phi(x) \cdot \underline{\mc{T}}^{-1}[\phi](x)\,\dd s  \bigg)^{\f{1}{2}} \;. \nonumber
\enq
The claim follows from H\"{o}lder's inequality and invoking the continuity of $\underline{\mc{T}}^{-1}$.

\section{Schwinger-Dyson equations and linear operators}
\label{S5}

\subsection{Hierarchy of Schwinger-Dyson equations}

To write the Schwinger-Dyson equations in a way amenable to asymptotic analysis, we require:
\begin{hypothesis}
\label{h1bis}Hypothesis~\ref{h1} and $T$ is holomorphic in a neighbourhood of $\mathsf{A}$.
\end{hypothesis}
Theorem~\ref{thregana} is therefore applicable. It is convenient for the asymptotic analysis to introduce:
\beq
\label{sigmadef}
\sigma_{ \text{hd} }(x) = \prod_{\alpha \in \partial\mathsf{S}\cap\partial\mathsf{A}} (x - \alpha)\;,\qquad \sigma_{\mathsf{S}}(x) = \prod_{\alpha \in \partial\mathsf{S}} (x - \alpha)\;,
\enq
and:
\beq
\sigma_{\text{hd}}^{[1]}(x,\xi) = \frac{\sigma_{\text{hd}}(x) - \sigma_{\text{hd}}(\xi)}{x - \xi}\;,\qquad
 \sigma_{\text{hd}}^{[2]}(x;\xi_1,\xi_2) = \frac{\sigma_{\text{hd}}^{[1]}(x,\xi_1) - \sigma_{\text{hd}}^{[1]}(x,\xi_2)}{\xi_1 - \xi_2}\;. \label{tyi}
\enq
Then, for any $n \geq 1$ and $I$ a set of cardinality $n-1$, 
the Schwinger-Dyson equations take the form:
\bea
\Big(1 - \frac{2}{\beta}\Big)\,\partial_{x} W_n(x,\bs{x}_I) + W_{n + 1}(x,x,\bs{x}_I) 
+ \sum_{J \subseteq I} W_{|J| + 1}(x,\bs{x}_J)W_{n - |J|}(x,\bs{x}_{I\setminus J}) & & \nonumber \\
- \frac{2}{\beta}\sum_{\substack{a \in \partial\mathsf{A} \\ \setminus \Dp{} \mathsf{S}   }} 
\frac{\sigma_{ \text{hd} }(a)}{\sigma_{ \text{hd} }(x)}\,\frac{\partial_{a}\,W_{n - 1}(\bs{x}_I)}{x - a}
 + \frac{2}{\beta}\,N^{2 - r}\Oint{\mathsf{A}^r}{} \frac{\dd^r\bs{\xi}}{(2{\rm i}\pi)^{r}}\,
\frac{\sigma_{ \text{hd} }(\xi_1)}{\sigma_{ \text{hd}  }(x)}\,\frac{\partial_{\xi_1} T(\xi_1,\ldots,\xi_r)}{(r - 1)!\,(x - \xi_1)}\,
\overline{W}_{r; n-1}(\xi_1,\ldots,\xi_r \mid \bs{x}_I) 
 & & \nonumber \\ 
+ \Big(1 - \frac{2}{\beta}\Big)\oint_{\mathsf{A}} \frac{\dd\xi}{2{\rm i}\pi}\,
\frac{\sigma_{ \text{hd}  }^{[2]}(x;\xi,\xi)}{\sigma_{ \text{hd}  }(x)}
\,W_n(\xi, \bs{x}_I) + \frac{2}{\beta}\sul{i \in I}{} \Oint{\mathsf{A}}{} \frac{\dd\xi}{2{\rm i}\pi}\,
 \frac{\sigma_{ \text{hd}  }(\xi)}{ \sigma_{ \text{hd}  }(x)}\,\frac{W_{n - 1}(\xi, \bs{x}_{I\setminus\{i\}})}{(x - \xi)(x_i - \xi)^2}  
&  &\label{TheSD} \\ 
- \Oint{\mathsf{A}}{} \frac{\dd^2\bs{\xi}}{(2{\rm i}\pi)^2}\,\frac{\sigma_{ \text{hd}  }^{[2]}(x;\xi_1,\xi_2)}{\sigma_{ \text{hd}  }(x)}
\Big\{ W_{n+1}(\xi_1,\xi_2, \bs{x}_I)  \; + \; \sul{J \subseteq I}{}  W_{|J|+1}\big(\xi_1,\bs{x}_J\big) 
 W_{n-|J|}\big(\xi_2,\bs{x}_{I\setminus J} \big)  \Big \}
 &= & 0\;. \nonumber 
\eea
There, $\bs{x}_I$ is as defined in \eqref{ecriture reduction T a somme pots k corps} and we have made use of the semi-connected correlators: 
\beq
\overline{W}_{r; n}(\xi_1,\ldots,\xi_r \mid x_1,\dots,x_n) \; = \; 
\Dp{t_1}\dots \Dp{t_n} \wt{W}_{r}(\xi_1,\ldots,\xi_r )\big[ T \tend  \wt{T}_{t_1\dots t_n} \big] \nonumber
\enq
where $\wt{T}_{t_1\dots t_n}$ is as defined in \eqref{definition tilde T}. For instance:
\bea
\overline{W}_{2;2}(\xi_1,\xi_2|x_1,x_2) & = & W_{4}(\xi_1,\xi_2,x_1,x_2) + W_{3}(\xi_1,x_1,x_2)W_1(\xi_2) \nonumber \\
& & + W_2(\xi_1,x_1)W_2(\xi_2,x_2) + W_2(\xi_1,x_2)W_2(\xi_2,x_1) + W_1(\xi_1)W_3(\xi_2,x_1,x_2)\;.
\eea
We also use the convention $W_0 = \ln Z$.

\subsection{The master operator}
\label{S53}
Upon a naive  expansion of the Schwinger-Dyson equation around $W_1 = NW_{\mathrm{eq}} + o(N)$, there arises a linear operator $\mathcal{K}\,:\,\mathscr{H}^m(\mathsf{A}) \rightarrow \mathscr{H}^1(\mathsf{A})$. This operator depends on the Stieltjes transform $W_{\mathrm{\rm eq }}$ of the equilibrium measure and on $T$ and is given by 
\begin{multline*}
\mathcal{K}[\varphi](x) \; = \;  2W_{\mathrm{\rm eq }}(x)\,\varphi(x)  
\; - \; 2 \Oint{ \mathsf{A}^2 }{} \f{ \dd^2 \bs{\xi}}{ (2i\pi)^2 }\,\f{ \sg_{\text{hd}}^{[2]}(x;\xi_1,\xi_2) }{ \sg_{\text{hd}}(x)  }
\,\vp(\xi_1) W_{\mathrm{\rm eq }}(\xi_2) \\
\; +  \; 
\frac{2}{\beta}\Oint{\mathsf{A}^r}{} \frac{\dd^{r}\bs{\xi}}{(2{\rm i}\pi)^r}\,\frac{\sigma_{\text{hd}}(\xi_1)}{\sigma_{\text{hd}}(x)}\,\frac{\partial_{\xi_1} T(\xi_1,\ldots,\xi_r)}{(r - 1)!\,(x - \xi_1)}
		\Big\{ \varphi(\xi_1)W_{\mathrm{\rm eq }}(\xi_2) + (r-1) W_{\mathrm{\rm eq }}(\xi_1)\varphi(\xi_2) \Big\}
			\Big[\prod_{i = 3}^{r} W_{\mathrm{\rm eq }}(\xi_i)\Big] \;. 
\end{multline*}
It is then necessary to invert $\mathcal{K}$ in a continuous way in order to study the corrections to the leading order of the correlators via the Schwinger-Dyson equations. The two lemmata below answer this question, and are the key to the bootstrap analysis of Section~\ref{S5}. Let us introduce the period map $\Pi\,:\,\mathscr{H}^m(\mathsf{A}) \rightarrow \mathbb{C}^{g+1}$ as:
\beq
\Pi[\varphi] \;  = \;  \Bigg(\Oint{\mathsf{A}_0}{} \frac{\dd\xi\,\varphi(\xi)}{2{\rm i}\pi} \, , \dots  \, ,
 \Oint{\mathsf{A}_g}{}  \frac{\dd\xi\,\varphi(\xi)}{2{\rm i}\pi} \Bigg) \;. \nonumber
\enq
We denote $\mathscr{H}^m_0(\mathsf{A}) = \mathrm{Ker}\,\Pi$. 

\begin{lemma}
\label{lemy}Assume the local strict convexity of Hypothesis~\ref{h2}, the analyticity of Hypothesis~\ref{h1bis}, and that $\mu_{\mathrm{eq}}$ is off-critical (Definition~\ref{offcsq}). Let $m \geq 1$. Then, the restriction of $\mathcal{K}$ to $\mathscr{H}^m_0(\mathsf{A})$ is invertible on its image $\mathscr{J}^{m}(\mathsf{A})=\mc{K}\big[ \mathscr{H}^m_0(\mathsf{A}) \big]$. 
\end{lemma}
\begin{lemma}
\label{lemy2}
$\mathscr{J}^2(\mathsf{A})$ is a closed subspace of $\mathscr{H}^1(\mathsf{A})$, and for any contour $\Gamma$ surrounding $\mathsf{A}$ in $\mathbb{C}\setminus\mathsf{A}$, and $\Gamma[1]$ exterior to $\Gamma$, there exists a constant $c > 0$ so that:
\beq
\forall \psi \in \mathscr{J}^2(\mathsf{A})\;,\qquad \p \mathcal{K}^{-1}[\psi] \p_{\Gamma[1]} \leq c\,\p \psi \p_{\Gamma}\;. \nonumber
\enq
\end{lemma}
The remaining of this section is devoted to the proof of those results.

\subsection{Preliminaries}
\label{SecP}
We remind that, in the off-critical regime, according to the definitions given in \eqref{sigmadef} and in virtue of Lemma~\ref{thregana},
one has the representation:
\beq
\frac{\dd\mu_{\mathrm{eq}}}{\dd x}(x) = \frac{\mathbf{1}_{\mathsf{S}}(x)}{2\pi}\,M(x)\,\Bigg|\frac{\sigma_{\mathsf{S}}^{1/2}(x)}{\sigma_{{\rm hd}}(x)}\Bigg| \nonumber
\enq
with $M$ holomorphic and nowhere vanishing in a neighbourhood of $\mathsf{A}$. Equivalently, in terms of the Stieltjes transform:
\beq
\label{defer2}W_{\text{eq};-}(x) \; - \; W_{\text{eq};+}(x) \; = \; \f{ M(x) \sg_{\mathsf{S};-}^{1/2}(x) }{ \sg_{\text{hd}}(x) } \;. 
\enq
And, from the formula \eqref{516} for the Stieltjes transform:
\bea
\label{defer}2W_{\mathrm{\rm eq }}(x) - V'(x) & = &   M(x)\,\frac{\sigma_{\mathsf{S}}^{1/2}(x)}{\sigma_{ \text{hd} }(x)}\;, \\
\label{defer1} V'(x) & = & -\frac{2}{\beta} \Oint{\mathsf{S}^{r - 1}}{} \frac{\dd^{r - 1}\bs{\xi}}{(2{\rm i}\pi)^{r - 1}}\,\frac{\partial_{x}T(x,\xi_2,\ldots,\xi_r)}{(r - 1)!}\,\prod_{i = 2}^r W_{\mathrm{\rm eq }}(\xi_i)\;, 
\eea
There $\sigma^{1/2}_{\mathsf{S}}(x)$ is the square root such that $\sigma_{\mathsf{S}}^{1/2}(x) \sim  x^{g+1}$ when $x \rightarrow \infty$. To rewrite $\mathcal{K}$ in a more convenient form, we introduce four auxiliary operators. Let $m \geq 1$:
\begin{itemize}
\item[$\bullet$] $\mathcal{O}\,:\,\mathscr{H}^m(\mathsf{A}) \rightarrow \mathscr{O}(\mathsf{A})$ is defined by:
$$
\mathcal{O}[\varphi](x) = \frac{2}{\beta} \Oint{\mathsf{A}^{r - 1}}{} \frac{\dd^{r - 1}\bs{\xi}}{(2{\rm i}\pi)^{r - 1}}\,
\frac{\partial_{x} T(x,\xi_2,\ldots,\xi_r)}{(r - 2)!}\,\varphi(\xi_2)\,\Big[\prod_{i = 3}^r W_{\mathrm{\rm eq }}(\xi_i)\Big]\;.
$$
\item[$\bullet$] $\mathcal{L}\,:\,\mathscr{H}^m(\mathsf{A}) \rightarrow \mathscr{H}^{g + 2}(\mathsf{A})$ is defined by:
$$
\mathcal{L}[\varphi](x) = \oint_{\mathsf{A}} \frac{\dd\xi}{2{\rm i}\pi}\,\frac{\sigma_{\mathsf{S}}^{1/2}(\xi)}{\sigma_{\mathsf{S}}^{1/2}(x)}\,\frac{\mathcal{O}[\varphi](\xi)}{2(x - \xi)}\;.
$$
As a matter of fact, since $\mathcal{O}[\varphi]$ is holomorphic in a neighbourhood of $\mathsf{A}$, the contour integral in the formula above can be squeezed to 
$\mathsf{S}$, and $\mathrm{Im}\,\mathcal{L} \subseteq \mathscr{H}^{g + 2}(\mathsf{S})$.
\item[$\bullet$] $\mathcal{P}\,:\,\mathscr{H}^m(\mathsf{A}) \rightarrow \mathscr{H}^m(\mathsf{A})$ is defined by:
$$
\mathcal{P}[\varphi](x) \;  = \; 
\mathop{\mathrm{Res}}_{\xi \rightarrow \infty} \frac{\sigma_{\mathsf{S}}^{1/2}(\xi)\,\varphi(\xi)\,\dd\xi}
 			{\sigma_{\mathsf{S}}^{1/2}(x)\,(x - \xi)}\;.
$$
By construction,  $\mathcal{P}$ is a projector with:
$$
\mathrm{Ker}\,\mathcal{P} = \mathscr{H}^{g + 2}(\mathsf{A})\;,\qquad
 \mathrm{Im}\,\mathcal{P} =\sigma_{\mathsf{S}}^{-1/2}\cdot\mathbb{C}_{g + 1 - m}[x] \subseteq \mathscr{H}^{m}(\mathsf{S})\;.
$$

\item[$\bullet$] $\mathcal{I}\,:\,\mathscr{H}^1(\mathsf{A}) \rightarrow \mathscr{H}^1(\mathsf{A})$ is defined by:
$$
\mathcal{I}[\psi](x) = \Oint{\mathsf{A}}{} \frac{\dd\xi}{2{\rm i}\pi}\,\frac{\sigma_{ \text{hd} }(\xi)\,\psi(\xi)}{M(\xi)\,(x - \xi)} 
\; .
$$
Its kernel is the space of rational functions with at most simple poles at hard edges (\textit{ie}. the zeroes of $\sigma_{ \text{hd} }$). 
Its pseudo-inverse $\mathcal{I}^{-1}\,:\,\mathscr{H}^1(\mathsf{A}) \rightarrow \mathscr{H}^1(\mathsf{A})/\mathrm{Ker}\,\mathcal{I}$ 
can be readily described:
$$
\mathcal{I}^{-1}[\varphi](x) = \Oint{\mathsf{A}}{} \frac{\dd\xi}{2{\rm i}\pi}\,
\frac{M(\xi)\,\varphi(\xi)}{\sigma_{ \text{hd} }(\xi)\,(x - \xi)} \; .
$$
\end{itemize}

\begin{lemma}
\label{invert} We have the factorization between operators in $\mathscr{H}^m(\mathsf{A})$:
\beq
\label{anti}\mathrm{id} + \mathcal{L} - \mathcal{P} = \sigma_{\mathsf{S}}^{-1/2}\cdot(\mathcal{I}\circ\mathcal{K})\;.
\enq
\end{lemma}
\vspace{0.2cm}
\noindent \textbf{Proof.} A sequence of elementary manipulations allows one to recast $\mc{K}$ in the form 
\beq
\mc{K}[\vp](x) \; = \; \Oint{\mathsf{A}}{} \f{\dd \xi }{2 {\rm i} \pi } \f{ \sg_{\text{hd}}(\xi) }{  \sg_{\text{hd}}(x) }
\f{1}{x-\xi} \Big\{ \big[ 2 W_{\text{eq}}(\xi) - V^{\prime}(\xi) \big]\vp(\xi)  \; + \; 
 W_{\text{eq}}(\xi) \mc{O}[\vp](\xi) \Big\}\;,
\label{reecriture noyau integral K}
\enq
where the definition of $V(x)$ was given in \eqref{defer1}. Using \eqref{defer} and the fact that $M$ is holomorphic and nowhere vanishing in a neighbourhood of $\mathsf{A}$, we find:
\beq
(\mc{I}\circ  \mc{K})[\vp](x) \; = \; \Oint{ \mathsf{A} }{} \f{ \dd \xi  }{ 2 {\rm i } \pi } 
\f{ \sg_{\mathsf{S}}^{1/2}( \xi ) \vp(\xi)  }{ x-  \xi } \;\;  + \; 
\Oint{ \mathsf{A} }{}  \f{\dd \xi }{ 2 {\rm i } \pi  }
\f{  \sg_{\text{hd} }( \xi )\,W_{\text{eq}}(\xi)\,\mc{O}[\vp](\xi)  }{ M(\xi) (x-\xi) }  \;.  \nonumber
\enq
The first integral can be computed by taking the residues outside of the integration contour, whereas 
the second integral can be simplified by squeezing the integration contour to $\mathsf{S}$ and then using \eqref{defer2}.  Coming back to a contour integral, we obtain:
\beq
(\mc{I}\circ  \mc{K})[\vp](x) \; = \; \sg_{\mathsf{S}}^{1/2}(x)\,\vp(x) \, - \, \sg_{\mathsf{S}}^{1/2}(x)\,\mc{P}[\vp](x)
\; + \;  \Oint{ \mathsf{A} }{} \f{ \dd \xi  }{ 2 {\rm i } \pi } 
\f{   \sg_{\mathsf{S}}^{1/2}(\xi) \mc{O}[\vp](\xi)  }{ 2 (x-\xi) } \;, \nonumber
\enq
which takes the desired form. \hfill $\Box$

\subsection{Kernel of the master operator}
\label{S54}
The factorization property of Lemma~\ref{invert} implies:
\begin{cor}
\beq
\mathrm{Ker}\,\mathcal{K} \subseteq \mathrm{Ker}(\mathrm{id} + \mathcal{L} - \mathcal{P})\;, \nonumber
\enq
with equality when there is no hard edge.
\end{cor}
We may give an alternative description of the kernel of $\mathcal{I}\circ\mathcal{K}$.
\begin{lemma}

The three properties are equivalent:

\begin{itemize}
\item[$(i)$] $\mathcal{K}[\varphi] \in \mathrm{Ker}\,\mathcal{I}$.
\item[$(ii)$] $\varphi \in \mathscr{H}^m(\mathsf{S})$, the function $\sigma_{\mathsf{S}}^{1/2}\cdot\varphi$ has continuous $\pm$ boundary values on $\mathsf{S}$, and for any $x \in \mathring{\mathsf{S}}$:
$$
\varphi_+(x ) + \varphi_-(x) + \mathcal{O}[\varphi](x) = 0\;.
$$
\item[$(iii)$] The expression:
$$
\dd\nu_{\varphi}(x) = \big(\varphi_-(x) \, - \,  \varphi_+(x) \big) \cdot \frac{\dd x}{2{\rm i}\pi}
$$
defines a complex measure supported in $\mathsf{S}$ with density in $L^{p}(\mathsf{S)}$ for $p<2$. It satisfies the singular integral equation: for any $x \in \mathring{\mathsf{S}}$,
$$
\beta \Fint{\mathsf{S}}{} \frac{\dd\nu_{\varphi}(\xi)}{x - \xi}
\,  + \, 
 \Int{\mathsf{S}^{r - 1}}{} \frac{\partial_{x} T(x,\xi_2,\ldots,\xi_r)}{(r - 2)!}\,\dd\nu_{\varphi}(\xi_2)\prod_{i = 3}^r \dd\mu_{\mathrm{\rm eq }}(\xi_i) = 0\;.
$$
\end{itemize}
\end{lemma}

\noindent \textbf{Proof.}
$\bullet$ $(i) \Rightarrow (ii)$ -- If $\varphi$ satisfies $(i)$, then:
\beq
\label{muda}\varphi(x) = (\mathcal{P} - \mathcal{L})[\varphi](x)\;.
\enq
From the definition of our operators, $\sigma_{\mathsf{S}}^{1/2}(x)\,(\mathcal{P} - \mathcal{L})[\varphi](x)$ is holomorphic on 
$\mathbb{C}\setminus\mathsf{S}$, and admits continuous $\pm$ boundary values on $\mathsf{S}$. So, 
equation \eqref{muda} ensures that $\varphi(x) \sg_{\mathsf{S}}^{1/2}(x)$ admits 
continuous $\pm$-boundary values on  $\mathsf{S}$.  Given the definition of $\mathcal{L}$, we have:
\beq
\label{Lplus}\forall x \in \mathsf{S},\qquad 
\mathcal{L}[\varphi]_+(x ) \, + \, \mathcal{L}[\varphi]_-(x) = \mathcal{O}[\varphi](x)\;.
\enq
Hence, the claim follows upon computing the sum of the + and - boundary values of $\varphi(x)$ expressed by \eqref{muda}.

\noindent  $\bullet \; (ii) \Rightarrow (i)$ -- Conversely, assume $\varphi$ satisfies $(ii)$. Then, the definition of $\mathcal{K}$ implies that 
$\sigma_{\text{hd}}(x)\,\mathcal{K}[\varphi](x)$ has continuous $\pm$ boundary values on $\mathsf{S}$. Let us compute the difference of those boundary 
values using \eqref{reecriture noyau integral K}. For $x \in \mathsf{S}$:
\bea
\mathcal{K}[\varphi]_-(x) -\mathcal{K}[\varphi]_+(x) & = & 
\big( W_{\mathrm{\rm eq };-}(x) - W_{\mathrm{\rm eq };+}(x)\big)\big(\varphi_+(x ) + \varphi_-(x ) 
\; + \; \mathcal{O}[\varphi](x) \big) \nonumber \\
&& + \big(W_{\mathrm{\rm eq };+}(x ) + W_{\mathrm{\rm eq };-}(x ) \, - \,  V'(x)\big) \big(\varphi_-(x ) - \varphi_+(x)\big)\;. \nonumber
\eea
Since $\varphi$ satisfies $(ii)$,  for any $x \in \mathsf{S}$, the second factor in the first
line vanishes. Moreover, the first factor in the second line vanishes as well by the characterization of the equilibrium measure. 
Hence, $\sigma_{\text{hd}}(x)\,\mathcal{K}[\varphi](x)$ has continuous and equal $\pm$ boundary values on $\mathsf{S}$.
As a consequence $\sigma_{\text{hd}}(x)\,\mathcal{K}[\varphi](x)$ is an entire function. Since $\mathcal{K}[\varphi](x)$ behaves as 
$O(1/x)$ when $x \rightarrow \infty$, we deduce that $\mathcal{K}[\varphi](x)$ is a rational function with at most simple poles at hard edges, \textit{ie}. that $\mc{K}[\vp]$ belongs to $\mathrm{Ker}\,\mathcal{I}$. \hfill $\Box$

\vspace{0.2cm}

\noindent $ \bullet \; (iii) \Leftrightarrow (ii)$ -- $(iii)$ is stronger than $(ii)$. Conversely, assume $\varphi$ satisfies $(ii)$. Since $\sigma_{\mathsf{S}}$ has simple zeroes, the information in $(ii)$ imply that $\dd\nu_{\varphi}$ is an integrable, complex measure, which   has a density which is $L^p(\mathsf{S})$ for any $p < 2$. By construction:
\beq
\varphi(x) = \Int{\mathsf{S}}{} \frac{\dd\nu_{\varphi}(\xi)}{x - \xi}\;. \nonumber
\enq
Then, the equation for the $\pm$ boundary values of $\varphi$ in $(ii)$ translates into the singular integral equation for the measure $\dd\nu_{\varphi}$. \hfill $\Box$

\vspace{2mm}

\noindent \textbf{Proof of Lemma~\ref{lemy}}. We need to show that the restriction of $\mathcal{K}$ to $\mathrm{Ker}\,\Pi$ is injective. Let $\varphi \in \mathrm{Ker}\,\mathcal{K}\cap\mathrm{Ker}\,\Pi$. The singular integral equation of $(iii)$ holds since $\mathcal{K}[\varphi]=0\in \mathrm{Ker}\mathcal I$. Let us integrate it: there exist constants $c_0,\dots,c_g$ such that, 
\beq
\forall \, x \in \mathsf{S}_h \qquad  \beta \Int{\mathsf{S}}{} \ln|x - \xi|\,\dd\nu_{\varphi}(\xi) +  \Int{\mathsf{S}^{r - 1}}{} \frac{T(x,\xi_2,\ldots,\xi_r)}{(r - 2)!}\,\dd\nu_{\varphi}(\xi_2) \prod_{i = 3}^r \dd\mu_{\mathrm{\rm eq }}(\xi_i) = c_h\; . 
\label{equation pour nu phi dans noyau K et Pi}
\enq
Now, let us integrate under  $\dd\nu_{\varphi}^*(x)$ (here $*$ denotes the complex conjugate) and integrate over $x \in \mathsf{S}$.
This last operation is licit since every term in \eqref{equation pour nu phi dans noyau K et Pi} belongs to 
$L^{\infty}\big( \mathsf{S},  \dd x\big)$. The right-hand side will vanish since:
\beq
\label{vanpoe}\int_{\mathsf{S}_h} \dd\nu_{\varphi}(x) = \oint_{\mathsf{S}_h} \frac{\varphi(x)\,\dd x}{2{\rm i}\pi} = 0\;.
\enq
We find:
\beq
\label{kfd}\mathcal{Q}[\mathrm{Re}\,\nu_{\varphi}] + \mathcal{Q}[\mathrm{Im}\,\nu_{\varphi}] = 0\;.
\enq
where $\mathcal{Q}$ is the quadratic form of Hypothesis~\ref{h2}. The vanishing of periods \eqref{vanpoe} implies \textit{a fortiori} that $\mathrm{Re}\,\nu_{\varphi}$ and $\mathrm{Im}\,\nu_{\varphi}$ are signed measures supported on $\mathsf{S} \subseteq \mathsf{A}$ with total mass zero. The assumption of local strict convexity (Hypothesis~\ref{h2}) states that for any such measure $\nu$ supported on $\mathsf{A}$, $\mathcal{Q}[\nu] \geq 0$, with equality iff $\nu = 0$. So, \eqref{kfd} implies $\mathrm{Re}\,\nu_{\varphi} = \mathrm{Im}\,\nu_{\varphi} = 0$, \textit{ie}. $\varphi = 0$. \hfill $\Box$

\subsection{Continuity of the inverse}

We will prove Lemma~\ref{lemy2} via a detour to Fredholm theory in $L^2$ spaces. We fix non-intersecting contours $\Gamma_h$ surrounding $\mathsf{A}_h$ in $\mathbb{C}\setminus\mathsf{A}$, all lying in $\Om$ such that $T$ is holomorphic on $\Om^r$. 
We denote $\Gamma = \bigcup_{h = 0}^{g} \Gamma_h$. 
Then, $\mc{L}$ can be interpreted as an integral operator on $L^{2}(\Ga)$
\bea
\mc{L}[\varphi](x) & = & \Oint{\Gamma}{} \frac{\dd y}{2{\rm i}\pi}\,\mathscr{L}(x,y) \,\varphi(y)\,,\nonumber
\eea
where the integral kernel $\mathscr{L}(x,y)$ is smooth on $\Gamma\times\Gamma$:
\beq
 \mathscr{L}(x,y) \; = \; 
 -\frac{2}{\beta \sigma_{\mathsf{S}}^{1/2}(x) }
 \Oint{(\Gamma[-1])^{r - 1}}{} \hspace{-2mm}\frac{\dd^{r - 1}\bs{\xi}}{(2{\rm i}\pi)^{r - 1}}\,
  \frac{\sigma_{\mathsf{S}}^{1/2}(\xi_1)\,\partial_{\xi_1} T(\xi_1,y, \xi_2,\ldots,\xi_{r-1})}{(r - 2)!\,(\xi_1 - x)}
  \,\prod_{i = 2}^{r-1} W_{\mathrm{eq}}(\xi_i)\;. \nonumber
\enq
Similarly, the operator $\mc{P}$ can be recast as 
\beq
\mc{P}[\vp](x) \; = \; \Oint{ \Ga }{} \frac{\dd y}{2{\rm i}\pi}\,\mathscr{P}(x,y)\,\vp(y)\;, \qquad \text{with}
\qquad \mathscr{P}(x,y) \;= \;  \f{ 1 }{ \sigma_{\mathsf{S}}^{1/2}(x) }
\mathop{\mathrm{Res}}_{\xi \rightarrow \infty} \f{ \sigma_{\mathsf{S}}^{1/2}(\xi)\,\dd\xi }{ (\xi- y)(x-\xi) } \; . \nonumber
\enq
This last kernel is smooth on $\Ga \times \Ga$ and of finite rank $g+1$. 

Since $\mc{L}$ and $\mc{P}$, as operators on $L^{2}(\Ga)$, have smooth kernels and $\Ga$ is compact, 
the operator $(\mathcal{L} - \mathcal{P}) : L^{2}(\Ga) \tend L^{2}(\Ga)$ is compact and trace class in virtue of the 
condition established in \cite{DudGB}. 
Finally, let $p_k$ be the unique polynomial of degree at most $g$ such that 
\beq
\forall h,k \in \intn{0}{g},\qquad \Oint{\Ga_h}{} \frac{\dd\xi}{2{\rm i}\pi}\,\frac{p_k (\xi)}{\sigma_{\mathsf{S}}^{1/2}(\xi)} \; = \; \de_{k,h}\;. \nonumber
\enq

Now consider the measure space $\mathsf{X} = \intn{0}{g} \cup \Ga$ endowed with the measure $\dd s$ that is the atomic measure
on $\intn{0}{g}$ and a curvilinear measure on $\Ga$. We shall constantly make the 
identification $L^2(\mathsf{X}) \simeq \Cx^{g+1} \oplus L^2( \mathsf{A})$.
It is then readily seen that the operator
\beq
\mc{N} \; : \; \ba{ccc} \Cx^{g+1} \oplus L^{2}(\Ga) &\longrightarrow & \Cx^{g+1} \oplus L^{2}(\Ga) \vspace{2mm}\\
						\big( \bs{v} , \vp \big)    & \longmapsto & \Big( -\bs{v} + \Pi[\vp] \, , \,  
	(\mathcal{L} - \mathcal{P})[\vp]  + \sigma_{\mathsf{S}}^{-1/2} \cdot \big(\sum_{h = 0}^{g} v_h\,p_h\big) \Big) \ea 	\nonumber		
\enq
is compact and trace class when considered as an integral operator $L^{2}(\mathsf{X}) \mapsto L^{2}(\mathsf{X}) $. The 
matrix integral kernel $\mathscr{N}$ of $\mathcal{N}$ has a block decomposition:
\beq
\mathscr{N} \; = \; \left( \ba{cc}   
								\mathscr{N}(h,k)  &   \mathscr{N}(h,y)    \\ 
								\mathscr{N}(x,k)  & \mathscr{N}(x,y) \ea \right)   \; = \; 
			\left( \ba{cc} 
						 	- \de_{j,k} &  \tf{ \bs{1}_{\Ga_h}(y) }{ 2{\rm i}\pi }  \\ 
			\sg_{\mathsf{S}}^{-1/2}(x) \cdot p_k(x) 
				&  (\mathscr{L}\, - \, \mathscr{P})(x,y) \ea  \right) 	\;, 			\nonumber	
\enq
The operator $\text{id}\, + \, \mc{N}$ is injective: indeed, if $\big( \bs{v} , \vp \big) \in \text{ker}\big( \text{id}\, + \, \mc{N} \big)$,
then 
\beq
\label{thy}\forall h \in \intn{ 0 }{ g },\qquad \Oint{\Ga_h}{}  \frac{\dd\xi\,\vp(\xi)}{2{\rm i}\pi} \; = \; 0 \qquad \text{and} \qquad 
\vp(x) \; = \; -\f{\sum_{h = 0}^{g} v_h\,p_h(x)}{ \sigma_{\mathsf{S}}^{1/2}(x)  } \; + \; (\mc{P}-\mc{L})[\vp](x) \;. 
\enq
The second equation implies that, in fact, $\vp \in \msc{H}^{1}(\mathsf{A})$. Further, since $\oint_{\; \Ga} \dd\xi\,\vp(\xi) = 0$, it follows that $\vp \in \msc{H}^{2}(\mathsf{A})$. Thus $ \vp(x)\dd x$ is a holomorphic differential 
all of whose $\Ga_h$ periods are zero. Hence $\vp=0$. Then  \eqref{thy} implies that $\bs{v} = 0$ as well. 
The Fredholm alternative thus ensures that $\text{id} + \mc{N}$ is continuously invertible. Furthermore,   
its inverse $\text{id} - \mc{R}_{\mc{N}}$ is given in terms of the resolvent kernel 
defined as in \eqref{definition noyau resolvant cas generique}. 
The integral kernel $\mc{N}$ falls into the class discussed in Appendix \ref{SousSection Inversion Fredholm}
with $f=1$. Hence, the kernel $\mathscr{R}_{\mathcal{N}}$ of $\mc{R}_{\mc{N}}$ belongs to 
$L^{\infty}\big( \mathsf{X}^2 \big)$.

We are now in position to  establish the continuity
of its inverse. The representation \eqref{reecriture noyau integral K} and the explicit form of the operators appearing 
there make it clear that $\mathcal{K}$ is a continuous operator for any norm $\p\cdot\p_{\Gamma}$ in the sense that 
\beq
\p \mc{K}[\vp] \p_{\Ga[1]} \; \leq \; \p  \vp \p_{\Ga} \;. \nonumber
\enq

We have already proved in Lemma~\ref{lemy} that the map 
\beq
\begin{array}{crcl}\widehat{\mathcal{K}}\,:\, &  \msc{H}^{1}(\mathsf{A}) & \longmapsto & \mathbb{C}^{g + 1} \oplus \mathscr{H}^1(\mathsf{A}) \\
& \varphi & \longrightarrow & \big(\Pi[\vp],\mc{K}[\vp]\big) \end{array} \nonumber
\enq
is injective. So, for any $\psi \in \msc{J}^{1}(\mathsf{A})$ there exists 
there exists a unique $\vp \in  \msc{H}^{1}(\mathsf{A})$ such that 
\beq
\mc{K}[\vp] \; = \; \psi \quad \text{and} \quad \Oint{ \mathsf{A}_h }{} \frac{\dd\xi\,\vp(\xi)}{2{\rm i}\pi} \; = \;0 \;. \nonumber
\enq
In other words, $(\bs{0},\varphi) \in \mathbb{C}^{g + 1}\oplus \mathscr{H}^1(\mathsf{A})$ does provide one with the unique solution to 
\beq
\big(\text{id} + \mc{N}\big)\big[(\bs{0},\vp)\big] \; = \; \big( \bs{0}, \mc{I}[\psi] \big) \;. \nonumber
\enq
Then, it readily follows that for any $\psi \in \msc{J}^{1}(\mathsf{A})$
\beq
\mc{K}^{-1}[ \psi ](x)  \; = \;  \mc{I}[\psi](x) 
\; - \; \Oint{ \Ga }{} \frac{\dd\xi}{2{\rm i}\pi}\,\f{ \mathscr{R}_{\mc{N}}(x,\xi)\,\sg_{\text{hd}}(\xi) }{M(\xi) } \,\psi(\xi)  \;, \nonumber 
\enq
where we have used that the resolvent kernel $\mathscr{R}_{\mc{N}}(x,\xi)$ is an analytic function on 
$ \big( \Cx \setminus \mathsf{A} \big) \times \Omega$, with $\Om$ the open neighbourhood of $\mathsf{A}$ such that $T$
is analytic on $\Om^r$. It is straightforward to establish the continuity of the inverse 
on the basis of the formula above.  As a consequence, it follows that $\msc{J}^{1}(A)$ (resp. $\msc{J}^{2}(A)$)
is a closed subspace of $\mathscr{H}^1(\mathsf{A})$ (resp $\mathscr{H}^2(\mathsf{A})$). \hfill $\Box$

\section{Asymptotics of correlators in the fixed filling fraction model}
\label{S6}

\subsection{More linear operators}
\label{S51zu}
Let us decompose:
\beq
W_1 = N(W_{\mathrm{eq}} + \Delta_{-1}W_1)\;.\nonumber
\enq
We define, with the notations of \eqref{tyi},
\bea
\mathcal{D}_1[\varphi](x_1,x_2) & = & \frac{2}{\beta}\oint_{\mathsf{A}} \frac{\dd\xi}{2{\rm i}\pi}\,\frac{\sigma_{{\rm hd}}(\xi)}{\sigma_{{\rm hd}}(x)}\,\frac{\varphi(\xi)}{(x_1 - \xi)(x_2 - \xi)^2}\;, \nonumber \\
\mathcal{D}_2[\varphi](x) & = & \varphi(x,x) - \oint_{\mathsf{A}^2} \frac{\dd\xi_1\dd \xi_2}{(2{\rm i}\pi)^2}\,\frac{\sigma_{{\rm hd}}^{[2]}(x;\xi_1,\xi_2)}{\sigma_{{\rm hd}}(x)}\,\varphi(\xi_1,\xi_2)\;, \nonumber
\eea
and:
\beq
\mathcal{T}[\varphi](x) = \frac{2}{\beta}\oint_{\mathsf{A}^r} \frac{\dd^{r}\bs{\xi}}{(2{\rm i}\pi)^r}\,\frac{\partial_{\xi_1} T(\bs{\xi})\,\varphi(\bs{\xi})}{(r - 1)!\,(x - \xi_1)}\;, \nonumber
\enq
\bea
\Delta\mathcal{K}[\varphi](x) & = & \frac{1 - 2/\beta}{N}\left(\partial_{x}\varphi(x) + \oint_{\mathsf{A}} \frac{\dd\xi}{2{\rm i}\pi}\,\frac{\sigma_{{\rm hd}}^{[2]}(x;\xi,\xi)}{\sigma_{{\rm hd}}(x)}\,\varphi(\xi)\right) + 2\,\mathcal{D}_2\big[(\Delta_{-1} W_1)(\bullet_1)\,\varphi(\bullet_2)\big](x) \nonumber \\
& & + \sum_{i = 1}^r \sum_{\substack{J \subseteq \intn{1}{r}\setminus\{i\} \\ J \neq \emptyset}} \mathcal{T}\Bigg[\varphi(\bullet_i) \prod_{j \in J} (\Delta_{-1}W_1)(\bullet_j) \prod_{j \notin J} W_{\mathrm{eq}}(\bullet_{j})\Bigg] \;. \nonumber
\eea
Above and in the following the notation $\bullet_j$  inside of the action of an operator denotes the $j^{\text{th}}$ running 
variable of the function on which the given operator acts. 
We also remind that if $\Gamma = \Gamma[0]$ is a contour surrounding $\mathsf{A}$, then we denote by $(\Gamma[i])_{i \geq 0}$ a family of nested contours such that $\Gamma[i + 1]$ is exterior to $\Gamma[i]$ for any $i$. There exists positive constants $c_1,c_2,\ldots$ which depend on the model and on the contours, so that: %
\bea
\p \mathcal{D}_1[\varphi] \p_{ ( \Gamma[1] )^2 } & \leq & c_1\,\p \varphi \p_{\Gamma}\;, \nonumber\\
\p \mathcal{D}_2 [\varphi] \p_{\Gamma} & \leq & c_2\,\p \varphi \p_{\Gamma^2}\;, \nonumber\\
\p \mathcal{T}[\varphi] \p_{ \Gamma[1] } & \leq & c_3\,\p \varphi \p_{\Gamma^r}\;, \nonumber\\
\label{theb}\p \Delta\mathcal{K}[\varphi] \p_{\Gamma[1]} & \leq & (c_4/N)\,\p \varphi \p_{\Gamma} + c_5 \p (\Delta_{-1}W_1)\p_{\Gamma}\,\p \varphi \p_{\Gamma[1]} \;.
\eea
Above, $\varphi$ belongs to the domain of definition of the respective operators. Notice that we have to push the contour towards the exterior in order to control the effect of the singular factors in $\mathcal{D}_1$ and $\mc{T}$. Further, 
we have also used the continuity of the derivation operator  $\p \Dp{x} \vp \p_{\Gamma[1]} \leq c \p \vp \p_{\Ga}  $. 
In order to gather all of the relevant operator bounds in one place, 
we remind the continuity of $\mathcal{K}^{-1}$: for $\varphi \in \mathscr{J}^2(\mathsf{A})$
\beq
\p \mathcal{K}^{-1}[\varphi] \p_{\Gamma[1]} \leq c_6\,\p \varphi \p_{\Gamma} \;. \nonumber
\enq
Besides, if $\varphi$ is holomorphic in $\mathbb{C}\setminus\mathsf{S}$ instead of $\mathbb{C}\setminus\mathsf{A}$, then $\mathcal{K}^{-1}[\varphi]$ is also holomorphic in $\mathbb{C}\setminus\mathsf{S}$.

\subsection{Improving concentration bounds using SD equations}

In order to improve the \textit{a priori} control on the correlators which follows from the concentration bounds 
\beq
\label{ewr} \p N\,\Delta_{-1}W_1 \p_{\Gamma} \leq c_1\,\eta_N\;,\qquad \p W_n \p_{\Gamma} \leq c_n\,\eta_N^{n}
\qquad \text{with} \qquad \eta_N = (N\ln N)^{1/2}\;, 
\enq
it is convenient to recast the Schwinger-Dyson equations.

The Schwinger-Dyson equation relative to $W_1$, \textit{ie}. \eqref{TheSD} with $n=1$,  takes, after some algebra,  the form:
\beq
\label{rhs1}\mathcal{K}[N\Delta_{-1}W_1](x) =  A_1(x) + \,B_1(x) - (\Delta\mathcal{K})[N\Delta_{-1}W_1](x)
\enq
with:
\begin{multline}
A_1(x) = -N^{-1}\mathcal{D}_2[W_2](x) + N\,\mathcal{D}_2\big[(\Delta_{-1}W_1)(\bullet_1)\,(\Delta_{-1}W_1)(\bullet_2)\big] \; - \; 
 \sum_{\substack{J \vdash \intn{1}{r} \\ [J] \leq r - 1}} N^{1 -r}\,\mathcal{T}\bigg[\prod_{i = 1}^{[J]} W_{|J_i|}(\bullet_{J_i})\bigg](x)  \\
 + N  \sum_{\substack{J \subseteq \intn{1}{r} \\ |J| \geq 2}}\,(|J| - 1)
\,\mathcal{T}\bigg[\prod_{j \in J} (\Delta_{-1} W_1)(\bullet_j)\prod_{j \notin J} W_{\mathrm{eq}}(\bullet_j)\bigg](x)
-\big( 1-\tf{2}{\be} \big) \cdot \bigg\{ \Dp{x} W_{\rm eq}(x)  \, + \, \Oint{ S }{} \f{\dd \xi }{2 {\rm i } \pi } 
	\f{ \sg_{\rm hd}^{[2]}(x;\xi,\xi) }{ \sg_{\rm hd}(x) } W_{\rm eq}(\xi) 		 \bigg\}     \;.
\label{definition A1}
\end{multline}
Above, we have introduced:
\beq
B_1(x) = \frac{2}{N\beta}\sum_{a \in \Dp{}\mathsf{A} \setminus \Dp{}\mathsf{S}    } \frac{\sigma_{{\rm hd}}(a)}{\sigma_{{\rm hd}}(x)}\,\frac{\partial_a \ln Z_{ \mc{S}} }{x - a} \;. \nonumber
\enq
Also, some notational clarifications are in order.  The symbol $J \vdash \intn{1}{r}$ refers to a sum over all partitions of the 
set $\intn{1}{r}$ into $[J]$ disjoint subsets $J_1,\dots, J_{[J]}$, with $[J]$ going from $1$ to $r$. 
%
%
%
%
%
%
%
%
In particular, the above sum is empty when $r = 1$. Finally, the summation arising in the second line of \eqref{definition A1} 
corresponds to a summation over all subsets $J$ of $\intn{1}{r}$ whose cardinality $|J|$ varies from $2$ to $r$.

More generally, for $n \geq 2$, the $n$-th Schwinger-Dyson equation takes the form:
\beq
\label{rhs2}\mathcal{K}\big[W_n(\bullet,\bs{x}_I)\big](x) = A_n(x;\bs{x}_I) + B_n(x;\bs{x}_I) - (\Delta\mathcal{K})\big[W_n(\bullet,\bs{x}_I)\big](x)
\enq
with:
\bea
A_n(x;\bs{x}_I) & = & -N^{-1}\mathcal{D}_2\Big[W_{n + 1}(\bullet_1,\bullet_2,\bs{x}_I) + \sum_{\substack{I' \subseteq I \\ I' \neq \emptyset,I}} W_{|I'| + 1}(\bullet_1,\bs{x}_{I'})\,W_{n - |I'|}(\bullet_2,\bs{x}_{I\setminus I'})\Big](x)  \nonumber \\
& & - \sum_{i \in I} N^{-1}\mathcal{D}_1\big[W_{n - 1}(\bullet,\bs{x}_{I\setminus\{i\}})\big](x,x_i) - 
\sum_{\substack{J \vdash \intn{1}{r} \\ I_{1} \sqcup \cdots \sqcup I_{[J]} = I }}^{*} 
N^{1 - r}\,\mathcal{T}\bigg[\prod_{i = 1}^{[J]} W_{|J_i| + |I_i|}(\bs{\bullet}_{J_i},{\bs x}_{I_i})\bigg](x) 
\label{definition An}
\eea
and
\beq
B_n(x;\bs{x}_I) =  \frac{2}{N\beta}\,\sum_{ a \in \Dp{}\mathsf{A} \setminus \Dp{}\mathsf{S}  } 
\frac{\sigma_{{\rm hd}}(a)}{\sigma_{{\rm hd}}(x)}\,\frac{\partial_a W_{n - 1}({\bs x}_I)}{x - a} \;. \nonumber
\enq
Finally, one has:
\beq
\sum_{\substack{J \vdash \intn{1}{r} \\ I_{1} \sqcup \cdots \sqcup I_{[J]} = I}}^{*} 
\pl{k=1}{[J]}  W_{|J_k|+|I_k|}\big( \bs{\xi}_{J_k}, \bs{x}_{I_k}  \big)  \; = \; 
\sul{  \substack{I_{1} \sqcup \cdots \sqcup I_{[J]} = I  \\ |I_k| < |I| }  }{} 
 \pl{k=1}{r}  W_{1+|I_k|}\big( \xi_{k}, \bs{x}_{I_k}  \big)  
 \;  +\;  \sum_{\substack{J \vdash \intn{1}{r} \\  \exists \ell \, : \, |J_{\ell}|\geq 2} }
 \sul{ I_{1} \sqcup \cdots \sqcup I_{[J]} = I     }{}  \pl{k=1}{[J]}  W_{|J_k|+|I_k|}\big( \bs{\xi}_{J_k}, \bs{x}_{I_k}  \big) \;.\nonumber 
\enq
The $\sqcup$ means that one should sum up over all decompositions of $I$ into $[J]$ disjoint subsets $I_1, \dots, I_{[J]}$
some of which can be empty. We stress that the order of dispatching the elements does count, \textit{viz}. the decompositions 
$I  \sqcup \{ \emptyset \}$ and $\{ \emptyset \}  \sqcup I $ differ. In other words, 
the $*$ label means that one excludes all terms of the form 
\mbox{$W_n(\bullet_i,\bs{x}_I)\prod_{j \neq i} W_1(\bullet_j)$}.

The above rewriting in basically enough so as to prove that the Schwinger-Dyson are rigid, in the sense that even a very rough 
\textit{a priori} control on the correlators allows one to establish that $W_n \in O(N^{2 - n})$. 
Indeed, 
\begin{prop}
\label{jfuew} There exist integers $p_{n}$ and positive constants $c_{n}'$ so that:
\beq
\p\Delta_{-1}W_1 \p_{\Gamma[p_{1}]} \leq c_{1}'\,N^{-1},\qquad \p W_n \p_{ (\Gamma[p_n])^n} \leq c_n'\,N^{2 - n} \;. \nonumber
\enq
\end{prop}

In order to prove the above proposition we, however, first need to establish a technical lemma emphasizing a one-step improvement of bounds.

\begin{lemma}
\label{jouh}Assume there exist positive constants $c_n$ so that:
\bea
\label{jiuj2}\p N\,\Delta_{-1}W_1 \p_{\Gamma} & \leq & c_1\big(\eta_N\,\kappa_N + 1 \big)\;, \\
\label{jiuj}\p W_n \p_{\Gamma^n} & \leq & c_n\big(\eta_N^n\,\kappa_N + N^{2 - n}\big)\;,\qquad (n \geq 2)
\eea
for $\eta_N \rightarrow \infty$ so that $\eta_N/N \rightarrow 0$, and $\kappa_N \leq 1$. Then, there exist positive constants $c_n'$ so that:
\bea
\p N\,\Delta_{-1}W_1 \p_{\Gamma[2]} & \leq & c_1'\big(\eta_N\,\kappa_N\,(\eta_N/N) + 1\big)\;, \nonumber \\
\p W_n \p_{ (\Gamma[2])^n } & \leq & c_n'\big(\eta_N^n\cdot\eta_N/N\cdot\kappa_N + N^{2 - n}\big)\;. \nonumber 
\eea
\end{lemma}
\noindent\textbf{Proof.} Hereafter, the values of the positive constants $c_1,c_2,\ldots$ may vary from line to line, and we use repeatedly the continuity of the auxiliary operators introduced in \S~\ref{S51zu}. We remind that in the fixed filling fraction model, we have a priori:
\beq
\Delta_{-1} W_1 \in \mathscr{H}_0^2(\mathsf{A})\;, \nonumber
\enq
and for any fixed $\bs{x}_I = (x_2,\ldots,x_r) \in (\mathbb{C}\setminus\mathsf{A})^{n - 1}$:
\beq
W_n(\bullet,\bs{x}_I) \in \mathscr{H}_0^2(\mathsf{A}) \;. \nonumber
\enq
Therefore, the right-hand side of \eqref{rhs1} or \eqref{rhs2} (seen as a function of $x$) belongs to $\mathscr{J}^2(\mathsf{A})$ and we can apply the inverse of $\mathcal{K}$:
\bea
\label{upq} \Delta_{-1}W_1(x) & = & \mathcal{K}^{-1}\Big[A_1 + B_1 - (\Delta\mathcal{K})[\Delta_{-1}W_1]\Big](x)\;, \\
\label{upq2} W_{n}(x,\bs{x}_I) & = &  \mathcal{K}^{-1}\Big[A_n(\bullet,\bs{x}_I) + B_n(\bullet,\bs{x}_I) - (\Delta\mathcal{K})[W_n(\cdot,\bs{x}_I)\big](\bullet)\Big](x) \;.
\eea

We start by estimating the various terms present in $A_1$. The terms not associated with the sum over $J$ in the first line are readily estimated by using the control on the correlators and the continuity of the various operators introduced at the beginning of the section. 
In what concerns the terms present in the sum, we bound them by using that the bound \eqref{jiuj} trivially holds for $n = 1$.
All in all this leads to 
\begin{multline*}
\p A_1 \p_{\Ga[1]} \; \leq \; \frac{c_2 \tilde{c}}{N} (\eta^2_N\kappa_N + 1) 
\, + \, \frac{c_1^2  \tilde{c}}{N} \big(\eta_N \kappa_N + 1\big)^2
\, + \, c^{\prime} \sum_{\substack{J \vdash \intn{1}{r} \\ [J] \leq r - 1}} N^{1 -r} \pl{a=1}{ [J] } \big( \eta_N^{|J_a|}\kappa_N \, + \, N^{2-|J_a|} \big) \\
\,+ \, \tilde{c} c_1^2\,  \big(\eta_N \kappa_N \, + \, 1 \big)\big(\eta_N \kappa_N/N + 1/N \big) \, + \, c^{\prime \prime}
\end{multline*}
Thus, it solely remains to obtain an optimal bound for the product 
\beq
\Pi_1([J]) \; = \; N^{1 -r}  \pl{a=1}{ [J] } \big(  \eta_N^{|J_a|}\kappa_N \, + \, N^{2-|J_a|} \big) \; = \; 
\sul{ \substack{ \a \sqcup \ov{\a} \\  = \intn{1}{[J]} } }{}  \kappa_N^{ |\a | } N^{1 -r}  
\pl{a \in \a}{} \eta_N^{|J_a| }{} \pl{a \in \ov{\a} }{} N^{2-|J_a|}\;.
\enq
Note that, because of the structure of the sum, there exists at least one $\ell$ such that $|J_\ell|\geq 2$.
There are two scenarii then. Either, $\ov{\a}=\intn{1}{[J]}$ or $|\ov{\a}|< [J]$. In the second case, 
the sum is maximised by the choice of partitions $\a$ such that $\ell \in \a$ since $\eta_N^2 >1$ for $N$ large enough. 
In such partitions, one bounds 
\beq
\pl{a \in \a}{} \eta_N^{|J_a| }{} \pl{a \in \ov{\a} }{} N^{2-|J_a|} \; \leq \; 
\eta_N^2\pl{a \in \a \setminus \{ \ell \} }{} N^{|J_a|} \pl{a \in \ov{\a} }{} N^{2 - |J_a|} \; \leq \;
\eta_N^2 N^{r-2}\,. \nonumber
\enq
So, taking this into account, one gets 
\beq
\Pi_1([J]) \; \leq \; c_1 N^{1-r + [J] } \; + \; c_2 \frac{\eta_N^2}{N}\,\kappa_N\;.\nonumber
\enq
It remains to recall that $[J]\leq r-1$ so as to get 
\beq
 \p A_1 \p_{\Ga[1] } \; \leq \; c_1^{\prime} \Big( \frac{\eta_N^2}{N}\,\kappa_N + 1 \Big) \;. \nonumber
\enq
It follows from the large deviations of single eigenvalues, Lemma \ref{uuu3}, that $\p B_1 \p_{\Ga} =  \text{O}\big(N^{-\infty}\big)$. 
Finally, the bound \eqref{jiuj2} for $\Delta_{-1}W_1$ leads to:
\beq
\p\Delta\mathcal{K}[\varphi]\p_{\Gamma[1]} \leq c\,\f{ \eta_N }{ N }\,\p \varphi \p_{\Gamma} \qquad
ie. \qquad \p (\Delta\mathcal{K})[N\,\Delta_{-1}W_1] \p_{\Gamma[1]}  \leq  \eta_N\cdot\frac{\eta_N}{N}\,\kappa_N + 1 \;.  
\label{ecriture bornage Delta K}
\enq
Note that, above, the shift of contour was necessary because of \eqref{theb}.
It solely remains to invoke the continuity of $\mathcal{K}^{-1}$ -- which, however, demands one additional shift of contour -- 
so as to obtain the claimed improvement of bounds relative to $N\,\Delta_{-1}W_1$.

We can now repeat the chain of bounds for the Schwinger-Dyson equation associated with the $n^{\text{th}}$ correlator with $n \geq 2$. Since 
\beq
\max_{j \in \intn{1}{n-2}} \bigg\{  \big( \eta_N^{j+1}\kappa_N  + N^{2-(j+1)} \big) 
\big( \kappa_N \eta_N^{n-j} + N^{2-n + j} \big)  \bigg\} \; \leq \;  
					c \big(\eta_N^n \cdot \frac{\eta_N}{N}\,\kappa_N  \, + \, N^{2-n}   \big) \;, \nonumber
\enq
and:
\beq
\frac{1}{N}(\eta_N^{n - 1}\kappa_N + N^{-(n - 1)}) \leq \eta_N^{n}\cdot \frac{\eta_N}{N}\,\kappa_N + N^{2 - n}
\enq
for $N$ large enough, one gets 
\beq
\p  A_n \p_{ (\Ga[1])^n} \; \leq \; c \Big( \eta_N^n \f{ \eta_N }{ N } \,\kappa_N + \, N^{2-n}   \Big) \; 
+ \;c^{\prime}  \de_N^{(1)} \; + \; c^{\prime \prime }  \de_N^{(2)}\;,
\label{ecriture bornage An etape intermediaire}
\enq
where 
\bea
\de_N^{(1)} \; &  = & \;  \sul{  \substack{I_{1} \sqcup \cdots \sqcup I_{[J]} = I  \\ |I_k| < |I| }  }{} N^{1-r}
 \pl{a=1}{[J]}  \big(   \eta_N^{|I_a|} \,\kappa_N + \, N^{1-|I_a|} \big)\;, \nonumber \\
\de_N^{(2)} \; & = & \;   \sum_{\substack{J \vdash \intn{1}{r} \\  \exists \ell \, : \, |J_{\ell}|\geq 2} }
 \sul{ I_{1} \sqcup \cdots \sqcup I_{[J]} = I     }{}
  \pl{a=1}{[J]}  \big(  \eta_N^{|I_a|+|J_a|} \,\kappa_N + \, N^{2-|I_a|-|J_a|} \big) \;.
\nonumber
\eea
issue from bounding the last term in \eqref{definition An}.
It is readily seen that 
\beq
\de_N^{(1)} \; \leq \;  \sul{  \substack{I_{1} \sqcup \cdots \sqcup I_{[J]} = I  \\ |I_k| < |I| }  }{} 
\sul{ \substack{ \a \sqcup \ov{\a} \\ = \intn{1}{r} }  }{ } \Big( \f{\eta_N  }{ N }\kappa_N \Big)^{ | \a | }
N^{2-n} \big( \eta_N N \big)^{m_{\a} }  \qquad \text{with} \quad  m_{\a} = \sul{ a \in \a }{} |I_a|  \;. \nonumber
\enq
Thus, the right-hand side is maximised by choosing $|\a|$ minimal and $m_{\a}$ maximal. However, if $|\a|=0$, then
$\ell_{\a}=0$ and one gets $N^{2-n}$ as a bound. When $|\a|=1$, due to the constraint $|I_k|< |I| = n-1$, one gets that 
$\max m_{\a} < n-2$. Finally, for $\a\geq 2$ one has that $\max m_{\a} = n-1$. A short calculation then
shows that 
\beq
\de_N^{(1)} \; \leq \; c \big( N^{2-n} \, + \,  \eta_N^n\cdot\frac{\eta_N}{N}\,\kappa_N  \big) \;. \nonumber
\enq
Likewise, one obtains,
\beq
\de_N^{(2)} \; = \;   \sum_{\substack{J \vdash \intn{1}{r} \\  \exists a \, : \, |J_{a}|\geq 2} }
 \sul{ I_{1} \sqcup \cdots \sqcup I_{[J]} = I     }{} \sul{ \substack{ \a \cup \ov{\a} \\ = \intn{1}{ [J] } }  }{ } 
 \kappa_N^{ | \a | } N^{2-n} \big( \eta_N N \big)^{ \ell_{\a} + m_{\a} } N^{2([J]-r-|\a |)}
\qquad \text{with} \quad
				 \left\{  \ba{ccc}     m_{\a} &=& \sul{ a \in \a }{} |I_a|   \\ 
									 \ell_{\a} &=& \sul{ a \in \a }{} |J_a|  \ea \right. 				  \;. \nonumber
\enq
Thus, the above summand will be maximised by taking the smallest possible value for $|\a|$ and the largest possible ones for 
$\ell_{\a} $, $m_{\a}$ and $[J]$. Note, however, that $[J]\leq r-1$ due to the constraint $\exists a \, : \, |J_a| \geq 2$.  If $|\a|=0$, then $\ell_{\a}=m_{\a}=0$ and one obtains a bound by $N^{-n} \leq N^{2 - n}$. If $|\a|>0$, then one has 
$\ell_{\a} \leq r-[J]+1 \, + \,  (|\a|-1)$, thus leading to 
\beq
\kappa_N^{ | \a | }\,  N^{2-n} \, \big( \eta_N N \big)^{ \ell_{\a} + m_{\a} }  \, N^{2([J]-r-|\a |)} \; \leq \; 
\kappa_N^{ | \a | } \,  N^{2-n}  \, \big( \eta_N N \big)^{ |\a | + m_{\a} } \, \Big( \f{\eta_N}{N} \Big)^{ r-[J]}\;.\nonumber
\enq
The right-hand side is maximised for $m_{\a}=n-1$, $[J]=r-1$ and $|\a|=1$, what leads to a bound by 
$(\eta_N^{n+1}/N)\kappa_N $. Hence, 
$ \p  A_n \p_{ (\Ga[1])^n} \; \leq \; c \big(\eta_N^n\cdot(\eta_N/N)\cdot\kappa_N  \, + \, N^{2-n}   \big) $. 
The remaining terms in \eqref{definition An} are bounded analogously to the $n=1$ case. Repeating then the steps of this 
derivation one, eventually, gets the sought bounds on $W_n$. \hfill $\Box$

\vspace{0.2cm}

\noindent\textbf{Proof of Proposition \ref{jfuew}.} The concentration results \eqref{ewr} provide us the bounds \eqref{jiuj2}-\eqref{jiuj} with $\eta_N = (N\ln N)^{1/2}$ and $\kappa_N = 1$. From Lemma~\ref{jouh}, we can replace $\kappa_N$ by $(\eta_N/N)^m = (\ln N/N)^{m/2}$ provided $\Gamma$ is replaced by $\Gamma[2m]$. In particular, for $n = 1$, we may choose $m = 2$ to have $\eta_N\,(\eta_N/N)^{m} \in o(1)$, and for $n \geq 2$, we may choose $m = 4n + 4$, so that $\eta_N^{n} (\eta_N/N)^{m} \in o(N^{2 - n})$. At this point, the remainder $N^{2 - n}$ in \eqref{jiuj2}-\eqref{jiuj} gives us the desired bounds. \hfill $\Box$

\subsection{Recursive asymptotic analysis using SD equations}

\label{therpre}

\begin{lemma}
\label{L62}There exist $W_n^{[k]} \in \mathscr{H}^2_0(\mathsf{S},n)$, positive integers $p_{n}^{[k]}$ and positive constants $c_n^{[k]}$, indexed by integers $n \geq 1$ and $k \geq n-2$, so that, for any $k_0 \geq -1$:
\beq
W_n = \sum_{k = n - 2}^{k_0} N^{-k}\,W_n^{[k]} + N^{-k_0}\,\Delta_{k_0}W_n\;,\qquad \p \Delta_{k_0}W_n \p_{\Gamma[p_{n}^{[k]}]} \leq c_{n}^{[k]}/N\;. \nonumber
\enq
By convention, the first sum is empty whenever $k_0< n-2$. 
\end{lemma}
\textbf{Proof.} The proof goes by recursion. Our recursion hypothesis at step $k_0$ is that we have a decomposition for any $n \geq 1$:
\beq
W_n = \sum_{k = n - 2}^{k_0} N^{-k}\,W_n^{[k]} + N^{-k_0}\,\Delta_{k_0} W_n,\qquad \p \Delta_{k_0} W_n \p_{\Gamma} \rightarrow 0\;, \nonumber
\enq
where $W_n^{[k]} \in \mathscr{H}^2_0(\mathsf{S},n)$ is known and the convergence holds without uniformity in $\Gamma$ and $n$. From Proposition~\ref{jfuew}, we know the recursion hypothesis is true for $k_0 = -1$. We choose not to specify anymore the shift of the contours which are necessary at each step of inversion of $\mathcal{K}^{-1}$, since this mechanism of shifting is clear from the Proof of Proposition~\ref{jfuew}.

The recursion hypothesis induces a decomposition:
\bea
A_n(x,\bs{x}_I) & = & \sum_{k = n - 2}^{k_0+1} N^{-k}\,A_n^{[k]}(x,\bs{x}_I) + N^{-(k_0+1)}\,\Delta_{(k_0+1)}A_n(x,\bs{x}_I)\;,\nonumber \\
(\Delta\mathcal{K})[\varphi](x) & = & 
\sum_{k = 1}^{k_0+1} N^{-k}\,\mathcal{K}^{[k]}[\varphi](x) + N^{-(k_0+1)}(\Delta_{(k_0+1)}\mathcal{K})[\varphi](x)\;.
\nonumber
\eea
We give below the expressions of those quantities for $k \in \intn{1}{k_0+1}$ (this set is empty if $k_0 = -1$ ).
\bea
\mathcal{K}^{[k]}[\varphi](x) & = & \delta_{k,1}(1 - 2/\beta)\left(\partial_{x}\varphi(x) + \oint_{\mathsf{A}} \frac{\dd\xi}{2{\rm i}\pi}\,\frac{\sigma_{{\rm hd}}^{[2]}(x;\xi,\xi)}{\sigma_{{\rm hd}}(x)}\,\varphi(\xi)\right) + 2\,\mathcal{D}_2\big[W_{1}^{[k - 1]}(\bullet_1)\,\varphi(\bullet_2)\big](x) \nonumber \\
\label{Kdev1}& & + \sum_{i = 1}^r \sum_{\substack{J \subseteq \intn{1}{r}\setminus\{i\} \\ J \neq \emptyset 0}} \sum_{\substack{k_1,\ldots,k_{|J|} \geq 0 \\ (\sum_j k_j) + |J| = k}} \mathcal{T}\bigg[\varphi(\bullet_i)\prod_{j \in J} W_1^{[k_j]}(\bullet_j) \prod_{j \notin J} W_{\mathrm{eq}}(\bullet_{j})\bigg](x)\;.
\eea
For $n \geq 2$, we have for $k \leq k_0+1$:
\bea
A_n^{[k]}(x,\bs{x}_I) & = & -\mathcal{D}_2\Big[W_{n + 1}^{[k - 1]}(\bullet_1,\bullet_2,\bs{x}_I) + 
\sum_{\substack{I' \subseteq I \\ I' \neq \emptyset,I}} \sum_{0 \leq k' \leq k-1}
W_{|I'| + 1}^{[k']}(\bullet_1,\bs{x}_{I'})W_{n - |I'|}^{[k - k' -1]}(\bullet_2,\bs{x}_{I \setminus I'})\Big] \nonumber \\
\label{Anka}& & - \sum_{i \in I} \mathcal{D}_1\big[W_{n - 1}^{[k - 1]}(\bullet,\bs{x}_{I\setminus\{i\}})\big](x,x_i)  
- \sum_{\substack{J \vdash \intn{1}{r} \\ I_{1} \sqcup \cdots \sqcup I_{[J]} = I}}^{*} 
\sum_{\substack{k_1,\ldots,k_{[J]} \geq -1 \\ (\sum_i k_i) + r - 1 = k}}  
\mathcal{T}\bigg[\prod_{i = 1}^{[J]} W_{|J_i| + |I_i|}^{[k_i]}(\bullet_{J_i},\bs{x}_{I_i})\bigg](x)\; . 
\eea
In the above expression, we agree that $W_m^{[\ell]}=0$ whenever $\ell  < m -2$. The main point is that
both $A_n^{[k]}(x,\bs{x}_I) $ and $\mathcal{K}^{[k]}[\varphi](x)$ only involve $W_{m}^{[k']}$ with $k' \leq k-1 \leq k_0$. 
This is a matter of simple reading off in the case of the expression for $\mathcal{K}^{[k]}[\varphi](x)$.
Likewise, in the case of $A_n^{[k]}(x,\bs{x}_I) $ this is clear in what concerns the first three terms, but the last one ought to be discussed. So, for a given term of the sum, let  $J^{[-1]}$ be the collection 
of singletons $\{i\}$ such that $k_i = -1$. Then, the $k_i$ associated with this precise term of the sum satisfy 
\beq
\Big(\sum_{p \not \in J^{[-1]}} k_p\Big) + r - 1 - |J^{[-1]}| = k \;. \nonumber
\enq
The restriction $\sum^{*}$ ensures that, in the non-vanishing terms,  $|J^{[-1]}| \leq r - 2$ that is to say that there is 
at most $r-2$ $k_p$'s which can be equal to $-1$. Since $k_p \geq 0$ for $p\not \in J^{[-1]}$, this implies $k_p \leq  k-1$ for 
any $p\not \in J^{[-1]}$. We now discuss the error terms at order $N^{-k_0-1}$. These take the form:
\bea
(\Delta_{(k_0+1)} A_n)(x,\bs{x}_I) & = & - \mathcal{D}_2\Big[(\Delta_{k_0}W_{n + 1})(\bullet_1,\bullet_2,\bs{x}_I) + 
 \sum_{\substack{I' \subseteq I \\ I' \neq \emptyset,I}} \sum_{k' = 0}^{k_0} 
(\Delta_{k'}W_{|I'| + 1})(\bullet_1,\bs{x}_{I'})\,(\Delta_{(k_0 - k')}W_{n - |I'|})(\bullet_2,\bs{x}_{I\setminus I'})\Big] \nonumber \\
& & - \sum_{i \in I} \mathcal{D}_1\big[(\Delta_{k_0}W_{n - 1})(\bullet,\bs{x}_{I\setminus\{i\}})\big](x,x_i) \nonumber \\
\label{Danka} & & 
- \sum_{\substack{J \vdash \intn{1}{r} \\ I_1 \sqcup \cdots \sqcup I_{[J]} = I}}^{*} \sum_{\substack{k_1,\ldots,k_{[J]} \geq -1 \\ (\sum_i k_i) + r - 1 = k_0+1}} \mathcal{T}\bigg[\prod_{i = 1}^{[J]} (\Delta_{k_i}W_{|J_i| + |I_i|})(\bullet_{J_i},\bs{x}_{I_i})\bigg](x)\;.
\eea
and
\bea
(\Delta_{(k_0+1)}\mathcal{K})[\varphi](x) & = & 2\,\mathcal{D}_2\big[(\Delta_{k_0}W_1)(\bullet_1)\,\varphi(\bullet_2)\big](x) \nonumber \\
& & + \sum_{i = 1}^r \sum_{\substack{J \subseteq \intn{1}{r}\setminus\{i\} \\ |J| \geq 1}} \sum_{\substack{k_1,\ldots,k_{[J]} \geq -1 \\ (\sum_j k_j) + |J| = k_0+1}} \mathcal{T}\bigg[\varphi(\bullet_i)\,\prod_{j \in J} (\Delta_{k_j} W_1)(\bullet_j) \prod_{j \notin J} W_{\mathrm{eq}}(\bullet_j)\bigg](x)\,.\nonumber
\eea
For the reason invoked above, these expressions only involve $\Delta_{k}W_{\ell}$ with $k \leq k_0$. Note that, for $r=1$,
the $\sum^{*}$ is empty. 
Further, one readily checks that the recursion hypothesis implies 
\beq
\p \Delta_{(k_0+1)}A_n  \p_{\Ga^{n}}  \; \leq \; \f{ c }{ N } \qquad \text{and} \qquad
\p \Delta_{(k_0+1)}\mathcal{K}[\varphi] \p_{\Ga[1]} \; \leq \; \f{c^{\prime} }{N}\,\p \vp \p_{\Ga} \;. \nonumber
\enq
%
%
One likewise obtains similar expressions at $n=1$, namely:
\begin{multline}
A_1^{[k]}(x) =  
-\de_{0,k} \big( 1-\tf{2}{\be} \big)\bigg\{ \Dp{x} W_{\rm eq}(x)  \, + \, \Oint{ S }{} \f{\dd \xi }{2 {\rm i } \pi }
	\f{ \sg_{\rm hd}^{[2]}(x;\xi,\xi) }{ \sg_{\rm hd}(x) } W_{\rm eq}(\xi) 		 \bigg\}
-\mathcal{D}_2[W_{2}^{[k - 1]}](x)  \\
\label{Anka1}+ \sum_{\substack{k_1,k_2 \geq 0 \\ k_1 + k_2 = k - 1}}  
\mathcal{D}_2\big[W_1^{[k_1]}(\bullet_1)\,W_1^{[k_2]}(\bullet_2)\big]\;  + \sum_{\substack{J \subseteq \intn{1}{r} \\ |J| \geq 2}} \sum_{\substack{k_1,\ldots,k_{|J|} \geq -1 \\ (\sum_j k_j) + |J| = k + 1}} (|J| - 1)\,\mathcal{T}\Bigg[\prod_{j \in J} W_1^{[k_j]}(\bullet_{j})\,\prod_{j \notin J} W_{\mathrm{eq}}(\bullet_j)\Bigg](x)\;, 
\end{multline}
and
\begin{align}
(\Delta_{(k_0+1)}A_1)(x)  = & \,\, - \mathcal{D}_2\big[(\Delta_{k_0 }W_2)\big](x) + \sum_{\substack{k_1,k_2 \geq 0 \\ k_1 + k_2 = k_0}} \mathcal{D}_2\big[(\Delta_{k_1}W_1)(\bullet_1)\,(\Delta_{k_2}W_1)(\bullet_2)\big] \nonumber \\
 & \,\,+ \sum_{\substack{J \subseteq \intn{1}{r} \\ |J| \geq 2}} \sum_{\substack{k_1,\ldots,k_{|J|} \geq -1 \\ (\sum_j k_j) + |J| = k_0+2}} (|J| - 1)\,\mathcal{T}\Bigg[\prod_{j \in J} (\Delta_{k_j}W_1)(\bullet_j)\,\prod_{j \notin J} W_{\mathrm{eq}}(\bullet_j)\Bigg](x) 
 \; \\ - \; 
 & \,\,\sum_{J \vdash \intn{1}{r}}^{*} \sum_{\substack{k_1,\ldots,k_{[J]} \geq -1 \\ (\sum_{i} k_i) + r - 1 = k_0 + 1}}\,\mathcal{T}\bigg[\prod_{i = 1}^{[J]} \Delta_{k_i}W_{|J_i|}(\bullet_{J_i})\bigg](x). \nonumber
\end{align}
One checks that likewise, $A_1^{[k]}$ only involves $W_1^{[k']}$ with $k' < k$, and $W_2^{[k - 1]}$. 
Similarly, $\Delta_{(k_0+1)}A_1$ only involves $\Delta_{k}W_1$ and $\Delta_{k}W_2$ with $k \leq k_0$
and satisfies to the bounds $\p \Delta_{(k_0+1)}A_1 \p_{\Ga} \leq \tf{ c }{ N } $. 

 By inserting the obtained expansions into the appropriate Schwinger-Dyson equations \eqref{rhs1}-\eqref{rhs2} 
one obtains for $n \geq 2$:
\begin{multline*}
\mc{K}[W_n(\bullet, \bs{x}_I)](x) \; = \; 
\sul{k=n-2}{k_0+1} N^{-k} \Big( A_n^{[k]}(x;\bs{x}_I) 
\; - \; \sul{\ell=n-2}{k-1} \mc{K}^{[k-\ell]}\big[W_n^{[\ell]}(\bullet, \bs{x}_I)\big](x)\Big)
\; + B_n(x;\bs{x}_I) \\ - N^{-(k_0+1)}\mathcal{K}\big[\De_{(k_0+1)}W_n(\bullet, \bs{x}_I)\big](x)
\; - N^{-(k_0+1)}  \sul{\ell=n-2}{k_0} \mc{K}^{[k_0+1-\ell]}\big[\De_{\ell}W_n(\bullet, \bs{x}_I)\big](x)
\; + \; N^{-(k_0+1)} \De_{(k_0+1)}A_n^{[k]}(x;\bs{x}_I)  \;. 
\end{multline*}
For $n \geq 1$, the right-hand side is similar but the left-hand side should be replaced by $\mc{K}[N\Delta_{-1}W_1]$.

Using the continuity of $\mathcal{K}^{-1}$ and of the other operators involved the above equations yields 
a system of equations which determine the $W_n^{[k]}$ recursively on $k$. In particular, this 
implies that $W_n$ admits an asymptotic expansion up to order $k_0+1$. By the recursion hypothesis at step $k_0$, the function to which 
we apply $\mathcal{K}^{-1}$ to get $W_n^{[k_0 + 1]}$ is holomorphic on $\mathbb{C}\setminus\mathsf{S}$. Therefore, $W_n^{[k_0 + 1]}$ is 
also holomorphic in $\mathbb{C}\setminus\mathsf{S}$, so that $W_n^{[k_0 + 1]} \in \mathscr{H}^2_0(\mathsf{S},n)$. We just proved that the 
recursion hypothesis holds at step $k_0 + 1$, so we can conclude by induction. To summarize, the recursive formula for the coefficient of expansion of the correlators is:
\beq
\label{corexpq}W_n^{[k]} = \mathcal{K}^{-1}\Big[A_n^{[k]}(\bullet;\bs{x}_I) - \sum_{\ell = n - 2}^{k - 1} \mathcal{K}^{[k- \ell]}[W_{n}^{[\ell]}(\cdot,\bs{x}_I)](\bullet)\Big](x)
\enq
with $A_n^{[k]}$ given by \eqref{Anka}-\eqref{Anka1} and $\mathcal{K}^{[\ell]}$ by \eqref{Kdev1}. 
\hfill $\Box$

\section{Partition function in the fixed filling fraction model}
\label{S7}
\subsection{Asymptotic expansion}

\begin{lemma}
\label{interp}Assume the local strict convexity of Hypothesis~\ref{h2}, the analyticity of Hypothesis~\ref{h1bis}, and that $\mu_{\mathrm{eq},\bs{\epsilon}}$ is off-critical. There exists a $1$-linear potential $\widehat{T}$ satisfying the same assumptions and so that, for any $k_0$, we have an asymptotic expansion of the form:
\beq
\frac{Z_{\mathsf{A}_{\mathbf{N}}}^{T}}{Z_{\mathsf{A}_{\mathbf{N}}}^{\widehat{T}}} = \exp\Big(\sum_{k = -2}^{k_0} N^{-k}\,G^{[k]}_{\bs{\epsilon}} + N^{-k_0} \Delta_{k_0} G_{\bs{\epsilon}}\Big)\;. \nonumber
\enq
$G_{\bs{\epsilon}}^{[k]}$ are smooth functions of the filling fractions $\bs{\epsilon}$ in a small enough domain where $\mu_{{\rm eq}}^{\bs{\epsilon}}$ remains off-critical, and for any fixed $k_0$, the error is uniform in $\bs{\epsilon}$ in such a compact domain.
\end{lemma}
Since the asymptotic expansion for $1$-linear potentials and its smoothness with respect to $\bs{\epsilon}$ have been established in \cite{BGmulti} under the same assumptions as here, we obtain automatically:
\begin{cor}
\label{cc1}Assume the local strict convexity of Hypothesis~\ref{h2}, the analyticity of Hypothesis~\ref{h1bis}, and that $\mu_{\mathrm{eq}}^{\bs{\epsilon}}$ is off-critical. The partition function with fixed filling fractions has an asymptotic expansion of the form
\beq
Z_{\mathsf{A}_{\mathbf{N}}}^{T} = N^{(\beta/2)N + \gamma} \exp\Big(\sum_{k = -2}^{k_0} N^{-k}\,F^{[k]}_{\bs{\epsilon}} + o(N^{-k_0})\Big) . \nonumber
\enq
$\gamma$ is a universal exponent depending only on $\beta$ and the nature of the edges and reminded in \S~\ref{gammaapp}. $F^{[k]}_{\bs{\epsilon}}$ are smooth functions of $\bs{\epsilon}$ and for any fixed $k_0$, the error is uniform in $\bs{\epsilon}$ as explained in Lemma~\ref{interp}.
\end{cor}

\subsection{Proof of Lemma~\ref{interp} -- except regularity}

The characterization \eqref{juje2} of equilibrium measure can be rephrased by saying that $\mu_{{\rm eq}}^{\bs{\epsilon},T} = \mu_{{\rm eq}}^{\bs{\epsilon},\widehat{T}}$ where:
\beq
\label{tdeum}\widehat{T}(x_1,\ldots,x_r) = (r - 1)!\,\sum_{j = 1}^{r} \widehat{T}_{1}(x_j) \nonumber
\enq
is the $1$-body interaction defined, if $x \in \mathsf{A}_h$, by:
\beq
\label{tdeum1}\widehat{T}_1(x) = \sum_{\substack{h' = 0 \\ h' \neq h}}^{g} \beta\Int{\mathsf{S}_{h'}}{} \ln|x - \xi|\,\dd\mu_{{\rm eq}}^{\bs{\epsilon},T}(\xi)\mathbf{1}_{\mathsf{S}_{h'}}(\xi) + \Int{\mathsf{S}^{r - 1}}{} \frac{T(x,\xi_2,\ldots,\xi_r)}{(r - 1)!}\prod_{i = 2}^r \dd\mu_{{\rm eq}}^{\bs{\epsilon},T}(\xi_i).
\enq
Since $\mathsf{A}_{h}$ and $\mathsf{S}_{h'}$ are disjoint if $h' \neq h$, $(x - y)$ keeps the same sign for $x \in \mathsf{S}_{h'}$, $\widehat{T}_{1}(x)$ has actually an analytic continuation for $x$ in a neighbourhood of each $\mathsf{A}_h$. Besides, the characterization of the equilibrium measure implies that:
\beq
\forall t \in [0,1],\qquad \mu_{{\rm eq}}^{\bs{\epsilon}, tT + (1 - t)\widehat{T}} = \mu_{{\rm eq}}^{\bs{\epsilon}, T}\;. \nonumber
\enq
So, if $T$ satisfies our assumptions, $tT + (1 - t)\widehat{T}$ satisfies it uniformly for $t \in [0,1]$. And, we have the general formula:
\beq
\label{count}\partial_{t} \ln Z^{tT + (1 - t)\widehat{T}}_{\mathsf{A}_{\mathbf{N}}} = \frac{N^{2 - r}}{r!}\,\Oint{\mathsf{A}^{r}}{} \frac{\dd^{r}\bs{\xi}}{(2{\rm i}\pi)^{r}}\,(T(\bs{\xi}) - \widehat{T}(\bs{\xi})\big)\,\widetilde{W}_{r}^{tT + (1 - t)\widehat{T}}(\bs{\xi})
\enq
in terms of the disconnected correlators introduced in \eqref{discon}. By Lemma~\ref{L62}, $\widetilde{W}_r^{tT + (1 - t)\widehat{T}}(\bs{\xi})$ has an asymptotic expansion in $1/N$, starting at order $N^{r}$, and if we truncate it to an order $N^{-k_0}$, it is uniform in $\bs{\xi}$ on the contour of integration of \eqref{count} and in $t \in [0,1]$. So, we can integrate \eqref{count} over $t \in [0,1]$, exchange the expansion and the integrations to obtain the expansion of the partition function. The remaining  Smoothness of the coefficients of expansion of the correlators with respect to filling fractions will be a consequence of Proposition~\ref{smoothen2} below. \hfill $\Box$

\subsection{Lipschitz dependence in filling fractions}
\label{S711}
We first show that the equilibrium measures depend on the filling fractions in a Lipschitz way.

\begin{lemma}
\label{smoothen}Assume Hypothesis~\ref{h1} in the unconstrained  model  and $T$ holomorphic in a neighbourhood of $\mathsf{A}$ in $\mathbb{C}$. Then, for $\bs{\epsilon}$ close enough to $\bs{\epsilon}^\star$, $\mathcal{E}$ has a unique minimiser over $\mathcal{M}^{\bs{\epsilon}}(\mathbf{A})$, denoted $\mu_{{\rm eq}}^{\bs{\epsilon}}$. Let $(g + 1)$ be the number of cuts of $\mu_{{\rm eq}}$, and $\epsilon^{\star} = \mu_{{\rm eq}}[\mathsf{A}_h]$ its masses. Assume $\mu_{{\rm eq}}$ is off-critical. Then, for $\bs{\epsilon}$ close enough to $\bs{\epsilon}^{\star}$, $\mu_{{\rm eq}}^{\bs{\epsilon}}$ still has $g + 1$ cuts, is off-critical, has the edges of the same nature which are Lipschitz functions of $\bs{\epsilon}$, and the density of $\mu_{{\rm eq}}^{\bs{\epsilon}}$ is a Lipschitz function of $\bs{\epsilon}$ away from its edges.
\end{lemma}

\noindent\textbf{Proof.}  
We remind (\S~\ref{energf}) that the level sets $E_{M} = \{ \mu \in \mathcal{M}^{1}(\mathsf{A}),\;\; \mathcal{E}[\mu] \leq M\}$ are compact. 
Therefore, $\mathcal{E}$ achieves its minimum on $\mathcal{M}^{\bs{\epsilon}}(\mathbf{A})$, and we denote $\mu_{{\rm eq}}^{\bs{\epsilon}}$ any such minimiser. It must satisfy the saddle point equation \eqref{juje2}. By assumption, the minimiser over $\mathcal{M}^{1}(\mathsf{A})$ is unique, it is denoted by  $\mu_{{\rm eq}}$, and its partial masses $\epsilon^{\star}_h = \mu_{{\rm eq}}[\mathsf{A}_h]$. In other words, the minimiser $\mu_{{\rm eq}}^{\bs{\epsilon}^{\star}}$ is unique and equal to $\mu_{{\rm eq}}$. We must prove that the minimiser is unique for $\bs{\epsilon}$ close enough to $\bs{\epsilon}^{\star}$. 

\begin{itemize}
\item We first show that any  $\mu_{{\rm eq}}^{\bs{\epsilon}}$ must belong to a ball $B(\mu_{\rm eq},\delta_{\bs{\epsilon}})$ around $\mu_{\rm eq}$  for the Vaserstein distance, with $\delta_{\bs{\epsilon}}$ going to zero when $\bs{\epsilon}$ goes to $\bs{\epsilon}^\star$. 

Let us first prove  that
$$\mathcal{E}^{\bs \epsilon}:=\inf_{\mu\in  \mathcal{M}^{\bs{\epsilon}}(\mathbf{A})}\mathcal{E}(\mu)\ra \mathcal{E}^{{\bs \epsilon}^\star}\qquad \mbox{ as } {\bs \epsilon}\ra {\bs \epsilon}^\star\,.$$
In fact, if we denote $\mu_{\rm eq}^h$ the probability on $\mathsf{A}_h$ so that $\mu_{\rm eq}=\sum_h \epsilon_h^\star \mu_{\rm eq}^h$ we have
$$\mathcal{E}^{{\bs \epsilon}^\star}\le \mathcal{E}^{\bs \epsilon}\le \mathcal{E}\Big(\sum_h \epsilon_h \mu_{\rm eq}^h\Big)\,.$$
But we have seen that $\int \log |x-y|\dd\mu_{\rm eq}^h(y)$ is uniformly bounded on $\mathsf A$ for all $i$ and therefore one easily sees that there exists a finite constant $C$ such that 
$$ \mathcal{E}\Big(\sum_h \epsilon_h\,\mu_{\rm eq}^h\Big)\le  \mathcal{E}\Big(\sum_h \epsilon_h^\star \mu_{\rm eq}^h\Big)+C\max|\epsilon_h-\epsilon_h^\star|$$
from which the announced continuity follows.

Let us deduce  by contradiction that there exists a sequence $\delta_\epsilon$ 
so
that $\mu_{{\rm eq}}^{\bs{\epsilon}}\in B(\mu_{\rm eq},\delta_\epsilon)$ for $|{\bs\epsilon}-{\bs\epsilon}^\star|$ small enough.
Otherwise, we can find a $\delta>0$ and a sequence $\mu_{{\rm eq}}^{\bs{\epsilon}_n} \notin B(\mu_{\rm eq},\delta)$  with ${\bs\epsilon}_n$ converging to ${\bs\epsilon}^\star$.
As we can assume by the above continuity that
 this sequence belongs to the level set  $E_{\mathcal{E}^{{\bs \epsilon}^\star}+1}$, this sequence is tight and we can consider a limit point $\mu$. But by lower semi-continuity of
 $\mathcal{E}$ we
 must
 have
 $$\liminf_{n\ra\infty}\mathcal{E}(\mu_{{\rm eq}}^{\bs{\epsilon}_n})\ge \mathcal{E}(\mu)$$
 whereas the previous continuity shows that the left-hand side is actually equal to $\mathcal{E}^{{\bs \epsilon}^\star}$. Hence $\mu$ minimises $\mathcal{E}$ on $ \mathcal{M}^{1}(\mathsf{A})$ which
 implies by Hypothesis ~\ref{h1} that $\mu=\mu_{\rm eq}$ hence yielding the announced contradiction.
 
 \item  We now show uniqueness of the minimiser of $\mathcal{E}$ on $\mathcal{M}^{\bs{\epsilon}}(\mathbf{A})$ for ${\bs \epsilon}$ close enough to ${\bs\epsilon}^\star$ by showing that
 the interaction keeps the property of local strict convexity. 
 
 Let us define, for any $\nu \in \mathcal{M}^{0}(\mathsf{A})$:
\beq
\mathcal{Q}^{\bs{\epsilon}}[\nu] = \beta\mathcal{Q}_{C}[\nu] - \Int{\mathsf{A}^{r - 2}}{} \frac{T(x_1,\ldots,x_r)}{(r - 2)!}\,\dd\nu(x_1)\dd\nu(x_2) \prod_{j = 3}^r \dd\mu_{{\rm eq}}^{\bs{\epsilon}}(x_j)\;. \nonumber
\enq
where $\mathcal{Q}_{C}$ was defined in \eqref{Qcdef}. For any probability measure $\mu$, we can write using Taylor-Lagrange formula at order $3$ around $\mu_{{\rm eq}}^{\bs{\epsilon}}$:
\beq
\label{equak} \mathcal{E}[\mu] = \mathcal{E}[\mu_{{\rm eq}}^{\bs{\epsilon}}] - \int T_{{\rm eff}}^{\bs{\epsilon}}(x)\dd\mu(x) + \frac{1}{2}\,\mathcal{Q}^{\bs{\epsilon}}[\mu - \mu_{{\rm eq}}^{\bs{\epsilon}}] + \mathcal{R}_{3}^{\bs{\epsilon}}[\mu - \mu_{{\rm eq}}^{\bs{\epsilon}}]\;,
\enq
where $T_{{\rm eff}}^{\bs{\epsilon}}$ is the effective potential \eqref{Teff} for $\mu_{{\rm eq}}^{\bs{\epsilon}}$. The remainder is:
\bea
\mathcal{R}_{3}^{\bs{\epsilon}}[\nu] & = & \Int{0}{1} \frac{\dd t(1 - t)^2}{2}\,\mathcal{E}^{(3)}[(1 - t)\mu_{{\rm eq}}^{\bs{\epsilon}} + t\mu]\cdot(\nu,\nu,\nu)\;.
\eea
where $\mathcal{E}^{(3)}$ was already defined in \eqref{Fre3}. If $\bs{\kappa} \in \mathbb{R}_+^{g + 1}$ so that $\sum_{h} \kappa_h = 1$, for any measure $\mu^{\bs{\kappa}} \in \mathcal{M}^{\bs{\kappa}}(\mathbf{A})$, we have:
\beq
\mathcal{E}[\mu^{\bs{\kappa}}_{{\rm eq}}] \leq \mathcal{E}[\mu^{\bs{\kappa}}]\;. \nonumber
\enq
We now use the equality \eqref{equak} for both sides, and assume the support of $\mu^{\bs{\kappa}}$ is included in the support of $\mu_{{\rm eq}}^{\bs{\epsilon}}$. Since $T_{{\rm eff}}^{\bs{\epsilon}}$ is non-positive, and equal to $0$ on the support of $\mu_{{\rm eq}}^{\bs{\epsilon}}$, we find:
\beq
\label{ines}\frac{1}{2}\,\mathcal{Q}^{\bs{\epsilon}}[\mu^{\bs{\kappa}}_{{\rm eq}} - \mu^{\bs{\epsilon}}_{{\rm eq}}] \leq \frac{1}{2}\,\mathcal{Q}^{\bs{\epsilon}}[\mu^{\bs{\kappa}} - \mu_{{\rm eq}}^{\bs{\kappa}}] + \mathcal{R}_3^{\bs{\epsilon}}[\mu^{\bs{\kappa}} - \mu_{{\rm eq}}^{\bs{\epsilon}}] - \mathcal{R}_3^{\bs{\epsilon}}[\mu_{{\rm eq}}^{\bs{\kappa}} - \mu_{{\rm eq}}^{\bs{\epsilon}}]\;.
\enq
We assume $\kappa_h \in [0,2\epsilon_h]$, and put $\mu^{\bs{\kappa}} = t\mu_{{\rm eq}}^{\bs{\epsilon}} + (1 - t)\mu_{{\rm ref}}$ with the choice:
\beq
\label{1moist}1 - t = \max_{0 \leq h \leq g} \frac{|\kappa_h - \epsilon_h|}{\epsilon_h} \in [0,1]\;,
\enq
and the choice of a probability measure $\mu_{{\rm ref}}$, whose support is included in that of $\mu_{{\rm eq}}^{\bs{\epsilon}}$, and with masses satisfying:
\beq
t\epsilon_h + (1 - t)\mu_{{\rm ref}}[\mathsf{A}_h] = \kappa_h\;. \nonumber
\enq
Note that we can assume $\bs{\epsilon}_h \neq 0$ since $\bs{\epsilon}^{\star}_h \neq 0$ follows from the assumption that $\mu_{{\rm eq}}$ is off-critical. We also require that $\mu_{{\rm ref}}$ is such that $\mathcal{Q}^{\bs{\epsilon}}[\mu_{{\rm ref}} - \mu_{{\rm eq}}^{\bs{\epsilon}}] < +\infty$, which is always possible, for instance by taking for $\mu_{{\rm ref}}$ the renormalised Lebesgue measure on the support of $\mu_{{\rm eq}}^{\bs{\epsilon}}$. 

Using $\mathcal{Q} = \mathcal{Q}^{\bs{\epsilon}^{\star}}$, we know from Lemma~\ref{l322} that the remainder can be bounded as:
\beq
\big|\mathcal{R}_{3}^{\bs{\epsilon}}[\nu]\big| \leq C^{\bs{\epsilon}}\,\|\nu\|\, \mathcal{Q}[\nu]\;. \nonumber
\enq
Therefore, we have:
\beq
\label{716}
\mathcal{Q}^{\bs{\epsilon}}[\mu^{\bs{\kappa}}_{{\rm eq}} - \mu^{\bs{\epsilon}}_{{\rm eq}}]  - C^{\bs{\epsilon}}\p\mu^{\bs{\kappa}}_{{\rm eq}} - \mu^{\bs{\epsilon}}_{{\rm eq}}\p\,\mathcal{Q}[\mu^{\bs{\kappa}}_{{\rm eq}} - \mu_{\rm eq}^{\bs{\epsilon}}]
\leq (1 - t)^2\mathcal{Q}^{\bs{\epsilon}}[\mu_{{\rm ref}} - \mu_{{\rm eq}}^{\bs{\epsilon}}] + (1 - t)^3\,C^{\bs{\epsilon}}\, \| \mu_{{\rm ref}} - \mu_{{\rm eq}}^{\bs{\epsilon}}\|\, 
\mathcal{Q}[\mu_{{\rm ref}} - \mu_{{\rm eq}}^{\bs{\epsilon}}]\;.
\enq

If we apply this inequality to $\bs{\epsilon} = \bs{\epsilon}^{\star}$, and ${\bs \kappa}$ close enough to ${\bs\epsilon}^{\star}$ so that 
$$ C^{\bs{\epsilon}^{\star}}\|\mu^{\bs{\kappa}}_{{\rm eq}} - \mu^{\bs{\epsilon}^{\star}}_{{\rm eq}}\|\le 1/2$$
(this is possible by the continuity previously established), we deduce from \eqref{1moist} that, for $\max_h |\kappa_h - \epsilon_h| < c'$ for some $c' > 0$ independent of $\bs{\epsilon}$:
\beq
\mathcal{Q}[\mu^{\bs{\kappa}}_{{\rm eq}} - \mu_{{\rm eq}}] \leq C\,\max_h |\kappa_h - \epsilon_h^{\star}|^2\;. \nonumber
\enq
Besides, we may compare $\mathcal{Q}^{\bs{\epsilon}}$ and $\mathcal{Q} = \mathcal{Q}^{\bs{\epsilon}^{\star}}$ by writing:
\beq
\mathcal{Q}^{\bs{\epsilon}}[\nu] - \mathcal{Q}[\nu] = \sum_{m = 3}^{r} \;\Int{\mathsf{A}^{r}}{} \frac{T(\xi_1,\ldots,\xi_r)}{(r - 2)!}\,\dd\nu(\xi_1)\dd\nu(\xi_2)\,\dd(\mu^{\bs{\epsilon}}_{{\rm eq}} - \mu_{{\rm eq}})(\xi_3)\prod_{i = 4}^{m} \dd\mu_{{\rm eq}}^{\bs{\epsilon}}(\xi_i) \prod_{j = m + 1}^{r} \dd\mu_{{\rm eq}}(\xi_{j})\;. \nonumber
\enq
Hence, from Lemma~\ref{l322}, there exists a constant $C > 0$, independent of $\bs{\epsilon}$ but depending on $T$, so that:
\beq
\big|\mathcal{Q}^{\bs{\epsilon}}[\nu] - \mathcal{Q}[\nu]\big| \leq C\,\mathcal{Q}[\nu]\,\mathcal{Q}^{1/2}[\mu_{{\rm eq}}^{\bs{\epsilon}} - \mu_{{\rm eq}}] \leq C'\,\mathcal{Q}[\nu]\,\max_{h} |\epsilon_{h} - \epsilon_{h}^{\star}|\;. \nonumber
\enq 
Therefore, for $\max_{h} C'|\epsilon_{h} - \epsilon_{h}^{\star}| < 1$, there exists constants $c_1,c_2 > 0 $ so that:
\beq
\label{comparis}\forall \nu \in \mathcal{M}^{0}(\mathsf{A}),\qquad c_1\,\mathcal{Q}[\nu] \leq \mathcal{Q}^{\bs{\epsilon}}[\nu] \leq c_2\,\mathcal{Q}[\nu]\;,
\enq
in particular $\mathcal{Q}^{\bs{\epsilon}}[\nu] \geq 0$ with equality iff $\nu = 0$. So, coming back to \eqref{716}, we deduce for $\bs{\epsilon}$ close enough to $\bs{\epsilon}^{\star}$ and $\bs{\kappa}$ close enough to $\bs{\epsilon}$, that there exists a constant $c'$ such that:
\beq
\mathcal{Q}[\mu_{{\rm eq}}^{\bs{\kappa}} - \mu_{{\rm eq}}^{\bs{\epsilon}}] \leq c'(1 - t)^2\,\mathcal{Q}[\mu_{{\rm ref}} - \mu_{{\rm eq}}^{\bs{\epsilon}}]  \nonumber
\enq
with $t$ as in \eqref{1moist}. This entails:
\beq
\label{pol}
\mathcal{Q}[\mu_{{\rm eq}}^{\bs{\kappa}} - \mu_{{\rm eq}}^{\bs{\epsilon}}] \leq C\,\max_{h} |\kappa_h - \epsilon_h|^2\;.
\enq
We can apply this relation with $\bs{\epsilon} = \bs{\kappa}$ but $\mu_{{\rm eq}}^{\bs{\kappa}}$ another minimiser of $\mathcal{E}$ over $\mathcal{M}^{\bs{\epsilon}}(\mathbf{A})$, which would tell us that the $\mathcal{Q}$-distance between two minimisers is $0$. So, the minimiser is unique for any $\bs{\epsilon}$ close enough to $\bs{\epsilon}^{\star}$.
\item We finally prove smoothness of the minimising measures and related quantities.
As a second consequence of \eqref{comparis} and \eqref{pol}, for any $m \geq 1$ and any smooth test function $\varphi$ of $m$ variables, we have a finite constant $C_\varphi$ so that 
\beq
\label{inuit}\Bigg|\int \varphi(\xi_1,\ldots,\xi_m)\,\Bigg(\prod_{j = 1}^m\dd\mu_{{\rm eq}}^{\bs{\kappa}}(\xi_j) -\prod_{j =1}^m \dd\mu_{{\rm eq}}^{\bs{\epsilon}}(\xi_j)\Bigg)\Bigg| \leq C_\varphi\, \mathcal{Q}^{1/2}[\mu_{{\rm eq}}^{\bs{\kappa}} - \mu_{{\rm eq}}^{\bs{\epsilon}}] \leq C\,\max_{h}\,|\kappa_h - \epsilon_h|\;.
\enq
In other words, the integrals of $m$-variables test functions in $\mathcal{H}(m)$ against $\mu_{{\rm eq}}^{\bs{\epsilon}}$ (called below $m$-linear statistics) are Lipschitz in the variable $\bs{\epsilon}$ close enough to $\bs{\epsilon}^{\star}$.

To extend this regularity result to the equilibrium measure, we consider the expression \eqref{vrar} for its density:
\beq
\label{rejdu}\frac{\dd\mu_{{\rm eq}}}{\dd x}(x) = \frac{\mathbf{1}_{\mathsf{S}_{\bs{\epsilon}}}(x)}{2\pi}\,\sqrt{R_{\bs{\epsilon}}(x)},\qquad R_{\bs{\epsilon}}(x) =  \frac{4\widetilde{P}_{\bs{\epsilon}}(x) - \sigma_{\mathsf{A}}(x)\big(V'_{\bs{\epsilon}}(x)\big)^2}{\sigma_{\mathsf{A}}(x)}\;.
\enq
The important feature of this formula is that $V'_{\bs{\epsilon}}(x)$ (resp. $\widetilde{P}_{\bs{\epsilon}}(x)$) defined in \eqref{vdef}-\eqref{pdef} are integrals against $\mu_{{\rm eq}}^{\bs{\epsilon}}$ of a holomorphic test function in a neighbourhood of $\mathsf{A}^{r - 1}$ (resp. $\mathsf{A}^r)$ which depends holomorphically in $x$ in a neighbourhood of $\mathsf{A}$. Thanks to \eqref{inuit}, $x \mapsto R_{\bs{\epsilon}}(x)$ is a holomorphic function when $x$ belongs to a compact neighbourhood $\mathsf{K}$ (independent of $\bs{\epsilon}$) of $\mathsf{A}$ avoiding the hard edges, which has a Lipschitz dependence in $\bs{\epsilon}$. Thus, the density itself is a Lipschitz function of $\bs{\epsilon}$ away from the edges. The edges of the support of $\mu_{{\rm eq}}^{\bs{\epsilon}}$ are precisely the zeroes and the poles of $R_{\bs{\epsilon}}(x)$ in $\mathsf{K}$. If we assume that $\mu_{{\rm eq}}^{\bs{\epsilon}}$ is off-critical, then these zeroes and poles are simple. So, they must remain simple zeroes (resp. simple poles) for $\bs{\epsilon}'$ close enough to $\bs{\epsilon}$, and their dependence in $\bs{\epsilon}'$ is Lipschitz. \hfill $\Box$
\end{itemize}
\subsection{Smooth dependence in filling fractions}

\begin{prop}
\label{smoothen2}Lemma~\ref{smoothen} holds with $\mathcal{C}^{\infty}$ dependence in $\bs{\epsilon}$.
\end{prop}

\begin{cor}
\label{oiuhas}
Under the same assumptions, the coefficients of expansion of the correlators $W_n^{[k];\bs{\epsilon}}(x_1,\ldots,x_n)$ depends smoothly on $\bs{\epsilon}$ for $x_1,\ldots,x_n$ uniformly in any compact of $\mathbb{C}\setminus\mathsf{A}$.
\end{cor}

\noindent\textbf{Proof.} The idea of the proof is again very similar to \cite[Appendix A.2]{BGmulti}. Let $\mathsf{E}$ be an open neighbourhood of $\bs{\epsilon}_{{\rm ref}}$ in $$\big\{\bs{\epsilon} \in [0,1]^{g + 1}\quad \sum_{h} \epsilon_h = 1\big\}$$ so that the result of Lemma~\ref{smoothen} holds. For any given $x$ in a compact neighbourhood $\mathsf{K}$ of $\mathsf{A}$ avoiding the edges, $R_{\bs{\epsilon}}(x)$ is Lipschitz function of $\bs{\epsilon}$, therefore differentiable for $\bs{\epsilon}$ in a subset $\mathsf{E}_{x}$ whose complement in $\mathsf{E}$ has measure $0$. It is not hard to see that $\mathsf{E}_{\infty} = \bigcap_{x \in \mathsf{K}} \mathsf{E}_{x}$ has still a complement of measure $0$, therefore is dense in $\mathsf{E}$. For any $\bs{\epsilon} \in \mathsf{E}_{\infty}$ and any $\bs{\eta} \in \mathbb{R}^{g + 1}$ so that $\sum_{h} \eta_h = 0$, we can then study the effect of differentiation at $\bs{\epsilon}$ in a direction $\bs{\eta}$ in the characterization and properties of the equilibrium measure. We find that $\dd\nu_{\bs{\eta}}^{\bs{\epsilon}} = \partial_{t = 0}\,\mu_{{\rm eq}}^{\bs{\epsilon} + t\bs{\eta}}$ defines a signed measure on $\mathsf{S}_{\bs{\epsilon}}$ which is integrable (its density behaves atmost like the inverse of a squareroot at the edges), gives a mass $\eta_h$ to $\mathsf{S}_{\bs{\epsilon},h}$, and satisfies:
\beq
\forall x \in \mathring{\mathsf{S}}_{\bs{\epsilon}},\qquad \beta \Fint{\mathsf{S}_{\bs{\epsilon}}}{} \frac{\dd\nu_{\bs{\eta}}^{\bs{\epsilon}}(\xi)}{x -\xi} + \Int{\mathsf{S}^{r - 1}_{\bs{\epsilon}}}{} \frac{\partial_{x}\,T(x,\xi_2,\ldots,\xi_r)}{(r - 2)!}\,\dd\nu_{\bs{\eta}}^{\bs{\epsilon}}(\xi_2)\prod_{j = 3}^{r} \dd\mu_{{\rm eq}}^{\bs{\epsilon}}(\xi_j) = 0\;.  \nonumber
\enq
We have seen in the proof of Lemma~\ref{lemy} that there is a unique solution to this problem. As a matter of fact, if we introduce the Stieltjes transform:
\beq
\varphi_{\bs{\eta}}^{\bs{\epsilon}}(x) = \Int{\mathsf{S}_{\bs{\epsilon}}}{} \frac{\dd\nu_{\bs{\eta}}^{\bs{\epsilon}}(\xi)}{x - \xi}\;, \nonumber
\enq
by construction of the operator $\mathcal{K}$ we have $\mathcal{K}[\varphi_{\bs{\eta}}^{\bs{\epsilon}}] = 0$ and the condition on masses is $\oint_{\mathsf{S}_{\bs{\epsilon},h}} \varphi_{\bs{\eta}}^{\bs{\epsilon}}(\xi)\,\frac{\dd\xi}{2{\rm i}\pi} = \eta_h$ for any $h$. So, the invertibility of $\mathcal{K}$ used towards the end of the proof of Lemma~\ref{lemy} gives:
\beq
\label{kuju}\varphi_{\bs{\eta}}^{\bs{\epsilon}}(x) =-  \sul{h=0}{g} \eta_h\,\mathscr{R}_{\mc{N}}(x,k)\;,
\enq
where $\mathscr{R}_{\mc{N}}$ is a block of the resolvent kernel of $\mathcal{N}$. Eventually, we observe that $\mathcal{K}$  depends on $\bs{\epsilon}$ only via the Stieltjes transform of $\mu_{{\rm eq}}^{\bs{\epsilon}}$, therefore is Lipschitz in $\bs{\epsilon}$. The expression \eqref{definition series de Fredholm op general} for the resolvent kernel implies that if $\mathcal{K}$ depends on a parameter (here $\bs{\epsilon}$) with a certain regularity, its inverse will depend on the parameter with the same regularity.  Therefore, the right-hand side of \eqref{kuju} is Lipschitz in $\bs{\epsilon}$, a fortiori continuous. To summarise, we have obtained that:
\beq
W_{{\rm eq}}^{\bs{\epsilon}}(x) = \Int{\mathsf{S}_{\bs{\epsilon}}}{} \frac{\dd\mu_{{\rm eq}}^{\bs{\epsilon}}(\xi)}{x - \xi}\nonumber
\enq
is differentiable at a dense subset of $\bs{\epsilon}$, and that the differential happens to be a continuous function of $\bs{\epsilon}$. Therefore, $W_{{\rm eq}}^{\bs{\epsilon}}$ is differentiable everywhere, and \eqref{kuju} can be considered as a differential equation, where the right-hand side is differentiable. Hence $W_{{\rm eq}}^{\bs{\epsilon}}$ is twice differentiable. This regularity then carries to the right-hand side, and by induction, this entails the $\mathcal{C}^{\infty}$ regularity of $W_{{\rm eq}}^{\bs{\epsilon}}$ -- and thus the density of $\mu_{{\rm eq}}^{\bs{\epsilon}}$ -- for any $x$ away from the edges of $\mathsf{S}_{\bs{\epsilon}}$. Therefore, $R_{\bs{\epsilon}}$ in \eqref{rejdu} was $\mathcal{C}^{\infty}$ in $\bs{\epsilon}$, and the result of Lemma~\ref{smoothen} gets improved to $\mathcal{C}^{\infty}$ regularity in $\bs{\epsilon}$. \hfill $\Box$

\noindent \textbf{Proof of Corollary~\ref{oiuhas}}. From the Proof of \S~\ref{therpre}, the coefficient of expansion of correlators $W_n^{[k]}(x_1,\ldots,x_n)$  (cf. \eqref{corexpq}) and the errors $\Delta_{k}W_n(x_1,\ldots,x_n)$ can be computed recursively, by successive applications of $\mathcal{K}^{-1}$ to combinations involving $W_{{\rm eq}}^{\bs{\epsilon}} = W_1^{[-1],\bs{\epsilon}}(x)$, $T$ and the $W_n^{[k'],\bs{\epsilon}}$ computed at the previous steps. As we have seen, $\mathcal{K}^{-1}$ and $W_{{\rm eq}}^{\bs{\epsilon}}$ depend smoothly on $\bs{\epsilon}$ under the conditions stated above, so the $W_n^{[k]}$ enjoy the same property, and similarly one can show that the bounds on the errors $\Delta_{k}W_n$ are uniform with respect to $\bs{\epsilon}$. \hfill $\Box$

\noindent \textbf{End of the proof of Lemma~\ref{interp}}. After the proof of the corollary, we just have to check that $T = \widehat{T}$ given by \eqref{tdeum}-\eqref{tdeum1} depends smoothly on $\bs{\epsilon}$. Since it is expressed as the integration of an analytic function against (several copies of) $\mu_{{\rm eq}}$, this follows from the proof of Section~\ref{S711} and its improvement in Proposition~\eqref{smoothen2}.

\subsection{Strict convexity of the energy}

We show that the value of the energy functional at $\mu_{{\rm eq}}^{\bs{\epsilon}}$ is a strictly convex function of the filling fraction in a neighbourhood of $\bs{\epsilon}^{\star}$. This property will be useful in the analysis of the unconstrained model in the multi-cut regime.

\begin{prop}
\label{Hessp}
Assume Hypothesis~\ref{h1} in the unconstrained model, $T$ holomorphic in a neighbourhood of $\mathsf{A}$ in $\mathbb{C}$, $\mu_{{\rm eq}}$ is off-critical. Denote $\epsilon_{h}^{\star} = \mu_{{\rm eq}}[\mathsf{A}_{h}]$. Then, for $\bs{\epsilon}$ in a neighbourhood of $\bs{\epsilon}^{\star}$, $\mathcal{E}[\mu_{{\rm eq}}^{\bs{\epsilon}}]$ is $\mathcal{C}^2$ and its Hessian is definite positive.
\end{prop}
\noindent\textbf{Proof.} Proposition~\ref{smoothen} ensures the existence of $\mu_{{\rm eq}}^{\bs{\epsilon}}$ for $\bs{\epsilon}$ close enough to $\bs{\epsilon}^{\star}$. It is characterised, for any $h \in \intn{0}{g}$ and $x \in \mathsf{S}_{\bs{\epsilon},h}$, by
\beq
\beta \Fint{\mathsf{S}_{\bs{\epsilon}}}{} \ln|x - \xi|\dd\mu_{{\rm eq}}^{\bs{\epsilon}}(\xi) + \Int{\mathsf{S}_{\bs{\epsilon}}^{r - 1}}{} \frac{T(x,\xi_2,\ldots,\xi_r)}{(r - 1)!}\,\prod_{j = 2}^{r} \dd\mu_{{\rm eq}}^{\bs{\epsilon}}(\xi_j) = C_{h}^{\bs{\epsilon}}\;.
\nonumber\enq
For any $x \in \mathsf{A}$, the left-hand side minus the right-hand side defines the effective potential $T_{{\rm eff}}^{\bs{\epsilon}}$, which is therefore $0$ for $x \in \mathsf{S}_{\bs{\epsilon}}$. The proof of Proposition~\ref{smoothen2} provides us, for any $\bs{\eta} \in \mathbb{R}^{g + 1}$ so that $\sum_{h = 0}^{g} \eta_h = 0$, with the existence of:
\beq
\nu_{\bs{\eta}}^{\bs{\epsilon}} = \partial_{t = 0}\,\mu_{{\rm eq}}^{\bs{\epsilon} + t\bs{\eta}}\;, \nonumber
\enq
as a signed, integrable measure on $\mathsf{A}$, so that $\nu_{\bs{\eta}}^{\bs{\epsilon}}[\mathsf{A}_h] = \eta_h$ for any $h$. It satisfies, for any $h \in \intn{0}{g}$ and $x \in \mathsf{S}_{\bs{\epsilon},h}$:
\beq
\label{ouhiuh}\beta \Fint{\mathsf{S}_{\bs{\epsilon}}}{} \ln|x - \xi|\dd\nu_{\bs{\eta}}^{\bs{\epsilon}}(\xi) + \Int{\mathsf{S}_{\bs{\epsilon}}^{r - 1}}{} \frac{T(x,\xi_2,\ldots,\xi_r)}{(r - 2)!}\,\dd\nu_{\bs{\eta}}^{\bs{\epsilon}}(\xi_2)\prod_{j = 3}^{r} \dd\mu_{{\rm eq}}^{\bs{\epsilon}} = \partial_{t = 0}\,C_{h}^{\bs{\epsilon} + t\bs{\eta}}\;.
\enq
Therefore, $\mathcal{E}[\mu_{{\rm eq}}^{\bs{\epsilon}}]$ is $\mathcal{C}^1$ and we have:
\beq
\partial_{t = 0}\,\mathcal{E}[\mu_{{\rm eq}}^{\bs{\epsilon} + t\bs{\eta}}] = - \Int{\mathsf{S}_{\bs{\epsilon}}}{} \Bigg(T_{{\rm eff}}^{\bs{\epsilon}}(x) + \sum_{h = 0}^{g} C_{\bs{\epsilon},h}\,\mathbf{1}_{\mathsf{S}_{\bs{\epsilon},h}}(x)\Bigg)\dd\nu_{\bs{\eta}}^{\bs{\epsilon}}(x) = - \sum_{h = 0}^{g} C_{h}^{\bs{\epsilon}}\,\eta_{h}\;. \nonumber
\enq
We can differentiate the result once more:
\beq
\label{ouheq} \mathrm{Hessian}_{\bs{\epsilon}}\,\mathcal{E}[\mu_{{\rm eq}}^{\bs{\epsilon}}]\cdot(\bs{\eta},\bs{\eta}') = \partial_{t' = 0}\,\partial_{t = 0}\,\mathcal{E}[\mu_{{\rm eq}}^{\bs{\epsilon} + t\bs{\eta} + t'\bs{\eta}'}] = -\sum_{h = 0}^{g} (\partial_{t' = 0}\,C_{h}^{\bs{\epsilon} + t'\bs{\eta}'})\,\eta_{h}\;.
\enq
Since the right-hand side of \eqref{ouhiuh} is constant on each $\mathsf{S}_{\bs{\epsilon},h}$, we integrate it against $-\dd\nu_{\bs{\eta}}^{\bs{\epsilon}}$ and find a result equal to the right-hand side of \eqref{ouheq}:
\bea
\mathrm{Hessian}_{\bs{\epsilon}}\,\mathcal{E}[\mu_{{\rm eq}}^{\bs{\epsilon}}]\cdot(\bs{\eta},\bs{\eta}') = \mathcal{Q}^{\bs{\epsilon}}[\nu_{\bs{\eta}}^{\bs{\epsilon}},\nu_{\bs{\eta}'}^{\bs{\epsilon}}]\;,
\eea
where we have recognised the bilinear functional $\mathcal{Q}^{\bs{\epsilon}}$ introduced in \S~\ref{s1u}. By Hypothesis~\ref{h2}, we deduce that the Hessian at $\bs{\epsilon} = \bs{\epsilon}^{\star}$ is definite positive, and \eqref{comparis} actually shows that this remains true for $\bs{\epsilon}$ in a vicinity of $\bs{\epsilon}^{\star}$. \hfill $\Box$

\section{Asymptotics in the multi-cut regime}
\label{S8}

\subsection{Partition function}

We have gathered all the ingredients needed to analyse the partition function in the $(g + 1)$ regime when $g \geq 1$, decomposed as:
\beq
\label{euh}Z_{\mathsf{A}^{N}} = \sum_{N_0 + \cdots + N_{g} = N} \frac{N!}{\prod_{h = 0}^{g} N_h!} Z_{\mathsf{A}_{\mathbf{N}}}\;. \nonumber
\enq
Given the large deviations of filling fractions (Corollary~\ref{c3}), the expansion in $1/N$ of the partition function at fixed filling fractions $\bs{\epsilon}$ close to $\bs{\epsilon}^{\star}$ (Corollary~\ref{cc1}), the smooth dependence of the coefficients in $\bs{\epsilon}$ (Proposition~\ref{smoothen2}) and the positivity of the Hessian of $F^{[-2]} = - \mathcal{E}[\mu_{{\rm eq}}^{\bs{\epsilon}}]$ (Lemma~\ref{Hessp}), the proof of the asymptotic expansion of $Z_{\mathsf{A}^N}$ is identical to \cite[Section 8]{BGmulti}. To summarise the idea of the proof, one first restricts the sum in \eqref{euh} over vectors\footnote{When necessary, we identify $\bs{\epsilon}^{\star}$ with an element of $\mathbb{R}^{g}$ by forgetting the component $\epsilon_0^{\star}$.} $\bs{N} = (N_1,\ldots,N_g)$ such that $|\bs{N} - N\bs{\epsilon}^{\star}| \leq \sqrt{N \ln N}$ up to exponentially small corrections thanks to the large deviations of filling fractions, \textit{cf.} Corollary \ref{c3}. Since the filling fractions $\bs{\epsilon}$ kept in this sum are close to $\bs{\epsilon}^{\star}$, one can write down the $1/N$-expansion of each term in the sum. Then, one performs a Taylor expansion around $\bs{\epsilon}^{\star}$ of the coefficients of the latter expansion, and one can exchange the finite (although large) sum over $\bs{N}$ with the Taylor expansion while controlling the error terms. Eventually, one recognise the answer as the general term (in $\bs{N}$) of an exponentially fast converging series, so we can actually lift the restriction $\bs{N} - N\bs{\epsilon}^{\star}$ to sum over all $\bs{N} \in \mathbb{Z}^{g}$ up to an error $o(e^{-cN})$. The result can be expressed in terms of:
\begin{itemize}
\item[$\bullet$] the Theta function:
$$\Theta_{\bs{\gamma}}(\bs{v}|\bs{T}) = \sum_{\bs{m} \in \mathbb{Z}^{g}} \exp\Bigg(-\frac{1}{2}\,(\bs{m} + \bs{\gamma})\cdot\bs{T}\cdot(\bs{m} + \bs{\gamma}) + \bs{v}\cdot(\bs{m} + \gamma)\Bigg)\;,
$$
where $\bs{T}$ is a symmetric definite positive $g \times g$ matrix, $\bs{v} \in \mathbb{C}^{g}$ and $\bs{\gamma} \in \mathbb{C}^{g}\,\,\mathrm{mod}\,\,\mathbb{Z}^{g}$. 
\item[$\bullet$] The $\ell$-th order derivative of $F^{[k]}_{\bs{\epsilon}}$ with respect to the filling fractions. For a precise definition, we consider the canonical basis $(\bs{e}^{h})_{0 \leq h \leq g}$ of $\mathbb{R}^{g + 1}$, and introduce $\bs{\eta}^{h} = \bs{e}^{h} - \bs{e}^0$ for $h \in \intn{1}{g}$. Then, we can define the tensor of $\ell$-th order derivatives as an element of $(\mathbb{R}^{g})^{\otimes \ell}$:
$$
F^{[k],(\ell)}_{\bs{\epsilon}} = \sum_{1 \leq h_1,\ldots,h_{\ell} \leq g} \Big(\partial_{t_1 = 0}\cdots\partial_{t_{\ell} = 0}\,F^{[k]}_{\bs{\epsilon} + \sum_{i = 1}^{\ell} t_{i}\,\bs{\eta}^{h_{i}}}\Big)\,\bigotimes_{i = 1}^{\ell} \bs{e}^{h_i}\;.
$$
\end{itemize}
When necessary, we identify $\bs{\epsilon}^{\star}$ with an element of $\mathbb{R}^{g}$ by forgetting the component $\epsilon_0$.

\begin{theorem}
\label{T1}Assume Hypothesis~\ref{h1}, $T$ holomorphic in a neighbourhood of $\mathsf{A}^r$, and $\mu_{{\rm eq}}$ off-critical. Then, for any $k_0$, we have an asymptotic expansion of the form:
\bea
Z_{\mathsf{A}^{N}} & = & N^{(\beta/2)N + \gamma}\,\exp\Bigg(\sum_{k = -2}^{k_0} N^{-k}\,F^{[k]}_{\bs{\epsilon}^{\star}} + o(N^{-k_0})\Bigg) \nonumber \\
& & \times \Bigg\{\sum_{m \geq 0} \sum_{\substack{\ell_1,\ldots,\ell_m \geq 1 \\ k_1,\ldots,k_r \geq -2 \\ \sum_{i = 1}^m \ell_i + k_i  > 0}} \frac{N^{-\sum_{i = 1}^{m} (\ell_i + k_i)}}{m!}\Bigg(\bigotimes_{i = 1}^m \frac{F_{\bs{\epsilon}^{\star}}^{[k_i],(\ell_i)}}{\ell_i!}\Bigg)\cdot \nabla_{\bs{v}}^{\otimes (\sum_{i = 1}^m \ell_i)}\Bigg\}\Theta_{-N\bs{\epsilon}^{\star}}\Big( F_{\bs{\epsilon}^{\star}}^{[-1],(1)}\,\Big|\,F_{\bs{\epsilon}^{\star}}^{[-2],(2)}\Big)\;.
\eea
\end{theorem}

\subsection{Fluctuations of linear statistics}
\label{fluctulin}
We mention that, along the line of \cite{Smulticut} and \cite[Corollary 6.4]{BGmulti}, it is possible to show Theorem~\ref{T1}
while allowing $T$ to contain $1/N$ complex-valued contribution on $\mathsf{A}$ -- still under the assumptions that $T$ is analytic. Then, for any test function $\varphi$ holomorphic\footnote{At this point, the regularity of $\varphi$ can be weakened by going to Fourier space, see \cite[Section 6.1]{BGmulti} for details.} is a neighbourhood of $\mathsf{A}$, it follows for the fluctuations of the linear statistics:
\beq
X_N[\varphi] = \sum_{i = 1}^N \varphi(\lambda_i) - N\int \varphi(x)\,\dd\mu_{{\rm eq}}(x)\nonumber
\enq
that for any $s \in \mathbb{R}$, we have:
\beq
\mu_{\mathsf{A}_{N}}\big[e^{{\rm is}X_N[\varphi]}\big] \mathop{=}_{N \rightarrow \infty} e^{{\rm i}s\,M_1[\varphi] - s^2\,M_{2}[\varphi]}\,\frac{\Theta_{-N\bs{\epsilon}^{\star}}\Big(F^{[-1],(1)}_{\bs{\epsilon}^{\star}} + {\rm i}s\,\bs{w}[\varphi]\,\Big|\,F^{[-2],(2)}_{\bs{\epsilon}^{\star}}\Big)}{\Theta_{-N\bs{\epsilon}^{\star}}\Big(F^{[-1],(1)}_{\bs{\epsilon}^{\star}}\,\Big|\,F^{[-2],(2)}_{\bs{\epsilon}^{\star}}\Big)} \nonumber
\enq
with:
\bea
M_1[\varphi] & = & \Oint{\mathsf{S}}{} \frac{\dd x}{2{\rm i}\pi}\,\varphi(x)\,W_{1;\bs{\epsilon}^{\star}}^{[0]}(x)\;, \nonumber \\
M_2[\varphi] & = & \Oint{\mathsf{S}^2}{} \frac{\dd x_1\dd x_2}{(2{\rm i}\pi)^2}\,\varphi(\xi_1)\varphi(\xi_2)\,W_{2;\bs{\epsilon}^{\star}}^{[0]}(x_1,x_2)\;, \nonumber \\
\bs{w}[\varphi] & = & \sum_{h = 1}^g \Bigg(\Int{\mathsf{S}}{} \varphi(x)\,\dd\nu_{\bs{\eta}^h}^{\bs{\epsilon}^{\star}}(x)\Bigg)\,\bs{e}^h = \sum_{h = 1}^{g} \partial_{t = 0}\Bigg(\int \varphi(x)\,\dd\mu_{{\rm eq}}^{\bs{\epsilon} + t\bs{\eta}^h}(x)\Bigg)\,\bs{e}^h\;,
\eea
where we recall that $\eta^h={\bs e}^h-{\bs e}^0$ and $(\bs{e}^h)_{0\le h\le g}$ is the canonical basis of $\mathbb R^{g+1}$. 
We deduce a central limit theorem when the contribution of the Theta function vanishes, namely 
\begin{prop}\label{tcl} 
For the codimension $g$ space of test functions $\varphi$ satisfying $\bs{w}[\varphi] = \bs{0}$,  $X_N[\varphi]$ converges in law to a random Gaussian $\mathscr{G}(M_{1}[\varphi],M_{2}[\varphi])$ with mean $M_{1}[\varphi]$ and covariance $M_{2}[\varphi]$. 
\end{prop}
For test functions so that $\bs{w}[\varphi] \neq \bs{0}$, the ratio of Theta functions is present, and we recognise it to be the Fourier transform of the law of a random variable which is the scalar product of a deterministic vector $\bs{w}[\varphi]$ with $\bs{\mathscr{D}}(\bs{\gamma}_N,\bs{T}^{-1}[\bs{v}],\bs{T}^{-1})$, where $\bs{\mathscr{D}}$ is the sampling on $\bs{\gamma}_N + \mathbb{Z}^{g}$ of a random Gaussian vector with $g$ components, with covariance matrix $\bs{T}^{-1}$, and mean $\bs{T}^{-1}[\bs{v}]$. The values of the various parameters appearing here is:
\beq
\bs{T} = F^{[-2],(2)}_{\bs{\epsilon}^{\star}}\;,\qquad \bs{\gamma}_N = -N\bs{\epsilon}^{\star}\,\,\mathrm{mod}\,\,\mathbb{Z}^{g}\;,\qquad \bs{v} = F^{[-1],(1)}_{\bs{\epsilon}^{\star}}\;. \nonumber
\enq
Therefore, we can only say that, along subsequences of $N$ so that $-N\bs{\epsilon}^{\star}\,\,\mathrm{mod}\,\,\mathbb{Z}^{g}$ converges to a limit $\bs{\gamma}^*$, $X_N[\varphi]$ converges in law to the independent sum $$\mathscr{G}(M_{1}[\varphi],M_2[\varphi]) + \bs{w}[\varphi]\cdot\mathscr{D}(\bs{\gamma}^*,\bs{T}^{-1}[\bs{v}],\bs{T}^{-1})\;.$$

\subsection*{Acknowledgments}

We thank Tom Claeys for providing references on biorthogonal ensembles. The work of G.B. is supported by Swiss NSF $200021-143434$, Fonds Europ\'een S16905 (UE7 - CONFRA), a Forschungsstipendium of the Max-Planck-Gesellschaft, and the Simons Foundation. The work of A.G is supported by the Simons Foundation and the NSF 1307704. 
K.K.K. is supported by CNRS, the ANR grant ``DIADEMS'',
the Burgundy region PARI 2013 FABER grant "Structures et asymptotiques d'int\'{e}grales multiples"
and PEPS-PTI "Asymptotique d'int\'{e}grales multiples" grant. K.K.K would like to thank
Katedra Method Matematycznych dla Fizyki, Warsaw University, for their hospitality.

\appendix

\section{Inversion of integral operators}
In this section, we study integral operators on the real line which parallel the operators defined on the complex plane defined in 
Section \ref{S5}. This is necessary to obtain the concentration bounds of Section \ref{S3.5}.

\subsection{Reminder of Fredholm theory}
\label{SousSection Inversion Fredholm}
\label{App1}

Let $(\mathsf{X},\dd s)$ be a measured space, so that $|s|(\mathsf{X}) < +\infty$. Let $K$ be an integral operator on $L^{p}\big(\mathsf{X}, \dd s \big)$, $p\geq 1$ 
with a kernel $\mathscr{K}(x,y) = f(x) \widetilde{\mathscr{K}}(x,y)$ such that
 $\widetilde{\mathscr{K}} \in L^{\infty}\big( \mathsf{X}\times \mathsf{X} , \dd^2 s\big)$ 
and $f \in L^p\big( \mathsf{X},\dd s \big)$. Then, the series of multiple integrals 
\beq
\det\big[  \text{id} \; +  \;  K \big] \; = \; \sul{ n \geq 0 }{} \frac{  1  }{  n!  } 
\Int{ \mathsf{X} }{} \det_n \Big[ \mathscr{K}(\la_a, \la_b) \Big]\,\pl{a=1}{n}\dd s(\la_a)  \;. 
\label{definition series de Fredholm op general}
\enq
converges uniformly and defines the so-called Fredholm determinant associated with the integral operator $\text{id}+ K $.
The convergence follows by means of an application of Hadamard's inequality
\beq
\bigg| \Int{ X }{} \det_n \Big[ \mathscr{K}(\la_a, \la_b) \Big]\,\dd^n s(\la)  \bigg|  \; \leq \; 
n^{ \frac{n}{2} } \cdot \Norm{ \widetilde{\mathscr{K}} }_{L^{\infty}\big( X\times X , \dd^2 s \big) }^{n} \cdot 
\norm{f}_{L^{1}\big( \mathsf{X},\dd s\big) }^n \;.  \nonumber
\enq
This operator is invertible if and only if $\det\big[  \text{id} \; +  \;  K \big] \not=0$
and its inverse operator $\text{id} \; -  \;  \mc{R}_K $ is described in terms of the resolvent kernel
given by the absolutely convergent series of multiple integrals: 
\beq
\mathscr{R}_{K}(x,y) \; = \; \frac{ 1 }{  \det\big[  \text{id} +  K  \big]  }  
\sul{ n \geq 0 }{} \frac{ 1 }{ n! } \Int{ \mathsf{X} }{} 
 \det_{n+1} \left[  \ba{cc}   \mathscr{K}(x,y) &  \mathscr{K}(x,\la_b)  \\ 
 							\mathscr{K}(\la_a, y) & \mathscr{K} (\la_a,\la_b)		  \ea   \right]  \cdot  \pl{a=1}{n}\dd \mu(\la_a)  \;. 
\label{definition noyau resolvant cas generique}
\enq
In particular, such a description ensures that the inverse operator is continuous as soon as the operator 
$\text{id} \; +  \; K $ is injective.  See \cite{GohbergGoldbargKrupnikTracesAndDeterminants} for a more detailed discussion. 
Note that the resolvent kernel $\mathscr{R}_{K}(x,y) $ satisfies to the bounds
\beq
\Big|  \msc{R}_{K}(x,y) \Big| \; \leq \; |f(x)| 
 \cdot \sul{ n \geq 0}{} \f{ (n+1)^{ \frac{n+1}{2} } }{n!} \cdot 
\f{ \Norm{ \widetilde{\mathscr{K}}\, }_{L^{\infty}\big( \mathsf{X}\times \mathsf{X} , \dd^2 s \big) }^{n+1} \cdot 
\norm{f}_{L^{1}\big( \mathsf{X},\dd s \big) }^n }{ \det\big[  \text{id} \; +  \;  K \big]  } 
\; \leq \; c_{K} \cdot |f(x)| \;,   \nonumber
\enq
with $c_K$ a kernel $K$-dependent constant.

\subsection{Inversion of $\underline{\mathcal{T}}$}
\label{SousSection Inversion operateur T cal}
\label{App2}

Let $\underline{\mathcal{T}}$ be the integral operator 
\beq
\underline{\mathcal{T}}[\phi](x) \; = \; - \Int{\mathsf{A}}{} \big[  \be \ln |x-y| \; + \;  \tau(x,y)  \big]\,\phi(y)\, \dd y 
\; + \; \Int{\mathsf{A}^2}{} \big[  \be \ln |x-y| \; + \;  \tau(x,y)  \big]\,\phi(y)\,\dd x\,\dd y  \nonumber
\enq
with $\tau $ defined by 
\beq
\tau(x,y) \; = \; \Int{}{} \frac{T(x,y, \xi_3,\dots, \xi_{r})}{(r - 2)!} \pl{i=3}{r-2} \dd \mu_{\text{eq}}(\xi_i) \;. \nonumber
\enq
Let\footnote{The operators in underline letters -- like $\underline{\mathcal{L}}$ -- are the analog on the real axis of the operators -- like $\mathcal{L}$ -- defined in Section~\ref{SecP} on spaces of analytic functions, the correspondence being given by the Stieltjes transform. It should therefore not be surprising that the computations in this Appendix are parallel to those of Section~\ref{S5}.} $\underline{\mathcal{L}}$ and $\underline{\mathcal{P}}$ be the integral operators on $L^{p}(\mathsf{A}, \dd x )$ for $1<p<2$, 
with respective integral kernels  
\beq
\underline{\mathscr{L}}(x,y) \; = \; \frac{1}{\be\,\pi^2} \Fint{\mathsf{A}}{} \dd\xi\,\frac{\sigma_{\mathsf{A};+}^{1/2}(\xi)}{ \sigma_{\mathsf{A};+}^{1/2}(x) }\,\frac{ \partial_{\xi}\tau(\xi,y)}{x-\xi},\qquad 
\underline{\mathscr{P}}(x,y) \; = \;   \f{1}{{\rm i}\pi\,\sigma_{\mathsf{A};+}^{1/2}(x) }\,
\mathop{\mathrm{Res}}_{\xi \rightarrow \infty}\, \bigg( \f{\sigma_{\mathsf{A}}^{1/2}(\xi)}{(x-\xi)(\xi-y)} \bigg)\nonumber
 \enq
where we remind $\sigma_{\mathsf{A}}(x) = \prod_{h = 0}^{g} (x - a_h^-)(x - a_h^+)$. Let $\mf{X}$ be the set $\intn{1}{g} \cup \mathsf{A}$ endowed with the measure $\dd s$ given by 
the atomic measure on $\intn{1}{g}$ and the Lebesgue measure on 
$\mathsf{A}$.  We shall make the identification 
$L^{p}_0( \mf{X}, \dd s ) \simeq \Cx^{g} \oplus L^{p}_0(\mathsf{A}, \dd x )$
where 
\beq
L^{p}_0(\mathsf{A}, \dd x ) \; = \; \bigg\{ \phi \in L^p(\mathsf{A}, \dd x )  \; : \; \int_{ \mathsf{A} }{} \phi(x)\,\dd x \; = \; 0 \bigg\}  \nonumber
\;. 
\enq
We define similarly a subspace $W_{0}^{1,q}(\mathsf{A})$ of the Sobolev space $W^{1,q}(\mathsf{A}) \subseteq L^q(\mathsf{A})$, 
and introduce the space:
\beq
W_{0}^{1,q}(\mathfrak{X}) = \Big\{(\bs{v},\phi) \in L^q_0(\mathfrak{X},\dd s),\qquad \big(\max_{1 \leq k \leq g} |v_k| \big) + \p \phi \p_{q} < +\infty\Big\} \nonumber
\enq
Let $\underline{\mathcal{N}}$ be integral operator 
\beq
\underline{\mathcal{N}} \; : \; \left\{ \ba{ccc}  \Cx^{g} \oplus L^{p}_0(\mathsf{A}, \dd x )  & \longrightarrow & 
								\Cx^{g} \oplus L^{p}_0(\mathsf{A}, \dd x )  \vspace{2mm} \\
					\big( \bs{v} , \phi \big) & \longmapsto & \Big( \underline{\Pi}[\phi] - \bs{v} , \big(\underline{\mathcal{L}} - \underline{\mathcal{P}}\big)[f]
					 \, + \, \sigma_{\mathsf{A};+}^{-1/2}\cdot Q_{\bs{v}}\Big) \ea \right.  \nonumber
\enq
where 
\beq
\underline{\Pi} \; : \; \left\{ \ba{ccc}   L^{p}(\mathsf{A}, \dd x )  & \longrightarrow & \Cx^{g}   \\
				 \phi & \longmapsto &  \Big( \Int{\mathsf{A}_1}{}\underline{\mathcal{T}}[\phi](\xi)\,\dd \xi, \ldots ,
				 \Int{\mathsf{A}_g}{}\underline{\mathcal{T}}[\phi](\xi)\,\dd \xi  \Big) \ea \right. \nonumber
\;. 				 
\enq
and $Q_{\bs{v}}$ is the unique polynomial of degree $g-1$ such that \mbox{$\int_{\mathsf{A}_k} \dd\xi\,\sigma_{\mathsf{A};+}^{-1/2}(\xi)\,Q_{\bs{v}}(\xi) = v_k$} for any $k \in \intn{1}{g}$. 
\begin{prop}\label{theop}
Let $1 < p < 2$ and $q > 2p/(2 - p)$. The integral operator $\underline{\mathcal{N}} \, : \, L^{p}_0(\mf{X}, \dd s) \tend W^{1,q}_0(\mf{X}, \dd s)$ is compact. The operator $\mathrm{id} + \underline{\mathcal{N}}$ is bi-continuous with inverse $\mathrm{id} - \mathcal{R}_{\underline{\mathcal{N}}}$.  Furthermore, the inverse of $\underline{\mathcal{T}}$ is expressed by 
\beq
\underline{\mathcal{T}}^{-1}[f](x) \; = \; 	\Xi[f^{\prime}](\xi) \, - \, \sul{k = 1}{g} \mc{R}_{\msc{N}}(x,k) \Int{\mathsf{A}_k}{} f(\xi)\,\dd\xi
\, -\, \Int{ \mathsf{A} }{} \mc{R}_{\underline{\mathcal{N}}}(x,\xi) \cdot \Xi[f^{\prime}](\xi)\,\dd\xi \;. 
\label{ecriture inverse de operateur mathcal T}
\enq
where  
\beq
\Xi[f](x) \; = \; \f{1}{\be\,\pi^2} 
\Fint{ \mathsf{A} }{} \frac{  \sigma_{\mathsf{A};+}^{1/2}(x)\,f(\xi) }{  \sigma_{\mathsf{A};+}^{1/2}(\xi)\,(x-\xi) }\,\dd \xi 
 \;. \nonumber
\enq
As a consequence, $\underline{\mathcal{T}}^{-1}$ extends to a continuous operator 
$ \underline{\mathcal{T}}^{-1}\; : \;  W^{1,q}_0(\mathsf{A}, \dd x ) \tend  L^{p}_0(\mathsf{A}, \dd x )$.
\end{prop}

\noindent {\bf Proof.} We first establish that $\underline{\mathcal{L}}\; = \; L^{p}_0(\mathsf{A}, \dd x ) \tend L^{p}_0(\mathsf{A}, \dd x )$, defined for $1<p<2$, is compact. It follows from 
\beq
\underline{\mathscr{L}}(x,y) \; = \; \frac{1}{\be\,\pi^2} \Int{\mathsf{A}}{} \dd\xi\,\frac{ \sigma_{\mathsf{A};+}^{1/2}(\xi) }{ \sigma_{\mathsf{A};+}^{1/2}(x) }\,\frac{\partial_{\xi}\tau(\xi,y) -\partial_{\xi}\tau(x,y)}{x-\xi} 
\; + \; \frac{\partial_{x} \tau(x,y) }{2\be\,\pi^2\,\sigma_{\mathsf{A};+}^{1/2}(x)}\,\Oint{ \Ga(\mathsf{A}) }{} \f{\dd\eta\,\sigma_{\mathsf{A}}^{1/2}(\eta)}{x-\eta}\,\;, \nonumber
\enq
where $\Ga(\mathsf{A})$ is a contour surrounding $\mathsf{A}$ with positive orientation, that $\underline{\mathscr{L}}(x,y) \; = \; \underline{\widetilde{\mathscr{L}}}(x,y)\,\sigma_{\mathsf{A};+}^{-1/2}(x)$ for a continuous function $\underline{\widetilde{\mathscr{L}}}(x,y)$
on $\mathsf{A}^2$. Furthermore, the relation 
\beq
\Fint{ \mathsf{A} }{}  \f{ \dd\xi }{ \sigma_{\mathsf{A};+}^{1/2}(\xi) (x-\xi) }    \; = \; 0 \nonumber
\enq
ensures that $\underline{\mathcal{L}}$ stabilises $L^{p}_0(\mathsf{A}, \dd x )$. 
Taking into account that $\mathsf{A}$ is compact, there exists a sequence of continuous functions $(\Phi_n, \Psi_n)_{n \geq 1}$ on $\mathsf{A}$ such that 
\beq
\underline{\widetilde{\mathscr{L}}}^{\; [n]}(x,y) \; = \; \sul{m=1}{n}\Phi_m(x)\Psi_m(y)  \nonumber 
\enq
converges uniformly on $\mathsf{A}^2$ to $\underline{\widetilde{\mathscr{L}}}(x,y)$. Let $\underline{\mathcal{L}}^{[n]}$ be the integral operator  on $L^p_0(\mathsf{A}, \dd x )$ with the integral kernel $\underline{\mathscr{L}}^{[n]}(x,y) = \sigma_{\mathsf{A};+}^{-1/2}(x)\,\underline{\widetilde{\mathscr{L}}}^{\; [n]}(x,y)$.  It follows from H\"{o}lder inequality that the $L^p_0(\mathsf{A}, \dd x )$ operator norm satisfies
\beq
\big|\big|\big|\,\underline{\mathcal{L}} - \underline{\mathcal{L}}^{[n]}\,\big|\big|\big| \; \leq \;   
\ell(\mathsf{A})^{ \f{p-1}{p} }\,\norm{\sigma_{\mathsf{A};+}^{-1/2} }_{L^p(\mathsf{A})}\,
\Norm{  \underline{\widetilde{\mathcal{L}}} - \underline{\widetilde{\mathcal{L}}}^{[n]}}_{L^{\infty}(\mathsf{A}^2)} \;. \nonumber
\enq
As a consequence, $\underline{\mathcal{L}}$ is indeed compact. An analogous statement is readily established for $\underline{\mathcal{P}}$ and hence $\underline{\mathcal{N}}$. 
We now establish that $\text{id} +\underline{\mathcal{N}}$ is injective. Let 
$(\bs{v},\phi \big) \in \text{ker}(\text{id}+ \underline{\mathcal{N}})$. Then one has  
\beq
\Fint{ \mathsf{A} }{}\Int{ \mathsf{A} }{}   \f{ \underline{\msc{L}}(x,y) \phi(y) }{s-x} \dd y \; = \; 0
\quad \Longrightarrow \quad \Int{ \mathsf{A} }{} \phi(s) \cdot \dd s  \; = \; 0 \;. \nonumber
\enq
By going back to the definition of a principal value integral, one gets 
\begin{multline}
-\be \Fint{ \mathsf{A} }{} \f{ \Xi[\phi](\xi) }{x-\xi } \dd \xi  \; = \; 
 \Int{ \mathsf{A}  }{} \dd \eta \f{ \phi(\eta)\,\sigma_{\mathsf{A};+}^{1/2}(\eta) }{ (2{\rm i}\pi)^2 } \Oint{ \Ga(\mathsf{A}) }{} \f{ 2\,\dd\xi }{\sigma_{\mathsf{A}}^{1/2}(\xi) (x-\xi)(\eta-\xi)}
\; - \; \lim_{\eps_1, \eps_2\tend 0^+} \bigg\{ \Int{\mathsf{A}^2}{}  \f{\phi(\eta)\,\sigma_{\mathsf{A};+}^{1/2}(\eta) }{ \sigma_{\mathsf{A};+}^{1/2}(\xi)} \\
\times \bigg[ \f{1}{x-\xi+{\rm i}\eps_1}\Big( \f{1}{\eta-\xi-{\rm i}\eps_2} \; - \; \f{1}{\eta-\xi+{\rm i}\eps_2}   \Big) 
\; + \;  \f{1}{x-\xi+{\rm i}\eps_1}\Big( \f{1}{\eta-\xi-{\rm i}\eps_2} \; - \; \f{1}{t-\xi+{\rm i}\eps_2}   \Big) \bigg]\,\dd \eta \dd \xi \bigg\}
\; = \; \phi(x) \;. 
\nonumber
\end{multline}
This ensures that 
\beq
-\be \Fint{ \mathsf{A} }{} \f{ \underline{\msc{L}}(\xi,y) }{ x-\xi }\,\dd \xi \; = \; (\partial_{x}\tau)(x,y) \;. \nonumber
\enq
In its turn this leads to $\partial_{\xi}\underline{\mc{T}}[f](\xi) = 0$ by acting with the principal value operator on $(\mathrm{id} + \underline{\mathcal{L}})[f]$ and using that:
\beq
\Fint{ \mathsf{A} }{} \f{\dd\xi\,Q(\xi)}{ \sigma_{\mathsf{A};+}^{1/2}(\xi)\cdot( x-\xi)  } \; = \; 0  \nonumber
\enq
for any polynomial $Q$ of degree at most $g$.
%
%
 In other words, there exist constants $c_k$ such that $\underline{\mc{T}}[\phi](\xi) \; = \; c_k$ on $\mathsf{A}_k$, $k=0,\dots, g$. Since, furthermore, 
\beq
\Int{ \mathsf{A}_k }{} \underline{\mc{T}}[\phi](x)\,\dd x  \; = \; 0 \; ,  \quad k=1,\dots, g \qquad \text{and}\; \text{by} \;\text{definition} \qquad 
\Int{ \mathsf{A} }{} \underline{ \mc{T}}[\phi](x)\,\dd x  \; = \; 0 \nonumber
\enq
it follows that, in fact, $\underline{\mc{T}}[\phi](x) = 0$. Therefore, 
\beq
\int_{\mathsf{A}}{} \phi(x) \cdot \underline{\mc{T}}[\phi](x)\,\dd x \; = \; \mc{Q}[\nu_{\phi}] =0  \quad \text{with} \quad 
\nu_{\phi}=  \phi(x)\,\dd x\, \in \, \mc{M}^{0}(\mathsf{A}) \;.  \nonumber
\enq
In virtue of the strict positivity of the functional $\mc{Q}$, it follows that $\nu_{\phi}=0$, \textit{viz}. $\phi=0$. 
This implies, in its turn, that $Q_{\bs{v}} = 0$, \textit{ie}. $\bs{v}=0$. 

 \vspace{3mm}

We now focus on the invertibility of the operator $ \underline{\mc{T}}$.
Hence, assume that one is given $f \in  \underline{\mc{T}}[L^p_0( \mathsf{A}, \dd x )] \cap W^{1;q}_0( \mathsf{A}, \dd x)$ with 
$1 < p < 2$ and $q> \tf{ 2p}{(2-p)}$. In other words that the function 
$\phi \in L^p_0( \mathsf{A}, \dd x )$, $1 < p < 2$ is a solution to $ \underline{\mc{T}}[\phi] = f$  for the given 
$f \in W^{1;q}_0( \mathsf{A}, \dd x )$. Since, 
the principal value operator is continuous on $L^{p}(\mathsf{A}, \dd x)$, $1<p<2$, it follows that 
one can differentiate both sides of the equality leading to 
\beq
\Fint{ \mathsf{A} }{} \frac{ \phi(s) }{x-\xi}\,\dd \xi \; = \; -\frac{f^{\prime}(x)}{\beta} \; + \; 
\frac{1}{\be} \Int{ \mathsf{A} }{} \partial_{x}\tau(x,\xi)\,\phi(\xi)\,\dd\xi  \; \equiv \; F(x)\;. \nonumber
\enq
The function 
\beq
\kappa[\phi](z) \; = \;  \sigma_{\mathsf{A}}^{1/2}(z) \cdot  \Int{ \mathsf{A} }{} \frac{ \phi(y) }{ z-y } \cdot \frac{ \dd y }{2{\rm i}\pi}  \;.  \nonumber
\enq
is holomorphic on $\Cx \setminus \mathsf{A}$, admits $L^p(\mathsf{A})$ $\pm$ boundary values on $\mathsf{A}$, 
and has the asymptotic behaviour at infinity 
\beq
\kappa[\phi](z) \; = \; 
\underbrace{ \mathop{\mathrm{Res}}_{\xi \rightarrow \infty}\, 
 \f{\sigma_{\mathsf{A}}^{1/2}(\xi) }{ x-\xi } \Bigg(\Int{ \mathsf{A} }{} \frac{ \dd y }{2{\rm i}\pi}\,\frac{ \phi(y) }{ \xi -y }\Bigg)  
}_{ P[\phi](z)/2 }\; + \; O(1/z) \;. \nonumber
\enq
Furthermore,
it satisfies to the jump conditions
\beq
\kappa[\phi]_+(x) \; - \;  \kappa[\phi]_-(x) \; = \; \sigma_{\mathsf{A};+}^{1/2}(x)\,\frac{ F(x) }{ {\rm i}\pi   } 
\; , \; x \in \overset{\circ}{\mathsf{A}}\;. \nonumber
\enq
Thus,
\beq
\kappa[\phi](z) \; = \; \frac{P[\phi](z)}{2}
 \; + \;  \Int{ \mathsf{A} }{} \frac{\dd\xi\,F(\xi)\,\sigma_{\mathsf{A};+}(\xi) }{ 2\pi^2\,(z - \xi) } \; . \nonumber
\enq
Note that $P[\phi]$ is at most of degree $g-1$ since $\phi \in L^p_0(\mathsf{A}, \dd x)$. Finally, it follows from the equation
\beq
\kappa[\phi]_+(x) \; + \;  \kappa[\phi]_-(x) \; = \; - \sigma_{\mathsf{A};+}^{1/2}(x)\,\phi(x) \nonumber
\enq
that $\phi$ solves the regular integral equation 
\beq
 \Xi[f^{\prime}](x)
\; = \; (\mathrm{id} + \underline{\mathcal{L}} - \underline{\mathcal{P}})[\phi](x) \nonumber
\enq
As a consequence, for any $f \in \underline{\mc{T}}\big[ L^p_0(\mathsf{A}, \dd x) \big]\cap W^{1;q}(\mathsf{A})$, there exists $\phi$
solving 
\beq
(\mathrm{id}  + \underline{\mathcal{N}})\big[ \big(\bs{0},\phi\big)\big] \; = \; 
\Bigg( \int_{\mathsf{A}_1} f(x)\,\dd x , \ldots, \int_{\mathsf{A}_g} f(x)\,\dd x , \Xi[f^{\prime}] \Bigg) \nonumber
\enq
Since $(\mathrm{id} + \underline{\mathcal{N}})$ is bijective, $\phi$ is necessarily unique and given by \eqref{ecriture inverse de operateur mathcal T}. The continuity of $\underline{\mc{T}}^{-1}$ is then obvious. 

\newpage

\providecommand{\bysame}{\leavevmode\hbox to3em{\hrulefill}\thinspace}
\providecommand{\MR}{\relax\ifhmode\unskip\space\fi MR }
\providecommand{\MRhref}[2]{%
  \href{http://www.ams.org/mathscinet-getitem?mr=#1}{#2}
}
\providecommand{\href}[2]{#2}

\end{document}